\documentclass[prl,showpacs,twocolumn,floats,10pt,aps,citeautoscript,longbibliography,superscriptaddress]{revtex4-2}
\usepackage{dcolumn}
\usepackage{placeins} 
\usepackage{anyfontsize} 
\usepackage{microtype} 
\usepackage{graphicx} 
\usepackage{xspace} 
\usepackage{blindtext}
\usepackage[usenames,dvipsnames]{xcolor} 

\usepackage{xparse} 

\usepackage{algorithm}
\usepackage{algpseudocode}

\usepackage[normalem]{ulem} 
\definecolor{darkgreen}{rgb}{0,0.5,0}
\definecolor{purple}{rgb}{0.6,0,0.5}
\definecolor{orange}{rgb}{1,0.5,0}
\definecolor{darkred}{rgb}{.7,0,0}
\definecolor{darkblue}{rgb}{0,0,.6}
\definecolor{grey}{rgb}{.6,.6,.6}
\definecolor{dimgreen}{rgb}{0.2,0.7,0.2}

\def\br{\mathbf{r}}
\def\bk{\mathbf{k}}
\def\bp{\mathbf{p}}
\def\bq{\mathbf{q}}

\newcommand{\CircleWhitePC}{\protect\raisebox{-0.5mm}{\protect\includegraphics[width=0.019\linewidth]{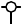}}}

\newcommand{\CircleGreyPC}{\protect\raisebox{-1mm}{\protect\includegraphics[width=0.019\linewidth]{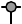}}}
\newcommand{\CircleWhiteC}{\protect\raisebox{-0.5mm}{\protect\includegraphics[width=0.037\linewidth]{Eq/CircleWhiteC}}}
\newcommand{\CircleGreyC}{\protect\raisebox{-1mm}{\protect\includegraphics[width=0.037\linewidth]{Eq/CircleGreyC}}}
\newcommand{\TriangleWhiteA}{\protect\raisebox{-0.5mm}{\protect\includegraphics[width=0.037\linewidth]{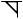}}}
\newcommand{\TriangleWhiteB}{\protect\raisebox{-0.5mm}{\protect\includegraphics[width=0.037\linewidth]{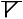}}}
\newcommand{\TriangleGreyA}{\protect\raisebox{-0.5mm}{\protect\includegraphics[width=0.037\linewidth]{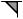}}}
\newcommand{\TriangleGreyB}{\protect\raisebox{-0.5mm}{\protect\includegraphics[width=0.037\linewidth]{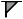}}}
\newcommand{\changed}[1]{{\color{black}{#1}}}

\newcommand{\jvdx}[1]{{\color{grey}{[\textbf{scrap]} #1}}}
\newcommand{\jvdomit}[1]{}

\usepackage{amsmath} 
\usepackage{amssymb}
\usepackage{mathrsfs} 
\usepackage{bbm}  
\renewcommand{\vec}[1]{{\boldsymbol{#1}}} 
\newcommand{\mr}[1]{\ensuremath{\mathrm{#1}}}
\newcommand{\mc}[1]{\ensuremath{\mathcal{#1}}}
\allowdisplaybreaks 
\usepackage{enumitem} 

\usepackage{scalefnt} 

\usepackage{tikz}
\usepackage{algorithm}
\usepackage{algpseudocode}

\usepackage[colorlinks, citecolor={blue!50!black}, urlcolor={blue!50!black}, linkcolor={red!50!black}]{hyperref}

\usepackage{scalerel,stackengine}  

\newcommand{\Eq}[1]{Eq.~\eqref{#1}}
\newcommand{\Eqs}[1]{Eqs.~\eqref{#1}}

\newcommand{\Sec}[1]{Sec.~\ref{#1}}

\newcommand{\Fig}[1]{Fig.~\ref{#1}}
\newcommand{\Figs}[1]{Figs.~\ref{#1}}

\def\Mt{\widetilde{M}}

\def\omegat{\widetilde{\omega}}

\def\main{{\mathrm{main}}} 
\def\satellite{{\mathrm{sat}}} 
\def\binned{{\mathrm{binned}}} 
\def\MPO{\mathrm{MPO}}
\def\lMPS{\mathrm{lMPS}}
\def\rMPS{\mathrm{rMPS}}
\def\numkr{\mathcal{N}_{\mathrm{kr}}}
\def\discrete{\mathrm{d}}
\def\broadened{\mathrm{b}} 

\def\doubleVonesite{\doubleV^\monesite}
\def\doubleVtwosite{\doubleV^\mtwosite}

\def\tU{{t\textrm{-}U}}
\def\teffU{{t_\textrm{eff}\textrm{-}U}}
\def\tUn{{t\textrm{-}t_n\textrm{-}U}}
\def\tUSigma{{t_\mathrm{eff}\textrm{-}U\textrm{-}\delta \Sigma}}
\newcommand\Vkq[1]{{V^{#1}_{\kk,\bq}}}

\newcommand{\tell}{{\tilde \ell}}
\newcommand{\ellp}{{\ell'}}
\newcommand{\tellp}{{{\tilde \ell}'}}
\newcommand{\neLL}{{\mathscr{L}}}  
\newcommand{\eLL}{{\mbox{\small$\mathscr{L}$}}}
\newcommand{\seLL}{{\scriptscriptstyle \! \mathscr{L}}}
\newcommand{\scripteLL}{{\scriptstyle \! \mathscr{L}}}

\def\Ab{{\overline{A}}}

\def\Bb{{\overline{B}}}

\def\Mt{\widetilde{M}}

\def\vt{\widetilde{v}}

\newcommand{\mi}{\mathrm{i}} 

\newcommand*{\ndots}{\kern-0.075em.\kern-0.05em.\kern-0.05em.}  
\newcommand*{\nidots}{.\kern-0.05em.\kern-0.05em.} 
\newcommand*{\ncdots}{\kern-0.15em\cdot\kern-0.2em\cdot\kern-0.2em\cdot\kern-0.15em}  

\newcommand{\bra}[1]{\ensuremath{\langle #1 |}}
\newcommand{\ket}[1]{\ensuremath{| #1 \rangle}}

\newcommand{\NKrylov}{N}
\newcommand{\pdag}{{\protect\vphantom{dagger}}}

\def\omegat{\widetilde{\omega}}

\newcommand{\Nkr}{\mathcal{N}_{\mathrm{kr}}}

\renewcommand{\Im}{\textrm{Im}}

\newcommand{\monesite}{\mathrm{1s}}
\newcommand{\mtwosite}{\mathrm{2s}}

\def\task{TaSK}

\def\Psig{\Psi_\mathrm{g}}
\def\Eg{E_{\mathrm{g}}} 
\def\PsiI{\Psi_0}
\def\Psie{\Psi_\mathrm{ex}}

\NewDocumentCommand{\doubleI}{O{}}{\mathbbm{1}_{#1}}
\NewDocumentCommand{\doubleIb}{O{}}{{\overline{\mathbbm{1}}_{#1}}}
\NewDocumentCommand{\doubleIk}{O{}}{\mathbbm{1}^\ks_{\! #1}}
\NewDocumentCommand{\doubleId}{O{}}{\mathbbm{1}^\ds_{\! #1}}
\NewDocumentCommand{\doubleIp}{O{}}{\mathbbm{1}^\ps_{\! #1}}
\NewDocumentCommand{\doubleV}{O{}}{\mathbbm{V}_{\! #1}}
\NewDocumentCommand{\doubleVk}{O{}}{\mathbbm{V}^\ks_{\! #1}}
\NewDocumentCommand{\doubleVd}{O{}}{\mathbbm{V}^\ds_{\! #1}}
\NewDocumentCommand{\doubleVp}{O{}}{\mathbbm{V}^\ps_{\! #1}}
\NewDocumentCommand{\doublev}{o}{{\mathbbm{v}_{#1}}}
\NewDocumentCommand{\doubleVb}{o}{{\overline{\mathbbm{V}}_{\! #1}}}
\NewDocumentCommand{\doubleVt}{o}{{\widetilde{\mathbbm{V}}_{\! #1}}}
\NewDocumentCommand{\doubleVh}{o}{\widehat{{\mathbbm{V}}_{\! #1}}}
\NewDocumentCommand{\doubleW}{o}{\mathbbm{W}_{\! #1}}
\NewDocumentCommand{\doubleWk}{o}{\mathbbm{W}^\ks_{\! #1}}
\NewDocumentCommand{\doubleWd}{o}{\mathbbm{W}^\ds_{\! #1}}
\NewDocumentCommand{\doubleWb}{o}{{\overline{\mathbbm{W}}_{\! #1}}}
\NewDocumentCommand{\doubleWt}{o}{{\widetilde{\mathbbm{V}}_{\! #1}}}
\NewDocumentCommand{\doubleWh}{o}{{\widehat{\mathbbm{V}}_{\! #1}}}

\def\kk{\mathbf{k}}

\def\D{{\scriptstyle {\rm D}}} 
\def\K{{\scriptstyle {\rm K}}} 
\def\P{{\scriptstyle {\rm P}}} 

\def\ds{{\scriptscriptstyle {\rm D}}}
\def\ks{{\scriptscriptstyle {\rm K}}}
\def\ps{{\scriptscriptstyle {\rm P}}}

\def\ps{{\scriptscriptstyle {\rm P}}}

\def\et{{\widetilde{e}}}

\NewDocumentCommand{\cor}{mod()}
{
	#1\IfValueTF{#2}{[#2]}{}\IfValueTF{#3}{(#3)}{}
}

\usepackage{mathtools}

\usepackage{xparse} 

\NewDocumentCommand{\xA}{
         O{fill=black}mm D<|{0} D|>{0} D//{0} O{}O{0}O{0}}  
{
	\draw [#1] (#2-00.135,#3) -- (#2,#3) -- (#2,-00.135+#3) -- cycle;
	\draw (#2-00.25-#4,#3) -- (#2-00.135,#3); 
	\draw (#2,#3) -- (#2+00.25+#5,#3); 
	\draw (#2,#3-00.135) -- (#2,#3-00.25-#6);
	\draw (#2+#8,#3+#9) node (X) {#7};
}

\NewDocumentCommand{\xAd}{
         O{fill=black}mm D<|{0} D|>{0} D//{0} O{}O{0}O{0}}  
{
	\draw [#1] (#2-00.135,#3) -- (#2,#3) -- (#2,#3+00.135) -- cycle;
	\draw (#2-00.25-#4,#3) -- (#2-00.135,#3); 
	\draw (#2,#3) -- (#2+00.25+#5,#3); 
	\draw (#2,#3+00.135) -- (#2,#3+00.25+#6);
	\draw (#2+#8,#3+#9) node (X) {#7};
}

\NewDocumentCommand{\xB}{
         O{fill=black}mm D<|{0} D|>{0} D//{0} O{}O{0}O{0}} 
{
	\draw [#1] (#2,#3) -- (#2+0.135,#3) -- (#2,-0.135+#3) -- cycle;
	\draw (#2-00.25-#4,#3) -- (#2,#3); 
	\draw (#2+0.135,#3) -- (#2+00.25+#5,#3); 
	\draw (#2,#3-0.135) -- (#2,#3-00.25-#6);
	\draw (#2+#8,#3+#9) node (X) {#7};
}

\NewDocumentCommand{\xBd}{
         O{fill=black}mm D<|{0} D|>{0} D//{0} O{}O{0}O{0}}  
{
	\draw [#1] (#2,#3) -- (#2+0.135,#3) -- (#2,#3+0.135) -- cycle;
	\draw (#2-00.25-#4,#3) -- (#2,#3); 
	\draw (#2+0.135,#3) -- (#2+00.25+#5,#3); 
	\draw (#2,#3+0.135) -- (#2,#3+00.25+#6);
	\draw (#2+#8,#3+#9) node (X) {#7};
}

\NewDocumentCommand{\xC}{
         O{fill=black}mm D<|{0} D|>{0} D//{0} O{}O{0}O{0}}  
{
	\draw [#1] (#2,#3) circle (0.065);
	\draw (#2-00.25-#4,#3) -- (#2-0.065,#3); 
	\draw (#2+0.065,#3) -- (#2+00.25+#5,#3); 
	\draw (#2,#3-0.065) -- (#2,#3-00.25-#6);
	\draw (#2+#8,#3+#9) node (X) {#7};
}

\NewDocumentCommand{\xCd}{
         O{fill=black}mm D<|{0} D|>{0} D//{0} O{}O{0}O{0}}  
{
	\draw [#1] (#2,#3) circle (0.065);
	\draw (#2-00.25-#4,#3) -- (#2-0.065,#3); 
	\draw (#2+0.065,#3) -- (#2+00.25+#5,#3); 
	\draw (#2,#3+0.065) -- (#2,#3+00.25+#6);
	\draw (#2+#8,#3+#9) node (X) {#7};
}

\NewDocumentCommand{\xW}{
         O{fill=black}mm D<|{0} D|>{0} D//{0} O{}O{0}O{0}}  
{
	\draw [#1] (#2-0.065,#3-0.065) rectangle (#2+0.065,#3+0.065);
	\draw (#2-00.25-#4,#3) -- (#2-0.065,#3); 
	\draw (#2+0.065,#3) -- (#2+00.25+#5,#3); 
	\draw (#2,#3-0.065) -- (#2,#3-00.25-#6);
	\draw (#2,#3+0.065) -- (#2,#3+00.25+#6);
	\draw (#2+#8,#3+#9) node (X) {#7};
}

\NewDocumentCommand{\lcurl}{mmm  O{}O{0}O{0} D<>{0}}   
{
	\draw (#1,#3) edge[out=180,in=-90] (#1-0.1+#7,#2*0.5+#3*0.5);
	\draw (#1-0.1+#7,#2*0.5+#3*0.5) edge[out=90,in=180] (#1,#2);
	\draw (#1+#5,0.5*#3+0.5*2+#6) node (X) {#4};
}

\NewDocumentCommand{\rcurl}{mmm  O{}O{0}O{0} D<>{0}}   
{
	\draw (#1,#3) edge[out=0,in=-90] (#1+0.1+#7,#2*0.5+#3*0.5);
	\draw (#1+0.1+#7,#2*0.5+#3*0.5) edge[out=90,in=0] (#1,#2);
	\draw (#1+#5,0.5*#2+0.5*#3+#6) node (X) {#4};
}

\NewDocumentCommand{\effL}{O{fill=black} mmm  O{$\,$}O{0}O{0} D<>{0}}   
{
	\draw (#2,#4) edge[out=180,in=-90] (#2-00.2+#8,#3*0.5+#4*0.5-0.1);
	\draw (#2-00.2+#8,#3*0.5+#4*0.5+0.1) edge[out=90,in=180] (#2,#3);
	\draw [#1](#2-00.2+#8,#3*0.5+#4*0.5+0.1) -- 
	(#2-00.2+#8,#3*0.5+#4*0.5-0.1) -- (#2-00.1+#8,#3*0.5+#4*0.5) -- cycle;
	\draw (#2-00.2+#8,#3*0.5+#4*0.5) -- (#2,#3*0.5+#4*0.5);
	\draw (#2+#6,#4+#7) node (X) {#5};
}

\NewDocumentCommand{\effR}{O{fill=black} mmm  O{$\,$}O{0}O{0} D<>{0}}   
{
	\draw (#2,#4) edge[out=0,in=-90] (#2+00.2+#8,#3*0.5+#4*0.5-0.1);
	\draw (#2+00.2+#8,#3*0.5+#4*0.5+0.1) edge[out=90,in=0] (#2,#3);
	\draw [#1] (#2+00.2+#8,#3*0.5+#4*0.5+0.1) -- 
	(#2+00.2+#8,#3*0.5+#4*0.5-0.1) -- (#2+00.1+#8,#3*0.5+#4*0.5) -- cycle;
	\draw (#2+00.2+#8,#3*0.5+#4*0.5) -- (#2,#3*0.5+#4*0.5);
	\draw (#2+#6,#4+#7) node (X) {#5};
}

\makeatletter
\def\maketitle{
\@author@finish
\title@column\titleblock@produce
\suppressfloats[t]}
\makeatother

\def\maintitle{Tangent-space Krylov computation of real-frequency spectral functions: \\
Influence of density-assisted hopping on 2D Mott physics}

\begin{document} 

\title{\maintitle}
\author{Oleksandra Kovalska}
\affiliation{Arnold Sommerfeld Center for Theoretical Physics, 
Center for NanoScience,\looseness=-1\,  and 
Munich Center for \\ Quantum Science and Technology,\looseness=-2\, 
Ludwig-Maximilians-Universit\"at M\"unchen, 80333 Munich, Germany}
\author{Jan von Delft}
\affiliation{Arnold Sommerfeld Center for Theoretical Physics, 
Center for NanoScience,\looseness=-1\,  and 
Munich Center for \\ Quantum Science and Technology,\looseness=-2\, 
Ludwig-Maximilians-Universit\"at M\"unchen, 80333 Munich, Germany}
\author{Andreas Gleis}
\affiliation{Department of Physics and Astronomy, Rutgers University, Piscataway, NJ 08854, USA}

\begin{abstract}
\begin{center}
(Dated: \today)
\end{center}

We present a tangent-space Krylov method for efficient computation of zero-temperature real-frequency spectral functions on top of ground-state matrix product states obtained from the density matrix renormalization group. It relies on projecting the Hamiltonian to the tangent space of the ground-state
matrix product states, where resolvents can be efficiently represented using Krylov space techniques. This
enables direct computations of excitation energies and the corresponding spectral weights. We demonstrate
the accuracy and efficiency of the tangent-space Krylov approach by showcasing spectral data for various
models. These include the 1D Haldane-Shastry and Heisenberg models as benchmarks. As an interesting application, we study the Hubbard model on a cylinder at half filling, augmented by a density-assisted hopping~(DAH) term. We find that DAH leads to particle-hole asymmetric single-particle mobilities and lifetimes in the resulting Mott insulator, and identify the responsible scattering processes.
Furthermore, we find that DAH influences the dispersion of Green's function zeros beyond its range, which has a frustrating effect on the Mott insulator studied here.
\\
\\
\noindent
\vspace{-2mm}
DOI: \vspace{-2mm}

\end{abstract}

\maketitle

\textit{Introduction.---} 
Numerical methods based on matrix product states (MPS) form a robust toolbox for studying strongly correlated systems.
A prime example is the density matrix renormalization group (DMRG)~\cite{White1992, Schollwoeck2011} 
for ground-state searches.
Numerous MPS tools are also available 
for computing ground-state dynamical correlators of the form 
$S[O_\bk](\omega) = \int d t \, e^{\mi \omega t}
\langle \Psi_\bk(0)|\Psi_\bk(t) \rangle,
$ 
with $|\Psi_\bk(t) \rangle = e^{-\mi  (H- \Eg) t} O_\bk^{\pdag} |\Psig \rangle$, 
relevant for experiments that probe
dynamical properties.
One strategy is to time-evolve $|\Psi_\bk(t) \rangle$  explicitly~\cite{Vidal2004,White2004,Daley2004,Barthel2013,Karrasch2013,Paeckel2019,Grundner2024,backwards_evolution}; but the inevitable growth of entanglement makes it difficult to reach the long times needed to access low frequencies. In a sense, 
when computing $|\Psi_\bk(t) \rangle$ 
 in full generality 
one works harder than necessary,
since only its overlap with $|\Psi_\bk(0) \rangle$
is required.

An alternative strategy is to compute dynamical 
correlators directly in frequency space~\cite{White1999,Jeckelmann2002,Raas2005,vonDelftHolzner2011,Wolf2014,Wolf2015,Alvarez2016,Alvarez2022,Paeckel2025}, 
where 
\begin{align}
    S[O^{\pdag}_\bk](\omega)
    = \langle \Psig |O^{\dag}_\bk\, \delta(\omega - H + 
    \Eg)\, O_\bk^{\pdag} |\Psig \rangle .
    \label{eq:SpectralFunction1}
\end{align}
An early approach, due to Hallberg~\cite{Hallberg1995}, employed a Lanczos-based~\cite{Koch2011} continued fraction expansion. It motivated further work~\cite{Dargel2011,Dargel2012,Wang2025,Dektor2025} on formulating a Lanczos scheme in the MPS context -- a task that proves challenging due to structural limitations. The core steps of any Lanczos implementation are applying the Hamiltonian to a quantum state, then orthogonalizing the result against other states. In the MPS framework, these steps increase the bond dimension and leave the manifold of MPSs of fixed bond dimension, which lacks a vector-space structure. To keep computational costs manageable, they must therefore be followed by compression. This introduces truncation errors that accumulate over iterations, leading to a global loss of orthogonality and convergence issues. These can be mitigated by reorthogonalization~\cite{Dargel2012} or restarting procedures~\cite{Wang2025}.

In this Letter, we introduce a novel tangent-space Krylov~(TaSK) method, which restricts computations to the \textit{tangent space} of the MPS manifold at the ground state, instead of the MPS manifold itself. Since the former has a \textit{vector-space} structure, it enables us to formulate a Lanczos scheme devoid of orthogonality issues. Our approach uses
a direct finite-size MPS adaptation~\cite{VanDamme2021,Gleis2022} of the quasiparticle ansatz~\cite{Ostlund1995,Rommer1997,Haegeman2012,Haegeman2013,Verstraete2013,Haegeman2019,Tu2021,White2026}, which 
has been successfully employed to compute spectral functions on top of uniform MPSs. We further establish the 2-site energy variance~\cite{Hubig2018,Gleis2022} as a novel means to estimate the projection errors due to the tangent space restriction.

We benchmark TaSK on the Haldane–Shastry and Heisenberg chains, finding excellent agreement with analytical results for the dynamical structure factor. As a challenging application offering new physical insight, we consider the Hubbard model on a cylinder, augmented by a density-assisted hopping (DAH) term~\cite{Hirsch1989a,Hirsch1989b,Schuttler1992,Simon1993,Aligia2000,Aligia2007,Yang2015,Spalek2017,Chen2023,White2023,Sun2024}. The spectral functions and self-energies obtained for the Mott insulating phase exhibit strong particle-hole asymmetry, which is absent in a Hartree-Fock treatment of DAH. We identify scattering processes causing the mobility and lifetime asymmetries and find an emergent next-nearest neighbor contribution to the Mott-pole dispersion.

\textit{Method.---}
Consider a Hamiltonian $H$ of an $\eLL$-site system, represented as a matrix product operator~(MPO)
\begin{align}
H =
     \raisebox{-3.0mm}{\includegraphics[width=0.75\linewidth]{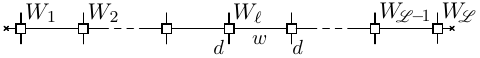}}  \, . &
\end{align}
Here, $d$ denotes the dimension of the local Hilbert space and $w$ the dimension of the internal virtual bond.
Its ground-state wavefunction $\Psig$,
\changed{found, e.g., via} DMRG~\cite{f:TaskInput}, can be expressed in site-canonical MPS form as
{\setlength{\abovedisplayskip}{1pt}
\setlength{\belowdisplayskip}{2pt}
\begin{align}
	\Psig &  =  \!\!
	\raisebox{-5.5mm}{\includegraphics[width=0.73\linewidth]{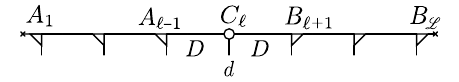}} \, ,
\end{align}}%
with internal bond dimension $D$, orthogonality center $C_{\ell}$, 
and left and right isometries $A_{\ell'} \,  (\TriangleWhiteA)$ and $B_{\ell''} \, (\TriangleWhiteB)$. The latter, 
together with their orthogonal complements, $\Ab^\pdag_{\ell'} (\TriangleGreyA)$ and $\Bb_{\ell''}^\pdag (\TriangleGreyB)$, can be used to explicitly construct \cite{Gleis2022} the tangent space $\doubleVonesite$ of the ground-state MPS, spanned by one-site variations of $\Psig$, 
and the one-site projector
{\setlength{\abovedisplayskip}{1pt}
\setlength{\belowdisplayskip}{4pt}
\begin{align}
\label{eq:P_1perp}
   \mathcal{P}^{1\perp} &  = \, \sum_{\ell = 1}^{\scripteLL}
\! \raisebox{-5mm}{
 \includegraphics[width=0.317\linewidth]{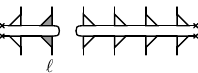}} 
\; ,
\end{align}}%
whose image $\doubleV^{1\perp} \!\! = \!\!\doubleVonesite/\mathrm{span}\bigl(\ket{\Psig}\bigr)$ is the space of MPS tangent vectors orthogonal to $\ket{\Psig}$. Its dimension, $\mr{dim} \, \doubleV^{1\perp}$, is equal to the number of independent variational parameters of $\Psig$ and scales as $\mc{O}(\eLL D^2 d)$.
Wavefunctions of states in $\doubleV^{1\perp}$ have the form
{\setlength{\abovedisplayskip}{1pt}
\setlength{\belowdisplayskip}{2pt}
\begin{align}
        \Psi^{1\perp} & = 
        \sum_{\ell=1}^{\scripteLL}  \!\!
        \raisebox{-0.45\height}{
        \includegraphics[width=0.322\linewidth]{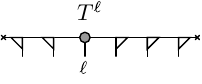}}
        \, , \;\; \mr{with} \; \;
        \raisebox{-5.8mm}{\includegraphics[width=0.14\linewidth]{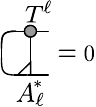}} \;.
        \label{eq:ExnAnsatz} 
\end{align}}%
In practice, the sum in Eq.~\eqref{eq:ExnAnsatz} is never performed explicitly, and the gauge choice constraint on $T_{\ell}$ ensures that all terms in this sum are mutually orthogonal. As a result, overlaps, sums, and scalar multiplication of tangent vectors can be performed as operations involving only the tensors $T_{\ell}$, at a computational cost of $\mc{O}(\eLL D^2 d)$.

The vector-space structure of $\doubleV^{1\perp}$ enables a controlled and efficient computation of spectral functions 
of the form \eqref{eq:SpectralFunction1}
via the Lanczos algorithm~\cite{Koch2011}, provided that we project all involved states to $\doubleV^{1\perp}$.
This projection is our main approximation (further motivated in Sec.~\ref{sec:SM_Method} of the Supplemental Material (SM)~\cite{supplement})\nocite{Gleis2022,Hubig2015,Gleis2023,Gleis2025b,vonDelftHolzner2011,Dickhoff2004,Wagner2023,Kugler2022,Pelz2025,Viswanath1994,Wagner2023}. Concretely, we use
\begin{align}
S\bigl[O^{\pdag}_\bk\bigr](\omega)
 &\approx \bra{\PsiI} \, \delta(\omega - H^{1\perp})\, \ket{\PsiI} + |\langle O_\bk \rangle|^2 \delta(\omega) \, ,
    \label{eq:SpectralFunction_TaSK}
\end{align}
with $\langle O_\bk \rangle = \bra{\Psig} O_\bk \ket{\Psig}$. Here, we have introduced the projected Hamiltonian~\cite{H_subtract} $H^{1\perp} = \mathcal{P}^{1\perp}(H-\Eg )\mathcal{P}^{1\perp}$
and initial state $\ket{\PsiI} = \mc{P}^{1\perp} O_\bk^{\pdag}\ket{\Psig}$. Within the scope of this work, $O_{\bk}$ is a sum of local operators
with $O_\bk^{\pdag}\ket{\Psig} \in \doubleV^{1s}$,
thus the initial state projection is exact \cite{f:CurrentOperator}.

Equation~\eqref{eq:SpectralFunction_TaSK} can be evaluated efficiently by means of a Lanczos scheme,
which iteratively generates an orthonormal basis of the $(\mathcal{N}_{\mathrm{kr}}+1)$-dimensional Krylov space 
\begin{align}
\mathcal{K}\left(\PsiI\right) 
&= \mathrm{span}\bigl\{|\PsiI\rangle,H^{1\perp}|\PsiI\rangle, \dots, \bigl(H^{1\perp})^{\Nkr}|\PsiI\rangle\bigr\} \, .
\label{eq:Krylov_space}
\end{align}
The resulting tridiagonal representation of $H^{1\perp}$ in $\mathcal{K}\left(\PsiI\right)$ is diagonalized to obtain the approximate eigenpairs 
$\bigl(\omega_{\alpha},|\Psi^{1\perp}_\alpha\rangle\bigr)$, which we then insert in Eq.~\eqref{eq:SpectralFunction_TaSK}, 
\setlength{\abovedisplayskip}{4pt}
\setlength{\belowdisplayskip}{2pt}
\begin{align}
\label{eq:SpectralFunction_TaSK_Nkr}
S\bigl[O^{\pdag}_\bk\bigr](\omega) \approx \sum_{\alpha = 0}^{\Nkr} S_{\alpha} \, \delta(\omega - \omega_{\alpha}) + |\langle O_\bk \rangle|^2 \delta(\omega) \, ,
\end{align}
with $S_{\alpha} = |\langle \Psi^{1\perp}_{\alpha} | \PsiI \rangle |^2$. 
For $\Nkr$ large enough, the set of real discrete
frequencies and spectral weights $\{\omega_\alpha,S_{\alpha}\}$ becomes an arbitrarily accurate approximation of Eq.~\eqref{eq:SpectralFunction_TaSK} and can be broadened if desired (see~\ref{subsec:CFE_discrete}). We find that $\Nkr < 150$ is sufficient for the cases studied in this work.

The main computational cost of \task\ is the application of $H^{1\perp}$ to a tangent vector, done once per Krylov step. 
It scales as of $\mc{O}(\eLL D^{3} d w)$, just as \textit{one} single-site DMRG or time-dependent variational principle (TDVP) sweep. In practice, \task\ is considerably cheaper than TDVP for computing spectral functions (see Sec.~\ref{sec:Comparison_to_TDVP} of (SM)~\cite{supplement}).

The projection of the Hamiltonian to $\doubleV^{1\perp}$, 
$H \to H^{1\perp}$, is the main source of error in \task. The distance of eigenstates $\ket{\Psi^{1\perp}_{\alpha}}$ of $H^{1\perp}$ from eigenstates of $H$ can be quantified through
their variance $\Delta_\alpha =  \| (H - E_{0} - \omega_{\alpha}) \ket{\Psi^{1\perp}_{\alpha}}\|^2$.
While evaluation of the full variance is challenging, its 2-site contribution~\cite{Hubig2018,Gleis2022},
{\setlength{\abovedisplayskip}{-1pt}
\setlength{\belowdisplayskip}{2pt}
\begin{align}
\label{eq:EnergyVariance}
\Delta_\alpha^{2\perp}  \! & = \| \mathcal{P}^{2\perp} H \ket{\Psi_\alpha^{1\perp}} \|^2 \!= \hspace{-1.3em}\raisebox{-0.45\height}{
\includegraphics[width=0.5\linewidth]{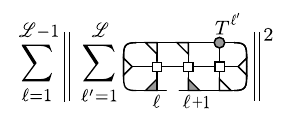}}\!\!\!\!,
\end{align}}%
can be computed at reasonable cost (\Sec{subsec:ErrorMeasure} of SM~\cite{supplement}). We will use $\delta_\alpha = (\Delta_\alpha^{2\perp})^{1/2}$ as error estimate for $\omega_\alpha$. \changed{Since $\mathrm{dim} \doubleV^{1\perp} \sim D^2$, increasing $D$ reduces the projection error.}

\textit{Spin Chains.---}
As the first benchmark, we consider the one-dimensional $\mathrm{SU}(2)$ Haldane-Shastry model (HSM) with periodic boundary conditions 
\begin{align}
    H_{\mathrm{HSM}} = \sum_{\ell<\ellp\leq\neLL} \frac{\pi^2\vec{\mathrm{S}}_\ell\cdot\vec{\mathrm{S}}_\ellp}{\eLL^2\sin^2{\frac{\pi}{\seLL}(\ell-\ellp)}}\, .
\end{align}
The $1/r^2$ spin-exchange yields a Luttinger liquid of free two-spinon excitations~\cite{Haldane1993,YamamotoPRL2000}. Unlike magnons, these fractionalized excitations are not naturally represented within $\doubleV^{1\perp}$, since this space is spanned by MPS that differ from the ground state at most on a single site.
Thus, HSM is a challenging test for TaSK, despite its integrability.

The ground state MPS is obtained using state-of-the-art single-site DMRG with controlled bond expansion \cite{Gleis2023}. Using the QSpace tensor library~\cite{Weichselbaum2012,Weichselbaum2020,Weichselbaum2024}, we exploit the non-Abelian $\mathrm{SU}(2)$ symmetry, retaining at most $D^{\ast} = 700$ multiplets. We consider the dynamical structure factor $S_k(\omega) = S\bigl[\mathrm{S}^{z}_k\bigr](\omega)$,
where $\mathrm{S}^{z}_{k} = \frac{1}{\sqrt{\scripteLL}} \sum_{\ell=1}^\scripteLL \exp{(\mathrm{i}\ell k)} \mathrm{S}^{z}_\ell$ is a spin-wave operator, with $k = m\pi/\eLL, m \in \mathbb{Z} \ \textrm{mod} \ \eLL$.
 \begin{figure}[t]
 \includegraphics[width=\linewidth]{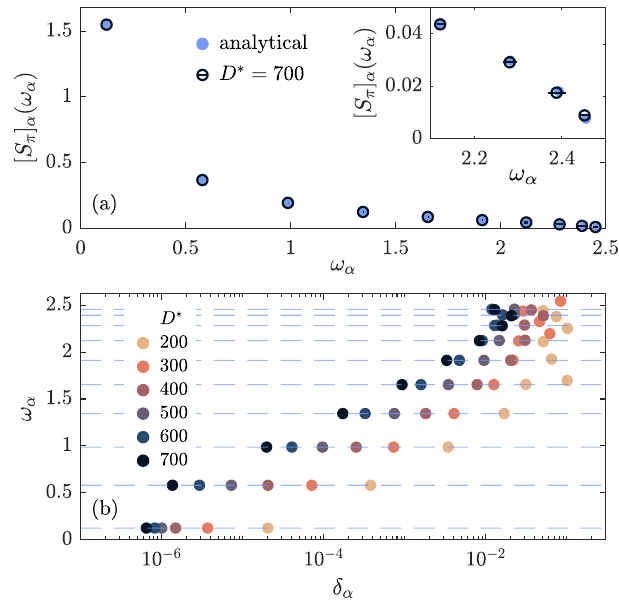}
 \vspace{-6mm}
\caption{
\label{fig:HaldaneShastry}
Dynamical structure factor of the $\eLL = 40$
Haldane-Shastry model at $k=\pi$ and $\Nkr=30$. 
(a) Comparison between analytical values (blue dots) and the TaSK result (black circles) for $D^* = 700$. Horizontal bars depict 
$\delta_\alpha$ as error estimate for $\omega_\alpha$. Inset: zoom to high frequencies. (b) Convergence trend across data sets 
with increasing $D^*$, relative to 
$\delta_\alpha$. Dashed lines indicate the analytical values from (a).}
\vspace{-6mm}
 \end{figure}
 
Figure~\ref{fig:HaldaneShastry}(a) displays $S_\pi(\omega)$ for $\eLL = 40$, computed with \task\ (black circles) and from exact formulas~\cite{YamamotoPRL2000, YamamotoPhysSocJ2000,Li2024} (blue dots). We use $\Nkr = 30$ (see Sec.~\ref{sec:Convergence} of the SM~\cite{supplement} for a convergence analysis).
The analytical and numerical results agree very well. Slight deviations appear only at high frequencies (inset), but remain within the error bars, confirming the reliability of $\delta_\alpha$ as an error estimate.

Figure~\ref{fig:HaldaneShastry}(b) demonstrates the convergence 
of low-lying excitations: their error estimates $\delta_\alpha$ 
decrease monotonically with increasing $D^*$. Thus, as expected, the projection error is reduced as $\mr{dim} \, \doubleV^{1\perp} \propto D^2$ grows. 

Next, we benchmark \task\ for the antiferromagnetic spin-$\tfrac{1}{2}$ Heisenberg chain $H_{\mathrm{Heis}} = \sum_{\ell = 1}^{\scripteLL-1} \vec{\mathrm{S}}_\ell \cdot \vec{\mathrm{S}}_{\ell+1}$. In contrast to the HSM, the spinons in this model interact, leading to multi-spinon contributions in $S_k(\omega)$ ($\sim 27\%$ of the total spectral weight~\cite{Caux2006}) and
a broad distribution of spectral weight over the entire frequency range. This poses a particular challenge for Lanczos-based methods~\cite{Hallberg1995,White1999, Dargel2012} as each new peak requires an additional iteration. 

In Fig.~\ref{fig:HeisenbergPowerLaw} we present $S_\pi(\omega)$ computed for $\eLL = 64$ and $\eLL = 128$ and compared with Bethe ansatz results~\cite{Caux2006} in the thermodynamic limit. For both system sizes, we keep $D^* = 512$ $\mathrm{SU}(2)$ multiplets and define the spin-wave operator $\mathrm{S}^{z}_{k}$ as above. To approximate $S_\pi(\omega)$ in the thermodynamic limit, the two data sets of discrete frequencies $\omega_\alpha$ are collapsed onto a single curve using the rescaling scheme described in Sec.~IV (D) of Ref.~\onlinecite{vonDelftHolzner2011} after preliminary binning (see Sec.~\ref{sec:Convergence} of Ref.~\onlinecite{supplement}).
The resulting data points closely follow the Bethe-ansatz curve corresponding to the full $2+4$-spinon contribution. 

\textit{Density-assisted hopping in the 2D Hubbard model.---} 
Due to its efficiency, \task\ can readily obtain spectra of interacting electrons on a $10 \times 4$ cylinder, where ground-state calculations are already challenging. Our goal is to study how density-assisted hopping~(DAH) influences the physics of the 2D Hubbard model at half-filling, where it describes a Mott insulator---the parent system from which superconductivity arises upon doping. Hirsch has argued that DAH is pivotal for high-$T_c$ superconductivity in cuprates ~\cite{Hirsch1989a,Hirsch1989b}, and understanding its physical consequences is an active area of research~\cite{Aligia2000,Aligia2007,Simon1993,Spalek2017,Sun2024,Chen2023,Schuttler1992,Yang2015,White2023}.

\begin{figure}[t]
 \includegraphics[width=\linewidth]{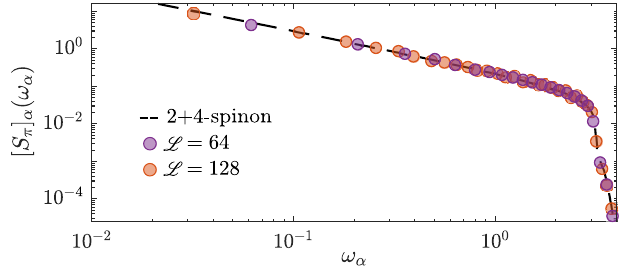} \vspace{-7mm}
 \caption{
 \label{fig:HeisenbergPowerLaw}
Dynamical structure factor of the Heisenberg chain at $k = \pi$. \changed{We have used $\Nkr = 100$ and $\Nkr = 140$ for $\eLL = 64$ and $\eLL = 128$, respectively}. Dashed line: the exact Bethe ansatz result for the full $2+4$-spinon contribution from Ref.~\onlinecite{Caux2006}. \changed{For other values of $k$, 
see Figs.~\ref{fig:HeisenbergPineapple} 
and \ref{fig:HeisenbergPihalf} in the SM \cite{supplement}.
}}
\vspace{-5mm}
 \end{figure}

A recent DMRG-based downfolding study of the Emery model~\cite{White2023} yielded a single-band Hubbard model augmented by a DAH term,
$H=H_{\tU} + H_n$, with
\begin{subequations}
\label{eq:H_full}
\begin{align}
     H_\tU & = -t \sum_{\langle \br \br' \rangle,\sigma} \bigl(c^\dagger_{\br \sigma}c^{\pdag}_{\br'\sigma} + \mathrm{h.c.} \bigr) + U\sum_\br n_{\br \uparrow}n_{\br \downarrow},
     \\    
     H_{n} & = -t_n \sum_{\langle \br \br'\rangle,\sigma} \bigl(c^\dagger_{\br \sigma}c^{\pdag}_{
    \br' \sigma} + \mathrm{h.c.} \bigr) (n_{\br \bar{\sigma}} + n_{\br' \bar{\sigma}}).
\end{align}
\end{subequations}
Here, $c^\dagger_{\br \sigma}$ is a hole creation operator, $n_{\br \sigma} = c^{\dag}_{\br\sigma} c^{\pdag}_{\br \sigma}$ the hole-number operator, $\bar{\sigma}$ denotes the spin projection opposite to $\sigma$, and $\langle \br \br' \rangle$ denotes nearest neighbors. The DAH hopping rate was found~\cite{White2023} to be substantial, $t_n \simeq 0.6 t$, significantly enhancing hole mobility \textit{and} pairing. The latter finding cannot be understood within a simple Hartree-Fock (HF) treatment of DAH, which effectively replaces $t$ by $t_{\mr{eff}} = t + 2 t_n \langle n_{\bar{\sigma}} \rangle$, thereby remaining a pure Hubbard model that does not exhibit superconductivity~\cite{Qin2020}.

\begin{figure*}[t]
\includegraphics[width=1.025\textwidth]{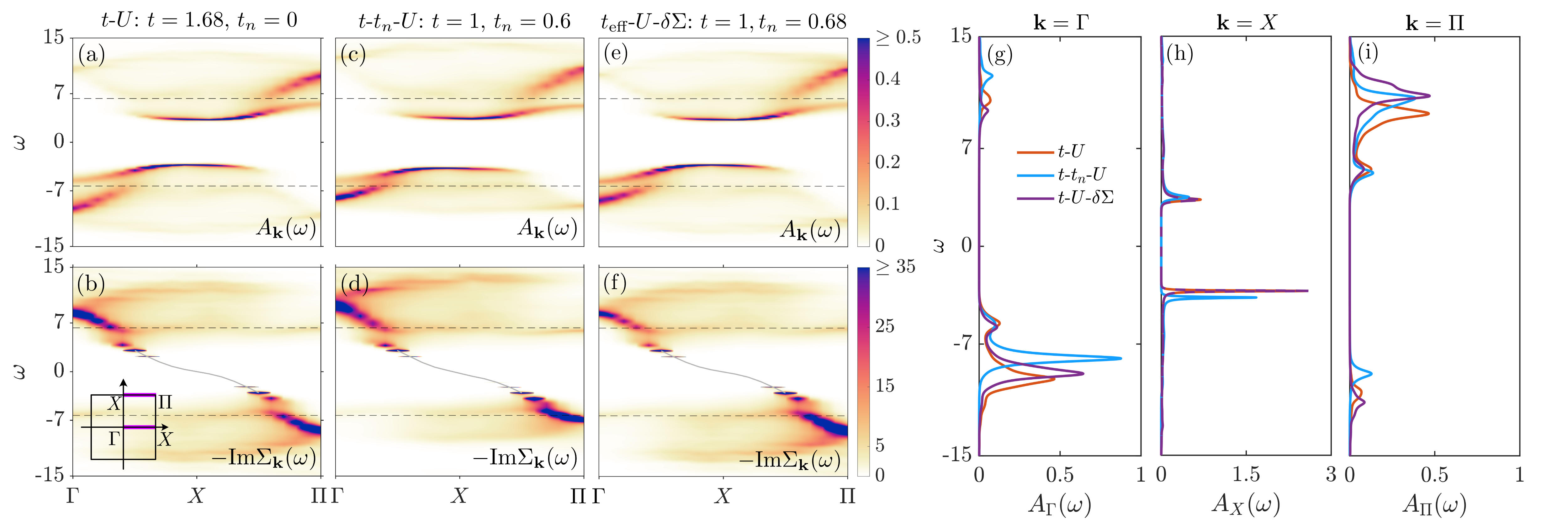} 
\vspace{-6mm}
 \caption{
\label{fig:CylinderPlots}
 Comparison of  $A_\kk(\omega)$ and $-\mathrm{Im}\Sigma_\kk(\omega)$ for the Hubbard model on a $10\times 4$ cylinder, 
 computed via \task\ $\mathrm{(a,b)}$ without and $\mathrm{(c,d)}$ with DAH, 
 and (e,f) computed by treating DAH simplistically via Eq.~\eqref{eq:SE_correction_term}. 
  Solid gray lines in (b,d,f) mark the self-energy pole where it becomes too sharp for the frequency resolution of our grid. Dashed gray lines indicate $\omega = \pm U/2$. Inset in (b): purple lines show the Brillouin zone paths 
 $\Gamma \to X$ and $X\to\Pi$. (g,h,i) Line cuts
 of $A_\bk(\omega)$ for $\bk=\Gamma,X,\Pi$.}
\vspace{-4mm}
 \end{figure*}

Here, we use \task\ to examine the (beyond-HF) effects of DAH on the single-particle spectral function, $A_\kk(\omega) = S\bigl[c_{\bk\sigma}^\dagger\bigr](\omega) + S\bigl[c^{\pdag}_{\bk\sigma}\bigr](-\omega)$, of the half-filled Mott insulator. We find a surprisingly strong particle-hole asymmetry in 
$A_\bk(\omega)$ and a next-nearest neighbor contribution to the dispersion of the self-energy pole.

We study the model~\eqref{eq:H_full} on a half-filled $10 \times 4$ cylinder, with  $U= 12.6$ and 
either $t = 1.68$, $t_n = 0$ (denoted ``$\tU$'') or $t = 1$, $t_n = 0.6$ (denoted ``$\tUn$''), matching the parameters of  Fig.~$4$ of Ref.~\onlinecite{White2023}. (For $\tUn$, 
a HF treatment of DAH yields  $t_{\mr{eff}} = 1.6$, similar to our $t=1.68$ for $\tU$.) We keep at least $D^* = 2000$ U(1)$\times$SU(2) multiplets. Figures~\ref{fig:CylinderPlots}(a,c)
show $A_\kk(\omega)$ for the Mott insulators obtained for $\tU$ or $\tUn$, respectively, and Figs.~\ref{fig:CylinderPlots}(b,d) the imaginary parts of the corresponding self-energies $\Sigma_\kk(\omega)$. To obtain smooth curves
from the discrete spectral data $\{\omega_\alpha,S_{\alpha}\}$, we used a continued-fraction-based broadening scheme that conserves the first few spectral moments of
$A_\bk(\omega)$. This scheme
and \task\ convergence properties are discussed in Secs.~\ref{sec:CFE} and~\ref{sec:Convergence} of the SM~\cite{supplement}, respectively.

For $\tU$, $A_\kk(\omega)$  [Fig.~\ref{fig:CylinderPlots}(a)] exhibits a Mott gap of size $6.67 < U$. Apart from broad incoherent features, the Hubbard bands exhibit dispersive coherent features, in particular at their inner edges.   
The self-energy [Fig.~\ref{fig:CylinderPlots}(b)] has a characteristic $S$-shaped spectrum~\cite{Pudleiner2016}, exhibiting broad horizontal bands 
connected by a sharp, dispersive pole. Both $A_\bk(\omega)$ and $\mathrm{Im} \Sigma_\bk(\omega)$  are particle-hole symmetric, as expected, i.e.\ invariant under $\omega \to -\omega$, $\kk \to \kk + 
\mathbf\Pi$, where $\mathbf \Pi = (\pi,\pi)$.

For $\tUn$, the corresponding results are shown 
in Figs.~\ref{fig:CylinderPlots}(c,d). The overall features are similar to those of 
$\tU$, but $t_n$, though small ($t_n < 0.05 \, U$), introduces 
a significant particle-hole asymmetry [see also Figs.~\ref{fig:CylinderPlots}(g-i)],
a dynamical effect that goes beyond HF. 
In particular, the particle part ($\omega < 0$) of $A_\kk(\omega)$  
is more coherent, i.e.\ dispersive features are sharper and exhibit smaller decay rates, and conversely for the hole spectrum ($\omega > 0$). $\mr{Im} \, \Sigma_{\kk}(\omega)$ shows corresponding
weight redistributions. Moreover, the effective bandwidth is
reduced or increased for particle or hole excitations, respectively, 
i.e.\ spectral features at $\omega < 0$ or $>0$ are ``squeezed'' 
or ``stretched'' in $\omega$. Thus,  $t_n$ dynamically enhances hole mobility compared to particle mobility (an effect also reported in Ref.~\onlinecite{White2023}) and reduces hole lifetimes compared to particle lifetimes.

Which scattering events generate the particle-hole asymmetric features? A  perturbative expansion of $\Sigma_\kk(\omega)$ 
in $t_n$ and $U$  shows that at second order, they arise from \textit{mixed} $t_n$-$U$ scattering
(see Sec.~\ref{sec:si_selfenergy2_dah} of SM~\cite{supplement}): 
a propagating particle or hole emits a particle-hole pair via the $t_n$ vertex and reabsorbs it via the $U$ vertex, or vice versa.
Using this insight, and the fact that the $U$ vertex and part of the 
$t_n$ vertex do not depend on transfer momenta, we 
can estimate the latter's contribution to the DAH-induced change in the self-energy, $\delta \Sigma^{t_n}_\bk$, 
as follows (see Sec.~\ref{sec:si_DAH_correction} of SM~\cite{supplement}): We use \task\ to compute the self-energy of the $\tU$ Hubbard model with hopping $t_{\mr{eff}}$, 
denoted $\Sigma^{\teffU}_{\kk}$; from the latter, we 
obtain $\delta \Sigma^{t_n}_\bk$ by replacing either the initial or final $U$  vertex by the transfer-momentum-independent part of the $t_n$ vertex.  This yields
\begin{align}
\label{eq:SE_correction_term}
\delta \Sigma^{t_n}_{\kk}(\omega) = -4 \frac{t_n \gamma_{\kk}}{U}  (\Sigma^{\teffU}_{\kk}(\omega) 
- \Sigma^{\mr{HF}}_{\kk}
) \, ,
\end{align}
where $\gamma_{\kk} = \cos k_x + \cos k_y$ and  $\Sigma^{\mr{HF}}_{\kk}$ is the HF self-energy.
This ``$\tUSigma$'' scheme yields
a total self-energy of $\Sigma^{\teffU}_{\kk}\! (\omega) + \delta \Sigma^{t_n}_{\kk}(\omega)$. It is shown in Fig.~\ref{fig:CylinderPlots}(f), computed using $t_\mathrm{eff} = t + t_n = 1 + 0.68$ (so that $\Sigma^{\teffU}_{\kk}(\omega)$ equals $\Sigma_\bk(\omega)$ of Fig.~\ref{fig:CylinderPlots}(b)),
almost matching the $\tUn$ settings of Fig.~\ref{fig:CylinderPlots}(d) (where $t_\mathrm{eff}=1+0.6$). 
The corresponding $\tUSigma$ spectral function [Fig.~\ref{fig:CylinderPlots}(e)] indeed reproduces
several properties of the coherent features of $\tUn$
in Fig.~\ref{fig:CylinderPlots}(c):  
reduced effective bandwidth and decay rate for particle excitations, and vice versa for hole excitations [see also
Figs.~\ref{fig:CylinderPlots}(g-i)].
 By contrast, $\delta \Sigma^{t_n}$ hardly affects the incoherent background; reproducing the latter's $\tUn$ properties thus requires a more elaborate treatment.

A hallmark feature of Mott insulators is their self-energy poles [grey lines in Figs.~\ref{fig:CylinderPlots}(b,d)], which cause Green's function zeros. 
Their generically dispersive nature, which has only recently been fully appreciated~\cite{Wagner2023}, is currently under active investigation,
with interesting results ranging from topology~\cite{Wagner2023,Setty2024,Wagner2024,Bollmann2024,Pangburn2025,Pangburn2025a} to possible connections to spinon physics~\cite{Fabrizio2020,Skolimowski2022,Fabrizio2022,Fabrizio2023}. 
If the interaction is purely Hubbard, it can be shown that the pole dispersion is directly inherited from the non-interacting dispersion~\cite{Wagner2023}. 

Thus, for $\tU$, the Mott pole dispersion $\xi_{\kk}$ should contain only nearest-neighbor terms. Indeed, using the ansatz $\xi_{\kk} = -2 \tau (\cos k_x + \cos k_y) - 4 \tau' \cos k_x \cos k_y$~\cite{PoleDispersion}, where $\tau$ and $\tau'$ are nearest and next-nearest neighbor amplitudes, we obtain a good fit with 
$\tau \simeq -2.03$ and $\tau' = 0$, confirming this expectation
(see Sec.~\ref{subsec:addingDAH} of SM~\cite{supplement}). 
By contrast, for $\tUn$ we find $\tau \simeq -1.86$ and $\tau' \simeq 0.17$. Hence \textit{nearest}-neighbor DAH generates a sizeable \textit{next}-nearest neighbor term $\tau'$—a remarkable result, arising from the fact that the DAH $t_n$ vertex, unlike the $U$ vertex, delocalizes electrons beyond single-particle hopping and HF (see Sec.~\ref{sec:SI_PH_asym_DAH} of the SM~\cite{supplement}).
This emergence of non-zero $\tau'$
suggests that $t_n$ may have a similar frustration effect as the much-studied next-nearest neighbor hopping $t'$ 
for the $t$-$t'$ Hubbard model,
in particular if $\xi_{\kk}$ can be interpreted as a spinon dispersion. Since frustration from $t'$ can stabilize superconductivity in the Hubbard model~\cite{Xu2024,Zhang2025}, a similar effect from $t_n$ may, together with the previously discussed mobility and lifetime effects, stabilize superconductivity in the $t$-$t_n$ Hubbard model~\cite{White2023}.

\textit{Outlook.---}
We showed that \task\ is an efficient and accurate method for computing real-frequency spectral functions. Its main source of error comes from the projection to the tangent space and can be estimated via the introduced two-site variance. Due to its relatively light-weight nature, it is especially useful in cases where ground-state calculations are already challenging.
These include strongly correlated electron systems on cylinders, geared, for instance, towards cuprates~\cite{Keimer2015,LeBlanc2015,Jiang2021,Jiang2024}, 
heavy-fermions~\cite{Si2001,Coleman2001,Senthil2003,Danu2021,Raczkowski2022,Raczkowski2024,Gleis2024,Gleis2025}, 
Moir\'e systems~\cite{Soejima2020,Zhou2022,Chatterjee2022}, 
or systems with strong electron-boson couplings~\cite{Kohler2021,Sous2018,Tang2023,Lau2025}.

As a first step in this direction, we have applied \task\ to a Hubbard cylinder exhibiting DAH, a problem relevant to cuprates. Our finding that mixed $t_n$-$U$ scattering leads to enhanced hole mobilities and a possibly frustrating next-nearest neighbor contribution to the dispersion of the self-energy pole may be relevant for high-$T_c$ superconductivity. A more detailed study of this, which would include doping and the calculation of other dynamical properties such as spin or pair spectra, is left for future work.

While \task\ is designed to tackle challenging strongly correlated systems, it is, in its current form, less effective for models with exact or near-exact MPS ground states. There, the restricted tangent-space dimension may not adequately capture multi-particle continua. A generalization of \task\ that aims to overcome this limitation is proposed in Sec.~S-1.F of the SM and will be pursued in future work.

\begin{acknowledgments}
\textit{Acknowledgments.---} 
We thank Seung-Sup Lee, Sebastian Paeckel, Felipe Picoli, Dai-Wei Qu, and Steven White for helpful discussions, and Jean-S\'ebastien Caux on providing the Bethe-ansatz data in Fig.~\ref{fig:HeisenbergPowerLaw}.
This work was supported in part by the Deutsche Forschungsgemeinschaft under grants INST 86/1885-1 FUGG,  LE 3883/2-2 and 
Germany's Excellence Strategy EXC-2111 (Project No.~390814868). It is part of the  Munich Quantum Valley, supported by the Bavarian state government with funds from the Hightech Agenda Bayern Plus. 
The National Science Foundation supported JvD in part under PHY-1748958. 
AG acknowledges support from the Abrahams Postdoctoral Fellowship of the Center for Materials Theory at Rutgers University.

\end{acknowledgments}

\textit{Data availability.---} The simulation data supporting these findings including plotting scripts are available via the zenodo repository at Ref.~\cite{data}.

\raggedbottom 

\bibliography{projected-krylov}

\clearpage

\title{Supplemental material: \\ \maintitle}

\date{\today}
\maketitle

\setcounter{secnumdepth}{2} 
\renewcommand{\thefigure}{S-\arabic{figure}}
\renewcommand{\thesection}{S-\arabic{section}}
\renewcommand{\theequation}{S\arabic{equation}}
\renewcommand{\thetable}{S-\arabic{table}}
\setcounter{figure}{0}
\setcounter{section}{0}
\setcounter{equation}{0}
\setcounter{table}{0}

This supplemental material provides additional details and results on the \task\ method, post-processing of discrete data, and theoretical considerations for the 2D Hubbard model with DAH. Section~\ref{sec:SM_Method} expands on the technical aspects of the \textit{Method} section in the main text, including a pseudocode. Section~\ref{sec:CFE} describes the continued fraction expansion~(CFE) broadening scheme used to obtain continuous spectral functions. In Sec.~\ref{sec:Convergence}, we demonstrate the convergence of \task\ with bond dimension $D^*$ and number of Krylov steps $\Nkr$. \changed{Section~\ref{sec:AdditionalBenchmarks} provides additional benchmark results for situations not analyzed in detail in the main text, showing that \task\ performs well in both cases: Section~\ref{sec:Heisenberg-k-not-pi/2} turns to  correlators whose spectrum is gapped by considering the dynamical structure factor~(DSF) of the Heisenberg model for  $k \neq \pi$. Section~\ref{sup:current-susceptibility} 
focuses on non-local (as opposed to 1-site) operators by considering the current spectrum for a non-interacting, gapped fermionic model in one and two dimensions.} In Sec.~\ref{sec:Comparison_to_TDVP}, we compare results from \task\ with those obtained using time-dependent variational principle~(TDVP) for the Heisenberg model. Sections~\ref{sec:si_selfenergy2_dah} to \ref{sec:SI_PH_asym_DAH} are devoted to our Hubbard cylinder study; they provide analytical details elucidating the origin of particle-hole asymmetry of our \task\ results for spectral quantities reported in the main text.
Section~\ref{sec:si_selfenergy2_dah} discusses the computation of the second-order self-energy for the $\tUn$ model and identifies scattering processes that lead to particle-hole asymmetry.
In Sec.~\ref{sec:si_DAH_correction}, we use the insights from Sec.~\ref{sec:si_selfenergy2_dah} to derive Eq.~\eqref{eq:SE_correction_term} of the main text;
 Sec.~\ref{sec:SI_PH_asym_DAH} shows how next-nearest neighbor terms arise in the dispersion of Green's function zeros from nearest-neighbor DAH.

\section{\task: technical details}
\label{sec:SM_Method}
Below, we detail the working procedure for \task: how to apply  the projected Hamiltonian 
$H^{1\perp}$, how to prepare the initial state $\ket{\PsiI} = \mc{P}^{1\perp} O_\bk^{\pdag}\ket{\Psig}$, and how to combine all this in the context of a Lanczos scheme. \changed{A QSpace-based implementation of the TaSK method is publicly available at Ref.~\cite{kovalska_task_code}}.

But before moving on to technical details, we offer some remarks on the rationale underlying our TaSK approach. The correlator defined in \Eq{eq:SpectralFunction1} can be expressed as
    \begin{align}%
    S[O^{\pdag}_\bk](\omega)
    & = 
    \sum_\alpha | \langle \Psi_\alpha  | \Psie \rangle |^2 \,  \delta(\omega - E_\alpha + 
    \Eg) ,  
    \label{eq:SpectralFunction1-SMb}
\end{align}
where $\ket{\Psie} = O_{\bk} \ket{\Psig}$ is the excitation whose dynamics is encoded in 
$S[O^{\pdag}_\bk](\omega)$, and $\{\ket{\Psi_\alpha}, E_\alpha\}$ is a complete set of
exact eigenpairs of $H$.
In this paper, we consider only operators $O_\bk$ for which $|\Psie \rangle$ lies in the tangent space of the ground state, $\doubleV^{1\perp}$. (Intuitively speaking, $O_\bk$ creates ``Bloch waves'' on top of the ground state, i.e.\ excitations that are just 
linear combinations of local perturbations of $\ket{\Psig}$.) 
Then, \textit{each overlap $| \langle \Psi_\alpha | \Psie  \rangle |^2$ will only receive a contribution from
the projection of the eigenstate $|\Psi_\alpha\rangle $
into the tangent space $\doubleV^{1\perp}$}. It is thus natural -- and this is the key idea
of TaSK (inspired by prior works on the quasiparticle ansatz~\cite{Ostlund1995,Rommer1997,Haegeman2012,Haegeman2013,Verstraete2013,Haegeman2019,Tu2021}) -- to approximate the eigenpairs 
$\{\ket{\Psi_\alpha}, E_\alpha\}$ of $H$ by those of $\mathcal{P}^{1\perp} H \mathcal{P}^{1\perp}$, say $\{\ket{\Psi_\alpha^{1\perp}}, E_\alpha^{1\perp}\}$.
The latter can conveniently be computed using a Lanczos scheme, exploiting the vector-space structure of $\doubleV^{1\perp}$. The TaSK approximation causes an error $E_\alpha^{1\perp}- E_\alpha$ for the eigenenergies contributing to 
\Eq{eq:SpectralFunction1-SMb}; the energy measure $\delta_\alpha = (\Delta_\alpha^{2\perp})^{1/2}$ 
introduced in the main text serves as an estimate of this error.

Note that it would not make sense to 
approximate the eigenstates $\ket{\Psi_\alpha}$
in \Eq{eq:SpectralFunction1-SMb} by their projections
into the tangent space $\doubleV{}_{\mr{ex}}^{1\perp}$ of $\ket{\Psie}$ (instead of the tangent space 
$\doubleV{}^{1\perp}$ of $\ket{\Psig}$), because those projections would by construction have zero overlap 
with $\ket{\Psie}$. (Intuitively speaking,
that would amount to computing ``Bloch waves'' on top of $\ket{\Psie}$ instead of $\ket{\Psig}$.)

\subsection{Ground State Isometries}
The ground-state MPS $\Psig$ is the centerpiece of the \task\ method. Besides serving as an input in Eq.~\eqref{eq:SpectralFunction_TaSK}, its constituent tensors form the building blocks of the tangent space projector~\eqref{eq:P_1perp} that enables this computation. 

Once $\Psig$ has converged to the desired accuracy, for instance via DMRG, the first step is to bring it to left- and right-canonical forms or equivalently to determine isometries $A_\ell \,  (\TriangleWhiteA)$ and $B_{\ell} \, (\TriangleWhiteB)$ for $\ell \in \bigl[1,\eLL\bigr]$.
They satisfy the following orthogonality relations
\vspace{-2mm}
\begin{flalign}
\label{eq:IsometricConditions}
& \raisebox{-6.5mm}{
 \includegraphics[width=0.866\linewidth]{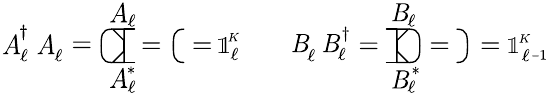}}  \, ,  \hspace{-1cm} &
 \end{flalign}
indicating that $A_\ell$~($B_\ell$) is a left~(right) isometry, mapping a larger \textit{parent} ($\P$) space of dimension $D_{\ell-1}d$ to a smaller $D_\ell$-dimensional \textit{kept} ($\K$) space (the definition is analogous for right isometries, with the roles of the legs reversed). For consistency, these tensors are computed once and stored in the memory to avoid possible issues arising due to the ambiguity of unitary transformations in the kept space. In the pseudocode below, the collections of all left and right isometries are denoted by $\lMPS$ and $\rMPS$, respectively. A more detailed discussion on these isometries can be found in Ref.~\onlinecite{Gleis2022}.

The orthonormality and completeness relations of the left and right isometries and their orthogonal complements, $\Ab^\pdag_{\ell} (\TriangleGreyA)$ and $\Bb_{\ell}^\pdag (\TriangleGreyB)$, 
\begin{subequations}%
\label{subeq:AdditionalOrthonormalityRelations}%
{%
  \addtolength{\abovedisplayskip}{3pt}%
  \addtolength{\abovedisplayshortskip}{3pt}%
  \addtolength{\belowdisplayskip}{6pt}%
  \addtolength{\belowdisplayshortskip}{6pt}%
\begin{flalign}
 & \hspace{-1mm} 
 \raisebox{-4mm}{
 \includegraphics[width=0.89\linewidth]{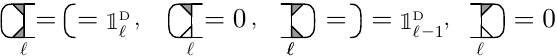}}
 \hspace{-1cm} &
\\
\label{eq:CompletenessMain}
 & \hspace{-2mm}  \raisebox{-3.5mm}{
 \includegraphics[width=0.886\linewidth]{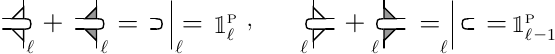}} \, ,
 \hspace{-2cm}  &
\end{flalign}%
}%
\end{subequations}
form the backbone of most subsequent derivations. We note that in this work, we never explicitly determine the orthogonal complements. To reduce the computational cost, any projection to the \textit{discarded} ($\D$) space, involving tensors $\Ab^\pdag_{\ell} (\TriangleGreyA)$ and $\Bb_{\ell}^\pdag (\TriangleGreyB)$, is performed implicitly via Eq.~\eqref{eq:CompletenessMain}. 
Furthermore, in the following sections, any white tensor $\CircleWhiteC$ is associated with the local kept space of the left isometry $A_\ell$ at $\ell \in \bigl[1,\eLL\bigr]$. Similarly, gray shaded tensors $\CircleGreyC$ have been projected to the local discarded space of $\Ab^\pdag_{\ell}$.

As emphasized in the main text, we limit our numerical computations to $\doubleV^{1\perp} \! = \!\doubleVonesite/\mathrm{span}\bigl(\ket{\Psig}\bigr)$, where wavefunctions have the form
\begin{align}
        \Psi^{1\perp} & = 
        \sum_{\ell=1}^{\scripteLL}  \!\!
        \raisebox{-0.45\height}{
        \includegraphics[width=0.322\linewidth]{Eq/1siteExcitationAnsatz}}
        \, , \;\; \mr{with} \; \;
        \raisebox{-5.8mm}{\includegraphics[width=0.14\linewidth]{./Eq/T_tensors_gauge2.pdf}} \;.
        \label{eq:ExnAnsatz_SM} 
\end{align}
The gauge choice guarantees that all terms in the sum are mutually orthogonal and can be written as
\begin{align}
    \label{eq:gauge_1perp}
\raisebox{-0.44\height}{
\includegraphics[width=0.22\linewidth]{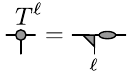}}\,.  
\end{align}
The presence of $\TriangleGreyA$ makes it explicit that $\ket{\Psi^{1\perp}} \in \doubleV^{1\perp}$ is orthogonal to $\ket{\Psig}$. 
In that way, we directly target excitations, and our Lanczos scheme avoids recomputing $\ket{\Psig}$ in cases when operator $O_{\kk}$ in Eq.~\eqref{eq:SpectralFunction_TaSK} has a non-zero ground-state expectation value.

\subsection{Application of the Hamiltonian}
\label{sec:applyHamiltonian}

\begin{table*}[t]
\begin{minipage}{\linewidth}
\begin{algorithm}[H]
  \caption{Application of the projected Hamiltonian $H^{1\perp}$}
  \label{alg:applyH}
   \begin{algorithmic}[1]
   \Require input state tensors $\{T_i\}$, $\MPO$, left isometries $\lMPS$, right isometries $\rMPS$, $L$ and $R$ environments defined in \eqref{eq:L-R-environments}, system size $\eLL$
  \Ensure resulting state tensors $\{T_{i+1}\}$
   \Procedure{applyH}{$\{T_i\}$,$\MPO$, $\lMPS$, $\rMPS$, $L$, $R$, $\eLL$} 

    \State Initialize left environment $\mathcal{L}$
    \State Compute the 1st term in the sum \eqref{eq:Exn_applyH_concept_compr}; project it \eqref{eq:TL1_ortho} to the discarded space 
    \For{$\ell=2,\ldots,\eLL$ }
    \State Update $\mathcal{L} \leftarrow \{T_{i}, \MPO, \lMPS, \rMPS, L\}$ 
    \State Compute temporary right environment $R' \leftarrow \{R, \rMPS,\MPO\} $
    \State Contract $\mathcal{L}$ and $R'$ $\rightarrow$ $T_{i+1}^{\mathbbm{1}}$
    \Comment{1st summand from Eq.~\eqref{eq:TL1_ortho}}
    \State Compute the kept $\K$ space contribution $T_{i+1}^{\mathbbm{\K}} \leftarrow \{T_{i+1}^{\mathbbm{1}}, \lMPS \}$ 
    \Comment{2nd summand from Eq.~\eqref{eq:TL1_ortho}}
    \State Compute the full projected $T_{i+1}^{\D} \leftarrow \{T_{i+1}^{\mathbbm{1}}, T_{i+1}^{\mathbbm{\K}} \}$ tensor using \eqref{eq:TL1_ortho}
    \EndFor
    
    \State Compute the 2nd term in the sum \eqref{eq:Exn_applyH_concept_compr}; project it \eqref{eq:TL1_ortho} to the discarded space 
     \For{$\ell=1,\ldots,\eLL$ }
    \State Compute temporary right environment $R' \leftarrow \{R, T_i,\MPO\} $
    \State Contract $L$ and $R'$ $\rightarrow$ $T_{i+1}^{\mathbbm{1}}$
    \State Compute the kept $\K$ space contribution $T_{i+1}^{\K} \leftarrow \{T_{i+1}^{\mathbbm{1}}, \lMPS \}$ 
    \State Update $T_{i+1}^{\D} \leftarrow \{T_{i+1}^{\mathbbm{1}}, T_{i+1}^{\mathbbm{\K}},T_{i+1}^{\D} \}$ 
    \EndFor
    \State Initialize right environment $\mathcal{R}$
     \State Compute the 3rd term in the sum \eqref{eq:Exn_applyH_concept_compr}; project it \eqref{eq:TL1_ortho} to the discarded space

      \For{$\ell=\eLL-1,\ldots,1$ }
    \State Update $\mathcal{R} \leftarrow \{T_{i}, \MPO, \lMPS, \rMPS, R\}$ 
    \State Compute temporary right environment $L' \leftarrow \{L, \lMPS,\MPO\} $
    \State Contract $\mathcal{R}$ and $L'$ $\rightarrow$ $T_{i+1}^{\mathbbm{1}}$
    \State Compute the kept $\K$ space contribution $T_{i+1}^{\mathbbm{\K}} \leftarrow \{T_{i+1}^{\mathbbm{1}}, \lMPS \}$ 
    \State Assign final value $T_{i+1} \leftarrow \{T_{i+1}^{\mathbbm{1}}, T_{i+1}^{\mathbbm{\K}},T_{i+1}^{\D} \}$ 
    \EndFor

   \State \Return $\{T_{i+1}\}$
  \EndProcedure
   \end{algorithmic}
\end{algorithm}
\end{minipage}
\caption{Pseudocode of the operation $\ket{\Psi_{i+1}} = H^{1\perp} \ket{\Psi_i}$ detailed in Sec.~\ref{sec:applyHamiltonian}. Numerically, it is performed in terms of operations that involve a collection of tensors $\{T_j\}$ describing a state \ket{\Psi_{j}}. Specifically, above $\{T_i\}$ denotes a collection of tensors corresponding to the input state \ket{\Psi_{i}} and $\{T_{i+1}\}$ -- the output state, \ket{\Psi_{i+1}}, respectively. To avoid index proliferation, here we drop the site labeling $\ell$ and instead simply use $T_i$ to denote an individual tensor. The site labeling can be inferred from the \texttt{for} loop structures and the environment diagrams introduced in Sec.~\ref{sec:applyHamiltonian}. The code is structured to compute each of the three main contributions \eqref{eq:Exn_applyH_concept_compr} one-by-one for \textit{all} new tensors $\{T_{i+1}\}$. We also introduce indices $\mathbbm{1}$, $\K$ and $\D$ to distinguish between the unprojected $\CircleWhitePC$ and projected $\CircleGreyPC$ value of $T_{i+1}$ which are related through Eq.~\eqref{eq:TL1_ortho}.}
\label{tab:ApplyH}
\end{table*}

The main step of the Lanczos scheme is applying the projected Hamiltonian $H^{1\perp} = \mathcal{P}^{1\perp}(H-\Eg)\mathcal{P}^{1\perp}$ to a state $\ket{\Psi_i} \in \doubleV^{1\perp}$. Assuming that $\ket{\Psi_i}$ is described by tensors $T^{\ell'}_i$, one wants
to compute tensors $T^{\ell}_{i+1}$ which describe the 
vector $\ket{\Psi_{i+1}} = H^{1\perp} \ket{\Psi_i}$. 
For a given site $\ell$, this amounts to evaluating a sum of the following diagrams:
\begin{align}
\label{eq:Exn_applyH_concept}
\raisebox{-0.3\height}{
 \includegraphics[width=0.1\linewidth]{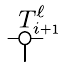}}
&\longleftarrow \!\sum_{\ell'=1}^{\scripteLL} \!
\raisebox{-0.40\height}{
 \includegraphics[width=0.38\linewidth]{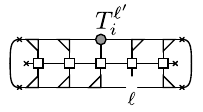}} \, , 
\end{align}
and projecting it to the discarded space using appropriate isometries
\begin{align}
\label{eq:TL1_ortho}
\raisebox{-0.35\height}{
 \includegraphics[width=0.47\linewidth]{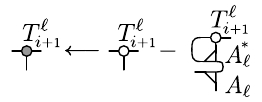}}
 \, .
\end{align}
As noted earlier, the latter is diagrammatically indicated by depicting the resulting tensors as gray. In the formulas above, the first projection with respect to $\mathcal{P}^{1\perp}$ in the definition of $H^{1\perp}$ is implicit since $\ket{\Psi_i} \in \doubleV^{1\perp}$.

The sum in Eq.~\eqref{eq:Exn_applyH_concept} can be subdivided into three terms 
\begin{align}
\label{eq:Exn_applyH_concept_result}
\nonumber
\raisebox{-0.3\height}{
 \includegraphics[width=0.1\linewidth]{Eq/t_tensor_v0.pdf}}
&\longleftarrow \sum_{\ell'=1}^{\ell-1} 
\! \raisebox{-0.49\height}{
 \includegraphics[width=0.35\linewidth]{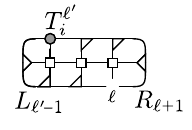}} + \! \raisebox{-0.44\height}{
 \includegraphics[width=0.24\linewidth]{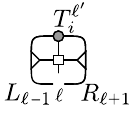}}
\hspace{-0.25cm}  
\\ 
& \; +
\sum_{\ell'=\ell+1}^{\eLL} 
\! \raisebox{-0.44\height}{
 \includegraphics[width=0.33\linewidth]{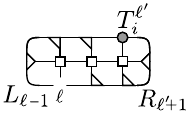}}
\hspace{-0.25cm}  
\, ,
\end{align}
where $L_\ell$ and $R_\ell$ indicate intermediate left and right environments defined as 
\begin{subequations}
	\label{eq:L-R-environments}
\begin{align}
	\label{eq:L-environments}
L_\ell & =  \raisebox{-5.3mm}{
 \includegraphics[width=0.45\linewidth]{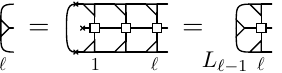}} \, ,
\\ 
	\label{eq:R-environments}
R_\ell & = \raisebox{-5.3mm}{
 \includegraphics[width=0.5\linewidth]{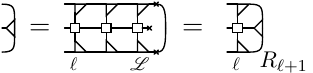}} \, .
\end{align}
\end{subequations}
They can be computed recursively by introducing the appropriate $A_\tell \,  (\TriangleWhiteA)$ and $B_{\tellp} \, (\TriangleWhiteB)$ tensors, starting from $L_0=1$, $R_{\seLL+1}= 1$.

In addition, to evaluate the terms corresponding to $\ell'<\ell$ and $\ell'>\ell$, we introduce a second type of environment, $\mc{L}_{\ell}$ and $\mc{R}_{\ell}$, containing $T^{\ell'}_i$ tensors of the initial state $\ket{\Psi_i}$ 
\begin{subequations}
 \hspace{-0.5cm}
\label{eq:RLc_def}
\begin{align}
\label{L_tp1_env}
 \mc{L}_{\ell}
 \,=\,
\!\raisebox{-0.36\height}{
 \includegraphics[width=0.0222\linewidth]{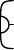}}
 &\,= 
\!
 \sum_{\ell'=1}^{\ell} \!\!\raisebox{-0.37\height}{
 \includegraphics[width=0.27\linewidth]{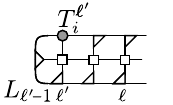}} \hspace{-0.2cm} =
\hspace{-0.4cm}
 \raisebox{-0.43\height}{
 \includegraphics[width=0.15\linewidth]{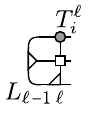}}
 \hspace{-0.15cm}
+
 \hspace{-0.3cm}
 \raisebox{-0.45\height}{
 \includegraphics[width=0.13\linewidth]{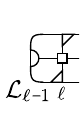}}
 \hspace{-0cm} \; ,  
 \\
 \mc{R}_{\ell}
 \,=\,
\!\raisebox{-0.36\height}{
 \includegraphics[width=0.0222\linewidth]{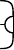}}
 &\,=  \,
 \hspace{-0.1cm}
 \sum_{\ell'=\ell}^{\scripteLL} 
\raisebox{-0.45\height}{
 \includegraphics[width=0.27\linewidth]{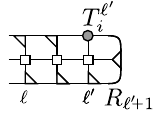}}
 \hspace{-0.3cm} = 
\hspace{-0.0cm}
 \raisebox{-0.43\height}{
 \includegraphics[width=0.15\linewidth]{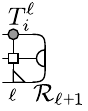}}
 \hspace{-0.3cm} \!\! + \hspace{-0cm}
 \raisebox{-0.43\height}{
 \includegraphics[width=0.17\linewidth]{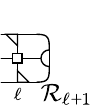}}
\hspace{-0.3cm}  
\, .
\hspace{-10cm}& &
\end{align}
 \end{subequations}
This allows us to rewrite Eq.~\eqref{eq:Exn_applyH_concept_result} in a compact form 
\begin{align}
\label{eq:Exn_applyH_concept_compr}
\raisebox{-0.3\height}{
 \includegraphics[width=0.1\linewidth]{Eq/t_tensor_v0.pdf}}
&\longleftarrow 
\!\!\!\! \hspace{-0.2cm} \raisebox{-0.4\height}{
 \includegraphics[width=0.22\linewidth]{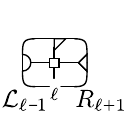}} \hspace{-0.25cm} + \!\!\!\raisebox{-0.45\height}{
\includegraphics[width=0.22\linewidth]{Eq/t_tensor_sum3_v0.pdf}}
\hspace{-0.25cm} + \!\!\!\! \raisebox{-0.45\height}{
\includegraphics[width=0.22\linewidth]{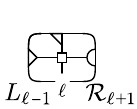}}
\hspace{-0.25cm}  
\, ,
\end{align}
which represents how $T^{\ell}_{i+1}$ is computed numerically. For the pseudocode version of this computation, see Tab.~\ref{tab:ApplyH}.

\subsection{Initialization}
\label{sec:Initialization}

To construct the initial state
$\ket{\PsiI} = \mc{P}^{1\perp} O_\bk^{\pdag}\ket{\Psig}$, we apply 
$O_{\kk} = \sum_{\ell =1}^{\seLL} \alpha_{\ell} O_{\ell} $, a linear combination of local operators, to $\Psig$ in the left-canonical form, 
\begin{subequations}%
\begin{align}%
\label{eq:Psi0DefinitionDiagram}
        \PsiI & = 
        \sum_{\ell=1}^{\scripteLL}   \sum_{\ell'=1}^{\scripteLL} 
        \raisebox{-0.6\height}{
        \includegraphics[width=0.3\linewidth]{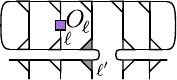}} 
        \\ 
        &= \sum_{\ellp=1}^{\scripteLL} \!\!\raisebox{-0.43\height}{
        \includegraphics[width=0.322\linewidth]{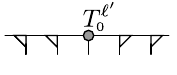}}
        \, .
\end{align}%
\end{subequations}%
In the above diagrams, the coefficients $a_\ell$ are absorbed into the tensor.
Using the orthogonality relations \eqref{subeq:AdditionalOrthonormalityRelations}, we can determine $T^{\ell'}_0$ through the expression
\noindent
\vspace{-0.5cm}
\begin{align}
    \label{eq:Psi1_sum}
    \raisebox{-0.3\height}{
     \includegraphics[width=0.08\linewidth]{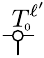}}
    &\longleftarrow \!\sum_{\ell'=1}^{\scripteLL} \!
    \raisebox{-0.40\height}{
     \includegraphics[width=0.24\linewidth]{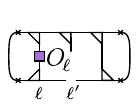}} \, , 
\end{align}
which can be computed using intermediate environments by analogy to Section \ref{sec:applyHamiltonian}, and then project to the discarded space as in Eq.~\eqref{eq:TL1_ortho}.

Since $O_\kk$ is a sum of local operators, the initial-state projection is exact, in the sense that $O_{\kk} \ket{\Psig} = \ket{\PsiI} + \langle O_{\kk} \rangle \ket{\Psig}$.

\subsection{Lanczos Scheme}
\label{sec:Lanczos}

\begin{table*}[t]
\begin{minipage}{\linewidth}
\begin{algorithm}[H]
  \caption{Lanczos}
  \label{alg:Lanczos}
   \begin{algorithmic}[1]
   \Require $\MPO$, left isometries $\lMPS$, right isometries $\rMPS$, initial state tensors $\{T_0\}$ (normalized), $L$ and $R$ environments defined in \eqref{eq:L-R-environments}, system size $\eLL$, number of orthogonalization steps $\mathrm{n\_orth}$ (default = 2)
  \Ensure Hamiltonian matrix in Krylov basis $H^{1\perp}$
   \Procedure{Lanczos}{$\MPO$, $\lMPS$, $\rMPS$, $\{T_0\}$, $L$, $R$, $\eLL$} 
  \State Initialize matrix $H^{1\perp}$
  \For{$i=0,\ldots,\numkr-1$}
   \Function{ApplyH}{$\{T_i\}$,$\MPO$, $\lMPS$, $\rMPS$, $L$, $R$, $\eLL$}
   \State \Return $\{T_{i+1}\}$
   \EndFunction
    
   \If{$i = \numkr-1$}
   \State Load $\{T_{i}\}$
   \State Compute overlap $c(i)$ \eqref{eq:lancz3}
   \State Store $c(i)$ as a diagonal element of $H^{1\perp}$
   \Else 
   
   \For{$n=1,\ldots,\mathrm{n\_orth}$}
   \State Initialize array $c$ containing vector overlaps 
   \For{$j=1,\ldots,i$}
   \State Load $\{T_{j}\}$
   \State Compute overlap $c(j)$ \eqref{eq:lancz3}
   \EndFor
    
   \If{n = 1}
   \State Store $c(i)$ as a diagonal element of $H^{1\perp}$
   \EndIf
   
   \State Orthogonalize $\{T_{i+1}\}$ with respect to all previous vectors 
   \eqref{eq:lancz2} 
   \State Using lMPS and rMPS, project each tensor from $\{T_{i+1}\}$ to the local discarded space \eqref{eq:lancz4}
   \State Compute the norm $b_{i+1}$ and normalize $\{T_{i+1}\}$ \eqref{eq:lancz5}
   \EndFor
   \EndIf
   \State Save $\{T_{i+1}\}$
   \State Store $b_{i+1}$ as a superdiagonal element of $H^{1\perp}$
   
  \EndFor
   \State \Return $H^{1\perp}$
  \EndProcedure
   \end{algorithmic}
\end{algorithm}
\end{minipage}
\caption{Pseudocode of the Lanczos algorithm in \task\ framework described in Sec.~\ref{sec:Lanczos}, \changed{assuming the ``heavy'' orthonormalization strategy}. Above, $\{T_{j}\}$ denotes a collection of tensors describing a state \ket{\Psi_{j}}. For details on function \texttt{ApplyH} see Tab.~\ref{tab:ApplyH}.}
\label{tab:LanczosAlgorithm}
\end{table*}

In this section, we present details on the Lanczos scheme implementation within \task. 
We start with a compact summary in equation format and then elucidate each step through corresponding tensor-network diagrams. In addition, we provide the corresponding pseudocode in Tab.~\ref{tab:LanczosAlgorithm}.

The Lanczos scheme involves the following steps,
repeated iteratively, starting from $i=0$:
\begin{equation}
\label{eq:Lanczos_eqs}
\begin{aligned}
\text{(i)}\quad & \ket{\Psi_{i+1}} \leftarrow H^{1\perp}\ket{\Psi_{i}} 
    && \text{apply $H^{1\perp}$}, \\
\text{(ii)}\quad & \,  
     c_j \leftarrow \langle\Psi_j|\Psi_{i+1}\rangle,  \;\; j \le i
    && \text{compute overlaps}, \\
    \text{(iii)}\quad & 
    \text{Set} \,\, a_i = c_i,\\
\text{(iv)}\quad & \ket{\Psi_{i+1}} 
    \leftarrow \ket{\Psi_{i+1}} - {\textstyle \sum_{j\changed{=0}}^\changed{i} c_j \ket{\Psi_{j}}}, 
    && \text{orthogonalize}, \\
\text{(v)}\quad & \text{Fix gauge to match~\eqref{eq:gauge_1perp}}, \\
\text{(vi)}\quad & b^2_{i+1} = \langle \Psi_{i+1} \ket{\Psi_{i+1}} && \text{compute norm}, \\
                & \ket{\Psi_{i+1}} \leftarrow \ket{\Psi_{i+1}}/b_{i+1} && \text{normalize}, \\
\text{(vii)}\quad & \text{Repeat steps (ii) and (iv)-(vi) once.} \hspace{-3cm}
\end{aligned}
\end{equation}
The reason for repeating steps {as specified in (vii)} is to reduce the round-off error noise, thereby improving numerical stability. For the same reason, 
the orthogonalization of $\ket{\Psi_{i+1}}$ w.r.t.\ to $\ket{\Psi_j}$ in step (iv) is performed for all $0 \le j \le i$. (With exact arithmetic, it would suffice to do this only for $j=i$ and $j=i-1$.)

In the Krylov basis  $\{\ket{\Psi_0}, \dots, \ket{\Psi_{\Nkr}}\}$
obtained after $\mathcal{N}_{\mathrm{kr}}$ iterations the matrix $H^{1\perp}$ is tridiagonal, 
{%
  \addtolength{\abovedisplayskip}{0pt}%
  \addtolength{\abovedisplayshortskip}{0pt}%
  \addtolength{\belowdisplayskip}{4pt}%
  \addtolength{\belowdisplayshortskip}{4pt}%
\begin{align}
H^{1\perp} = 
\begin{pmatrix}
a_{0} & b_{1} & 0 & \cdots & 0 & 0 \\
b_{1} & a_{1} & b_{2} & \cdots & 0 & 0 \\
0 & b_{2} & a_{2} & \cdots & 0 & 0 \\[-1mm]
\vdots & \vdots & \vdots & \ddots & \vdots & \vdots \\[-1mm]
0 & 0 & 0 & \cdots & a_{\Nkr-1} & b_{\Nkr} \\
0 & 0 & 0 & \cdots & b_{\Nkr} & a_{\Nkr}
\end{pmatrix} \! .
\end{align}}
Diagonalizing it yields a set of real discrete
frequencies and corresponding spectral weights, $\{\omega_\alpha,S_{\alpha}\}$. 

We now elaborate on Lanczos steps (i)-(vi) in diagrammatic format.
The Hamiltonian $H^{1\perp}$ is applied to $\ket{\Psi_i}$ according to the procedure described in Section \ref{sec:applyHamiltonian}:
\begin{align}
\label{eq:lancz1}
\raisebox{-0.35\height}{
 \includegraphics[width=0.3\linewidth]{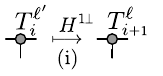}}
 \, .
\end{align}
Next, the overlaps  $c_j = \langle \Psi_{j} | \Psi_{i+1} \rangle$, $j\le i$  are computed,
\begin{align}
\label{eq:lancz3}
\raisebox{-0.5\height}{
 \includegraphics[width=0.29\linewidth]{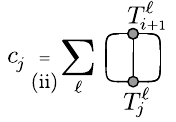}}
 \, ,
\end{align}
and used to orthogonalize
the new Krylov vector, $\ket{\Psi_{i+1}}$ 
w.r.t.\  
previous Krylov vectors,
\begin{align}
\label{eq:lancz2}
\raisebox{-0.35\height}{
 \includegraphics[width=0.45\linewidth]{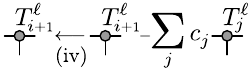}}
 \, ,
\end{align}
Then,  $\ket{\Psi_{i+1}}$ is projected to the discarded space to ensure the desired gauge~\eqref{eq:gauge_1perp} of its constituent tensors
\begin{align}
\label{eq:lancz4}
\raisebox{-0.35\height}{
 \includegraphics[width=0.5\linewidth]{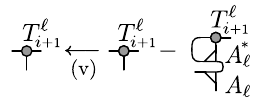}}
 \, .
\end{align}
This step prevents numerical noise from spoiling the gauge condition from Eq.~\eqref{eq:gauge_1perp} and would not be needed with exact arithmetic. Finally, the new Krylov vector is normalized,
\begin{align}
\label{eq:lancz5}
\raisebox{-0.35\height}{
 \includegraphics[width=0.75\linewidth]{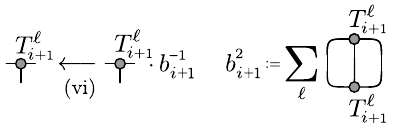}}
 \, .
\end{align}
Here, $b_{i+1}$ corresponds to the superdiagonal element of $H^{1\perp}$ in the Krylov basis. 

As mentioned above in step (vii) of Eq.~\eqref{eq:Lanczos_eqs}, we find that performing the orthonormalization steps \eqref{eq:lancz2}, \eqref{eq:lancz4}, and \eqref{eq:lancz5} a second time improves numerical stability by reducing roundoff-error noise. This does not introduce significant overhead, since orthonormalization scales as $\mc{O}(\eLL D^2 d)$ and is therefore much cheaper than the application of the Hamiltonian ($\mc{O}(\eLL D^3 w d)$).

\changed{Apart from
the ``heavy'' orthonormalization strategy described above,
we have also explored two cheaper options,
which we call ``standard'' and ``light''.
According to (vii), steps (ii) and (iv) are passed twice for any given $i$. The three strategies differ as to whether these passes involve full orthonormalization of $|\Psi_{i+1}\rangle$ w.r.t.\ to all previous states $|\Psi_{j\le i}\rangle$, or partial orthonormalization w.r.t.\ 
only the previous two states, $|\Psi_i\rangle$ and
$|\Psi_{i-1}\rangle$. 
The ``heavy'' strategy uses  
full orthonormalization for both passes through (ii)+(iv), as described above; ``standard'' uses partial or full orthonormalization for the first or second passes through (ii)+(iv), respectively; and ``light'' uses 
partial orthonormalization for both passes through (ii)+(iv). Notably, the ``light'' strategy requires the least memory, since it does not require storing all Krylov vectors throughout.

We have compared the performance of the orthonormalization strategies (i) to (iii) for one-dimensional spin models (specifically, $\eLL = 64$ Heisenberg chain at $k=\pi/2$). We found that 
even for discrete spetral data their results are all in remarkable agreement, with differences on the order of $10^{-10}$. Furthermore, we expect that for the computation of broadened spectral functions all three strategies will
yield essentially equivalent results.
This is in accord with the conclusion of Ref.~\cite{Greenbaum1989} that spectral functions are much less sensitive to loss-of-orthogonality issues than individual Krylov states.}

\subsection{Error Measure}
\label{subsec:ErrorMeasure}

In the main text, we proposed an error measure $\delta_\alpha = 
(\Delta_\alpha^{2\perp})^{1/2}$ based on the
2-site variance, $\Delta_\alpha^{2\perp} = \| \mathcal{P}^{2\perp} H \ket{\Psi_\alpha^{1\perp}} \|^2$ [cf.~\Eq{eq:EnergyVariance}], where $\ket{\Psi_\alpha^{1\perp}}$ is an eigenstate of the Krylov Hamiltonian. To compute the latter, we have to apply $H$ to a tangent vector, followed by a projection to $\doubleV^{2\perp}$ using the projector
\begin{align}
\mc{P}^{2\perp} =\raisebox{-0.42\height}{
\includegraphics[width=0.35\linewidth]{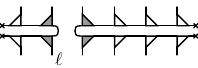}} \, .
\end{align}
This can be achieved with a small modification of the algorithm to apply $H^{1\perp}$, described in Sec.~\ref{sec:applyHamiltonian}. 

Consider for now a generic state $\ket{\Psi^{1\perp}} \in \doubleV^{1\perp}$, described by the tensors $T^{\ell'}$ with gauge constraint as in Eq.~\eqref{eq:gauge_1perp}. Reusing the environments defined in Sec.~\ref{sec:applyHamiltonian}, we obtain
{%
  \addtolength{\abovedisplayskip}{0pt}%
  \addtolength{\abovedisplayshortskip}{0pt}%
  \addtolength{\belowdisplayskip}{6pt}%
  \addtolength{\belowdisplayshortskip}{6pt}%
\begin{align}
\label{eq:T2s}
\raisebox{-0.33\height}{
\includegraphics[width=0.11\linewidth]{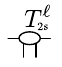}} &\longleftarrow \!\! \raisebox{-0.42\height}{
\includegraphics[width=0.35\linewidth]{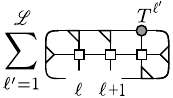}}
\\ \nonumber
&= \raisebox{-0.42\height}{
\includegraphics[width=0.8\linewidth]{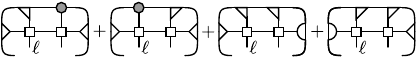}} \, .
\end{align}}
By projecting $T^{\ell}_{\mtwosite}$ to the discarded spaces with $\overline{A}_{\ell}$ and $\overline{B}_{\ell + 1}$ using Eq.~\eqref{subeq:AdditionalOrthonormalityRelations}, we get
{%
  \addtolength{\abovedisplayskip}{-6pt}%
  \addtolength{\abovedisplayshortskip}{-6pt}%
  \addtolength{\belowdisplayskip}{6pt}%
  \addtolength{\belowdisplayshortskip}{6pt}%
\begin{align}
\label{eq:T2perp}
\raisebox{-0.33\height}{
\includegraphics[width=0.11\linewidth]{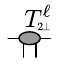}} &\longleftarrow \!\! \raisebox{-0.43\height}{
\includegraphics[width=0.32\linewidth]{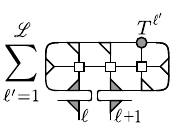}} \, .
\end{align}}
With that, we finally obtain its 2-site variance,
\begin{align}
\label{eq:Delta2perp}
\Delta^{2\perp} \! & = \| \mathcal{P}^{2\perp} H \ket{\Psi^{1\perp}} \|^2 = \sum_{\ell} \| T_{2\perp}^{\ell} \|^2 \, .
\end{align}

 We note that the best practice is to compute $\Delta_{\alpha}^{2\perp}$ for chosen eigenstates $\ket{\Psi^{1\perp}_\alpha}$ of the Krylov Hamiltonian, specifically those corresponding to the most prominent spectral weights.
If $\Delta_{\alpha}^{2\perp}$ is required for \textit{all} eigenstates, we can instead compute $T_{2\perp}^{\ell}$ for each Krylov basis vector $\ket{\Psi_i}$ on the fly, while applying $H^{1\perp}$ inside the Lanczos loop~\eqref{eq:Lanczos_eqs}.
This allows us to directly reuse the environments computed during the Hamiltonian-application step (see Sec.~\ref{sec:applyHamiltonian}) to also compute $T_{\mtwosite}^{\ell}$ via Eq.~\eqref{eq:T2s}, followed by the projection Eq.~\eqref{eq:T2perp}. Then, after terminating the Krylov scheme and determining the eigenbasis of the Krylov Hamiltonian, we subject the collected $T_{2\perp}^{\ell}$ to the
transformation to the Krylov eigenbasis before evaluating Eq.~\eqref{eq:Delta2perp}. 

\changed{
\subsection{Generalizations beyond 1-site TaSK}
\label{app:TaSK-limitations}

To end this section, we discuss two possible limitations which arise from describing excitations purely within the 1-site tangent space $\doubleV^{1\perp}$ and ways of generalizing \task\ to mitigate them.

(i) It may happen that excitations involving several contiguous sites play a crucial role
(e.g.\ if the Hamiltonian contains no terms creating purely single-site excitations). 
To address such cases, the 1-site scheme presented in \Sec{sec:applyHamiltonian} can be readily generalized to 
higher-degree excitations that live beyond $\doubleV^{1\perp}$. For example, one may include excitations living in $\doubleV^{n\mathrm{s}}$, 
the vector space of dimension $\mathcal{O}\bigl((\eLL-n+1)  D^2 d^n\bigr)$, spanned by states differing from the reference state on at most $n$ contiguous sites (see Sec.~V of Ref.~\onlinecite{Gleis2022} on $n$-site excitation Ansatz). This would reduce the tangent space projection error significantly, albeit at an increased computational cost that scales as $\mc{O}(D^3 d^n w)$. For the models considered in this work, we found it sufficient to restrict our attention to the computationally cheapest case of $n=1$.

(ii) Some models have 
ground states that are only weakly entangled, in that they admit exact or near-exact MPS representations with low bond dimension.  A canonical example is the AKLT model, whose ground state is exactly captured by an MPS with the minimal nontrivial bond dimension $D = 2$. In such cases, 
the single-site tangent space $\doubleV^{1\perp}$ used by \task\ may be too small to accurately capture the dynamical properties.

A possible strategy for lifting this limitation is to use, instead of $\mc{P}^{1\perp}$, an expanded projector $\mc{P}^{\mathrm{ex}}$, built from isometries whose bond dimensions have been expanded relative to those of the ground state, $\ket{\Psig}$. The organizing principle for 
constructing the expanded isometries 
is that the projector should satisfy 
$\mc{P}^\mathrm{ex} H^n \ket{\Psi_0} = H^{n} \ket{\Psi_0}$ (with $\ket{\Psi_0} = O_\bk^{\pdag}\ket{\Psig}$) for $n=0, 1, \dots, \bar n$,
so that its image space, ${\doubleV}^{\mathrm{ex}}$, contains the first $\bar n+1$ \textit{exact} Krylov vectors. 
(Note: if $O_\bk$ is a single-particle operator, the $\bar n = 0$
version of this condition is satisfied by the usual 1s tangent space projector,  $\mathcal{P}^{1\perp}$.)  We expect this approach to be  more reliable and efficient than approach (i), but leave its precise formulation and benchmarking  for the future.}

\section{Broadening of discrete spectra}
\label{sec:CFE}

 The spectra obtained through \task\ are discrete. 
 In this section, we describe the broadening strategy we employ to obtain continuous spectral functions. 
Our broadening scheme can be used for generic discrete spectral functions with $N+1$ spectral peaks.
 Within \task, we have $N = \Nkr$ (or $\Nkr+1$ for non-zero $\langle O_{\kk} \rangle$ in Eq.~\eqref{eq:SpectralFunction_TaSK_Nkr} of the main text).
 In the following, 
 \begin{align}
 \label{eq:DiscreteSpectralFunction}
     A^\discrete(\omega) = \sum_{\alpha=0}^N  A_\alpha \delta(\omega - \omega_\alpha)
 \end{align}
 refers to a discrete spectral function with (real, positive) spectral weights $A_\alpha$ and discrete frequencies $\omega_\alpha$.
The continuous, broadened spectral function is referred to as $A^\broadened(\omega)$. 
 If an expression does not specify a superscript $\discrete$ or $\broadened$, it holds for both discrete and broadened data. Superscripts are also specified for quantities derived from the spectra, such as the Green's function
 \begin{align}
 \label{eq:G(z)-appendix}
     G(z) = \int_{-\infty}^{\infty} \mr{d}\omega \, \frac{A(\omega)}{z - \omega} \, , 
     \quad z \in \mathbbm{C} ,
 \end{align}
 or its retarded version, $G^R(\omega) =  G(\omega + \mi 0^+)$.

\subsection{Gaussian broadening}
 \label{subsec:Gaussian_kernel}
\changed{Gaussian broadening uses the kernel}
 \begin{align}
    g_{\sigma}(\omega,\omega') = \frac{1}{\sqrt{2\pi} \sigma} \exp\left(-\frac{(\omega - \omega')^2}{2\sigma^2}\right) 
 \end{align}
\changed{to convert a discrete spectral function 
$A^\discrete (\omega)$ into a broadened, 
continuous one, $A^\broadened (\omega)$, defined as}   
\begin{samepage}
\begin{align}
\label{eq:GaussianBroadening_direct}
    A^\broadened(\omega) &= \int_{-\infty}^{\infty} \mr{d}\omega'  g_{\sigma}(\omega,\omega') A^\discrete(\omega')
    \\ \nonumber
    &= \sum_{\alpha=0}^{N} g_{\sigma}(\omega,\omega_\alpha) A_\alpha \, .
\end{align}
\end{samepage}
The corresponding real part can be computed via Kramers-Kronig relations.

Let $\mu_n$ denote the $n$-th moment of the spectral function:
\begin{align}
\label{eq:moment-mu_n}
\mu_n = \int_{-\infty}^{\infty} \mr{d}\omega \, \omega^n \, A (\omega) \, . 
\end{align}
Gaussian broadening preserves the weight, $\mu^\broadened_0 = \mu^\discrete_0$, and first moment,  $\mu^\broadened_1 = \mu^\discrete_1$,
of the discrete spectral function $A^\discrete(\omega)$, but not the higher moments, 
e.g.\ $\mu_2^\broadened  = \mu_2^\discrete + \sigma^2 \mu_0^\discrete$.
It is of course unavoidable (and intended) that high-order moments of the discrete and broadened data eventually differ. However, it is desirable for the broadening scheme to conserve some low-order moments beyond $\mu_0$ and $\mu_1$, since low-order moments control the short-time dynamics of the retarded Green's function, 
$G^R(t) = 
    \int_{-\infty}^{\infty} \frac{\mr{d}\omega}{2 \pi} \, G^R(\omega) \, e^{-\mr{i}\omega t}
    $:
\begin{subequations}
\label{subeq:G(t)-appendix}
\begin{align}
\label{eq:G(t)-appendix-a}
    \mu_n &=  - \mathrm{Im} \Bigl[
     (\mr{i}\partial_t)^n G^R(t)\Bigr]_{t={0^+}} 
    \\
    \label{eq:G(t)-appendix-b}
    G^R(t) & = - \mi \theta (t) \int_{-\infty}^{\infty} \mr{d}\omega \, A(\omega) \, e^{-\mr{i}\omega t} \, .
\end{align}
\end{subequations}
This short-time dynamics is controlled by physics on short distances, for which finite-size effects or MPS truncation errors should be small. A broadening strategy that preserves low-order moments is described in the next section. It is based
on representing the discrete Green's function $G^{\rm{d}}(z)$ through a continued fraction expansion (CFE) of finite depth, then using the tail-end part of the expansion to define a discrete residual function, and finally applying Gaussian broadening to the latter.

\subsection{CFE broadening}
\label{subsec:CFE_discrete}

It is a well-known fact \cite{Koch2011} that any Green's function of the form \eqref{eq:G(z)-appendix}
can be represented via a CFE of the form
\begin{align}
\label{eq:G_CFE}
    G(z) = \cfrac{t^2_0}{z - \epsilon_0 - \cfrac{t_1^2}{z - \epsilon_1 - \;\raisebox{-0.5em}{\ensuremath{\ddots}}}} \; \; .
\end{align}
If the spectral function $A(\omega)$ of $G(z)$ is known, the corresponding CFE coefficients can be obtained using a Lanczos scheme. Here, we describe this scheme for the case, relevant for this paper, of a discrete spectral function $A^{\rm{d}}(\omega)$ of the form \eqref{eq:DiscreteSpectralFunction}, with $\alpha=0, \dots, N$. 
Then, the discrete Green's function can be expressed as 
\begin{subequations}
\begin{align}
    G^\discrete(z) & = \sum_{\alpha = 0}^N \frac{A_\alpha}{z - \omega_\alpha}
 \\    & =  t_0^2 \sum_{\alpha=0}^{N} \langle v_0 | e_\alpha \rangle \bigl[(z - \Omega)^{-1} \bigr]_{\alpha\alpha}
     \langle e_\alpha | v_0 \rangle  \, ,
\label{eq:G-discrete}
\end{align}
\end{subequations}
where $\Omega$ is a diagonal matrix with elements 
$\Omega_{\alpha \alpha'} = \langle e_\alpha| \Omega |e_{\alpha'}\rangle = \delta_{\alpha \alpha'} \omega_\alpha$, and the normalized vector 
 $\ket{v_0}$ has components $\langle e_\alpha| v_0\rangle = \sqrt{A_\alpha}/t_0$, 
with $t_0^2 = \sum_{\alpha} A_\alpha$. Since $G^\discrete(z)$ has $N\!+\!1$ poles, the corresponding CFE \eqref{eq:G_CFE} will terminate at depth $N\!+\!1$. Given the weights and frequencies  $\{A_\alpha,\omega_\alpha\}$, the CFE coefficients 
for $G^{\rm{d}}(z)$ can be computed as follows: Use the Lanczos algorithm,
initialized with $\ket{v_0}$, to construct a set of orthonormal Krylov vectors
$\{\ket{v_0} , \ket{v_1}, \dots , \ket{v_N}\}$ 
that span the Krylov space $K_N = \mathrm{span} \{\ket{v_0} , \Omega \ket{v_0}, \dots, \Omega^N \ket{v_0}\}$. When $\Omega$ is expressed in the Krylov basis 
it takes the form of a  matrix $M$, with elements 
$M_{nn'} = \sum_\alpha \langle v_n |e_\alpha\rangle \omega_s \langle e_\alpha | v_{n'} \rangle$, 
that has a tridiagonal form:   
\begin{align}
\label{eq:Omega-to-M}
    M = 
    \begin{pmatrix}
\epsilon_{0} & t_{1} & 0 & 
\cdots & \hspace{-2mm} 0 & \hspace{-1mm} 0  \\
t_{1} & \epsilon_{1} & t_{2} & 
\cdots & \hspace{-2mm}0 & \hspace{-1mm} 0  \\
0 & t_{2} & \epsilon_{2} &   
\cdots & \hspace{-2mm}0 & \hspace{-1mm} 0  \\[-1mm]
\vdots & \vdots & \vdots & 
\ddots & \hspace{-2mm} \vdots & \hspace{-1mm} \vdots \\[-1mm] 
0 & 0 & 
\cdots & \hspace{-2mm} \epsilon_{\NKrylov-2} 
& \hspace{-1mm} t_{\NKrylov-1}^\pdag & 0  \\
0 & 0 & 
\cdots &  \hspace{-1mm} t_{\NKrylov-1}^\dagger & \hspace{-1mm} \epsilon_{\NKrylov-1} & t_\NKrylov \\ 
0 & 0 & 
\cdots & \hspace{-2mm} 0 & \hspace{-1mm} t_\NKrylov & \epsilon_\NKrylov
\end{pmatrix} \! .
\end{align}
Using a standard formula for the inverse of a block matrix, the Green's function \eqref{eq:G-discrete} can thus be expressed 
through the top-left element of the matrix $(z-M)^{-1}$, as
\begin{align}
\label{eq:G_tridiag}
    G^\discrete(z) = t_0^2\bigl[(z-M)^{-1}\bigr]_{00}  
    \, .
\end{align}
Equation~\eqref{eq:G_tridiag} immediately 
implies a depth-$(N+1)$ CFE of the form~\eqref{eq:G_CFE}, with CFE coefficients 
$\{\epsilon_n, t_n\}$ given 
by the diagonal and next-to-diagonal elements of  $M$, and terminated by $-t_N^2/(z-\epsilon_N)$, see for instance Eqs.~(23-25) of Ref.~\onlinecite{Koch2011}. This CFE gives an exact representation of $G^\discrete(z)$ and hence still represents a discrete Green's function. A scheme for broadening it is described next. 

The CFE coefficients are related to the moments $\mu_n = \bra{v_0} \Omega^{n} \ket{v_0}$. 
In particular, if the  CFE coefficients up to depth $m+1$ (i.e.\ $\epsilon_0,t_0$ to $\epsilon_m, t_m$) of two Green's functions are identical, their first $2m+2$ moments, $\mu_0,\dots, \mu_{2m+1}
=\bra{v_0} \Omega^{m} \cdot \Omega \cdot \Omega^{m}\ket{v_0}$, are also identical.
The reason is the matrix representations of the $\Omega$s of both Green's functions coincide within their respective depth $m+1$ Krylov spaces $K_m = \mr{span} (\ket{v_0},\Omega \ket{v_0}, \dots,\Omega^{m} \ket{v_0})$.

\subsection{Broadening the residual function}
\label{eq:residual-broadening}

We seek a scheme for broadening $G^\discrete(z)$ that preserves the first $2m+2$ moments, with $m$ typically much smaller than $N$. To achieve this, the low-order CFE coefficients with $n \le m$ of the broadened Green's function $G^\broadened(z)$ should coincide with those of the discrete $G^\discrete(z)$. We therefore express the latter as
\begin{flalign}
\label{eq:G_CFE_R}
    & G^{\rm{d}} (z) = \cfrac{t^2_0}{z - \epsilon_0 - \cfrac{t_1^2}{z - \epsilon_1 - \;\raisebox{-0.8em}{\ensuremath{\ddots
    \raisebox{-0.5em}{\ensuremath{ - \cfrac{t_{m}^2}{z - \epsilon_m - R^\discrete_{m}(z)}
    }}}}}} \; , \hspace{-1cm} &
\end{flalign}
where we have used the CFE coefficients $\{\epsilon_n, t_n\}$ with $n > m$ to define the 
residual function 
\begin{align}
\label{eq:define-residual-R-CFE}
    R^\discrete_{m}(z) = \cfrac{t^2_{m+1}}{z - \epsilon_{m+1} - \cfrac{t_{m+2}^2}{z - \epsilon_{m+2} - \;\raisebox{-0.8em}{\ensuremath{\ddots
    \raisebox{-0.5em}{\ensuremath{-  \cfrac{t_{N}^2}{z - \epsilon_N}}}}}}} \; .
\end{align}
This function, being defined through a finite-depth CFE, itself represents a discrete Green's function. The key idea is now to broaden 
this residual function (using the Gaussian kernel of Sec.~\ref{subsec:Gaussian_kernel}) to obtain a continuous version, $R^\broadened_{m}(z)$. Inserting the latter into the CFE \eqref{eq:G_CFE_R} then yields the desired continuous, broadened Green's function
\begin{flalign}
\label{eq:G^c-broadened-R}
     & G^{\mathrm{c}}(z) = \cfrac{t^2_0}{z - \epsilon_0 - \cfrac{t_1^2}{z - \epsilon_1 - \;\raisebox{-0.8em}{\ensuremath{\ddots
    \raisebox{-0.5em}{\ensuremath{
    - \cfrac{t_m^2}{z - \epsilon_{m} - R^\broadened_{m}(z)}
    }}}}}} \; . \hspace{-1cm} &
\end{flalign}
Since $G^{\mathrm{c}}(z)$ and $G^{\mathrm{d}}(z)$ have the same CFE coefficients 
for $n\leq m$, their first $2m+2$ moments are equal, too.

The CFE form \eqref{eq:define-residual-R-CFE} of  $R^\discrete(z)$ is not convenient for computing its convolution 
with a Gaussian kernel. Instead, we first express it in the form of a spectral representation analogous to \Eq{eq:G-discrete}. To this end, we reverse the
steps leading from \Eq{eq:G-discrete} to \eqref{eq:G_tridiag}.
We express $R^\discrete_{m}(z)$ as 
\begin{align}
    R^\discrete_{m}(z) &= t_{m+1}^2 \bigl[(z - \Mt)^{-1}\bigr]_{00} ,
\end{align}
involving the top-left element of the matrix $(z-\Mt)^{-1}$, 
where $\Mt$ is the matrix
\begin{align}
    \Mt &= 
\begin{pmatrix}
\epsilon_{m+1} & t_{m+2} & 0 & 
\cdots & \hspace{-2mm} 0 & \hspace{-1mm} 0  \\
t_{m+2} & \epsilon_{m+2} & t_{m+3} & 
\cdots & \hspace{-2mm}0 & \hspace{-1mm} 0  \\
0 & t_{m+3} & \epsilon_{m+3} &   
\cdots & \hspace{-2mm}0 & \hspace{-1mm} 0  \\[-1mm]
\vdots & \vdots & \vdots & 
\ddots & \hspace{-2mm} \vdots & \hspace{-1mm} \vdots \\[-1mm] 
0 & 0 & 
\cdots & \hspace{-2mm} \epsilon_{\NKrylov-2} 
& \hspace{-1mm} t_{\NKrylov-1}^\pdag & 0  \\
0 & 0 & 
\cdots &  \hspace{-1mm} t_{\NKrylov-1}^\dagger & \hspace{-1mm} \epsilon_{\NKrylov-1} & t_\NKrylov \\ 
0 & 0 & 
\cdots & \hspace{-2mm} 0 & \hspace{-1mm} t_\NKrylov & \epsilon_\NKrylov
\end{pmatrix} \! .
\end{align}
In other words, $\Mt$ is a submatrix of $M$, with elements  $\Mt_{nn'} = \bra{\vt_n} M \ket{\vt_{n'}}$,
where we defined $\ket{\vt_n} = \ket{v_{m+1+ n}}$ for $n=0, \dots, N-m-1$. We diagonalize this matrix
to find its eigenvalues $\omegat_\alpha$ and
eigenvectors $\ket{\et_\alpha}$, so that its elements can 
be expressed as 
$\Mt_{nn'}  = \sum_{\alpha=0}^{N-m-1}
\langle \vt_n | \et_\alpha \rangle \omegat_\alpha \langle \et_\alpha | \vt_{n'} \rangle . 
$ 
By analogy to \Eq{eq:G-discrete}, $R^\discrete_{m}(z)$ thus has a spectral representation of the form
\begin{align}
    R^\discrete_{m}(z) &= t_{m+1}^2\sum_{\alpha = 0}^{N-m-1} 
  \frac{|\langle \et_\alpha | \vt_0 \rangle|^2}{z - \omegat_\alpha}  \, ,
\end{align} 
from which we can read off the discrete weights and frequencies defining
its spectral function.

Broadening the latter through convolution with a Gaussian  using 
\Eq{eq:GaussianBroadening_direct}, as discussed in Sec.~\ref{subsec:Gaussian_kernel}, we obtain the desired continuous, broadened 
residual function $R^\broadened_{m}(z)$, to be inserted 
into \Eq{eq:G^c-broadened-R} for $G^\broadened(z)$.
Since Gaussian broadening conserves the zeroth \textit{and} first moments of the residual function, the above scheme conserves the first $2m +4$ moments of $A(\omega)$, $\mu_0,\dots, \mu_{2m+3}$. 

In subsequent sections on numerical results, we will denote the depth of the CFE up to the residual function by $\mathcal{N}_\mathrm{CFE} = m+1$;
the corresponding number of conserved moments is $2\mathcal{N}_\mathrm{CFE} + 2$. For $\mathcal{N}_\mathrm{CFE} = 0$, i.e.\  $m = -1$, 
the residual function \eqref{eq:define-residual-R-CFE} is
the Green's function itself, $R^\discrete_{-1}(z) = G^\discrete(z)$;
then no CFE is involved and we in effect apply Gaussian broadening to the spectral function, which preserves $\mu_0$ and $\mu_1$.

\section{TaSK convergence}
\label{sec:Convergence}

\task\ method is subject to two sources of error: (i) the projection error arising from 
the tangent-space projection of the Hamiltonian, $H \to H^{1\perp}$, and (ii) an error due to an insufficient number of Krylov steps $\Nkr$. Since the dimension of $\doubleV^{1\perp}$ scales as $\mc{O}(\eLL  D^2 d)$, we can address the former by performing a more precise ground-state~(GS) calculation with larger bond dimension $D$, while the latter can be reduced by simply increasing $\Nkr$.

To control the projection error, we fix a bond-dimension threshold for the GS and expand $D$ during the DMRG GS search using DMRG3S-like mixing~\cite{Hubig2015} together with controlled bond expansion~\cite{Gleis2023} (see specifically Sec.~3 of the supplemental material of Ref.~\onlinecite{Gleis2023} and also Ref.~\onlinecite{Gleis2025b}) until this threshold is saturated. None of the models tested here have an exact MPS representation of the ground state. However, for the $\eLL = 64$ site Heisenberg model, our choice $D^{\ast} = 512$ is so large that the ground state energy is converged to machine precision.

In the following sections, we explore \task's convergence with bond dimension $D$ (or, in our case, the number of SU(2) multiplets $D^*$) and the number of Krylov steps $\Nkr$. 
In Sec.~\ref{sec:Convergence_Spins}, we discuss how the discrete spectra of the two benchmark spin-chain models, the Haldane-Shastry and the Heisenberg model, converge with $\Nkr$. Section~\ref{sec:Convergence_Fermions} discusses the convergence of continuous spectral functions for the Hubbard model on a cylinder with both $D^*$ and $\Nkr$ and shows how their resolution changes depending on the choice of broadening parameters introduced in Sec.~\ref{sec:CFE}. 
 
\begin{figure}[h]
\includegraphics[width=1.02\linewidth]{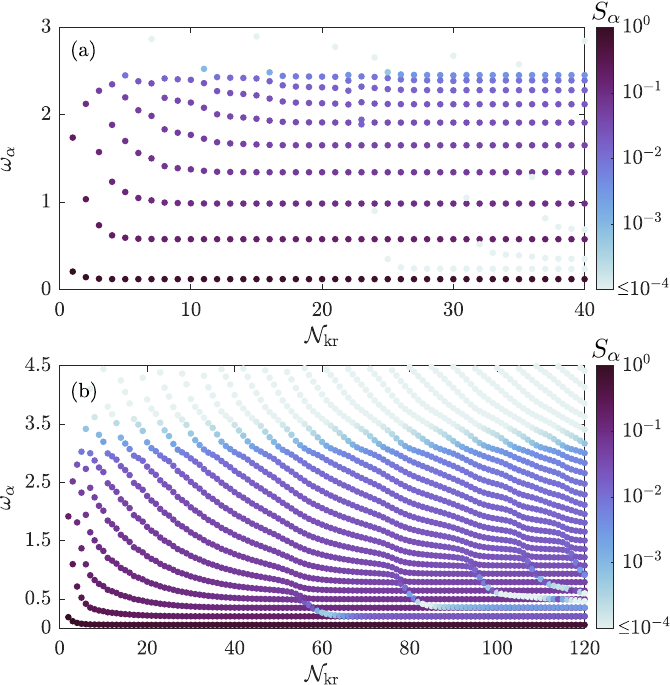} \vspace{-6mm}
 \caption{
 \label{fig:Convergence_Krylov}
Convergence of $\omega_\alpha$ with the number of Krylov steps $\mathcal{N}_{\mathrm{kr}}$ at $k=\pi$ for (a) the Haldane-Shastry model on a ring of size $\eLL = 40$ with $D^* = 700$, and (b) the Heisenberg chain of length $\eLL = 64$ with $D^* = 512$. The color scale indicates spectral weights $S_\alpha$, represented on a logarithmic scale. Spectral weights $\leq 10^{-4}$ are clipped to the minimum color value to highlight dominant contributions to the DSF.
}
\vspace{-7mm}
 \end{figure}

\subsection{Spin Models}
\label{sec:Convergence_Spins}

Figure~\ref{fig:Convergence_Krylov} demonstrates convergence of discrete frequencies $\omega_\alpha$ with the number of Krylov steps $\Nkr$ for (a) the Haldane-Shastry model~(HSM) at $D^*=700$, and $\eLL=40$ and (b) the Heisenberg chain at $D^*=512$ and $\eLL=64$. As shown in Fig.~\ref{fig:Convergence_Krylov}(a), the HSM exhibits rapid convergence across the entire spectrum, saturating around $\Nkr = 30$. In the low-frequency region, we observe three peaks that do not correspond to any physical excitations (when compared with analytical results for the full \textit{unprojected} Hamiltonian). However, their weight $[S_{\pi}]_\alpha$ is negligible ($\sim \mc{O}(10^{-12})$) and their error estimate $\delta_\alpha$ is at least three orders of magnitude higher than that of the remaining peaks on \textit{the same energy scale} -- an important distinction, since the error monotonically increases towards the high-frequency region, as shown in Fig.~\ref{fig:HaldaneShastry}(b). 
 
We conjecture that (i) these low-weight peaks are very likely \textit{not} ghost states as they often appear in Lanczos schemes, but (ii) rather a result of the projection to the tangent space, which modifies the Hamiltonian.
The reason for conjecture (i) is that we are working with full orthonormalization (see Sec.~\ref{sec:Lanczos}). 
As a result, the eigenstates generated by our Lanczos scheme are mutually orthogonal within machine precision (overlaps $<10^{-16})$, i.e.\ they are indeed independent eigenstates of $H^{1\perp}$.
Conjecture (ii), that these states 
result from the tangent space projection, is based on the fact that they have a large 2-site variance $\Delta_\alpha^{2\perp} $ relative to other states on the same energy scale. This means that these states have a substantial projection error, i.e.\ reducing the projection error, for instance by expanding $\doubleVonesite$ or by working with $\doubleVtwosite$ (the space
of two-contiguous-site variations w.r.t.\ the ground state \cite{Gleis2022})
will substantially affect these states. By contrast, the small variance of the high-weight states (see Fig.~\ref{fig:HaldaneShastry} of the main text and its discussion) indicates that the same reduction of the projection error hardly affects these states.

For the Heisenberg chain, the convergence of the dynamical structure factor $S_{\pi}(\omega)$ 
with $\Nkr$ is slower than for the HSM, but systematic, as seen in Fig.~\ref{fig:Convergence_Krylov}(b). This is expected, since the spinon excitations of the Heisenberg model are more complex and lead to dense, broadly distributed spectral weight. 
Nonetheless, the low-frequency part of the spectrum converges rapidly. 
By contrast, the high-frequency part is already \textit{qualitatively} well resolved at moderate $\Nkr$ (see Fig.~\ref{fig:HeisenbergPowerLaw} of the main text), whereas a \textit{quantitative} resolution of the entire discrete spectrum would require more iterations.

\begin{figure}[t]
\includegraphics[width=1\linewidth]{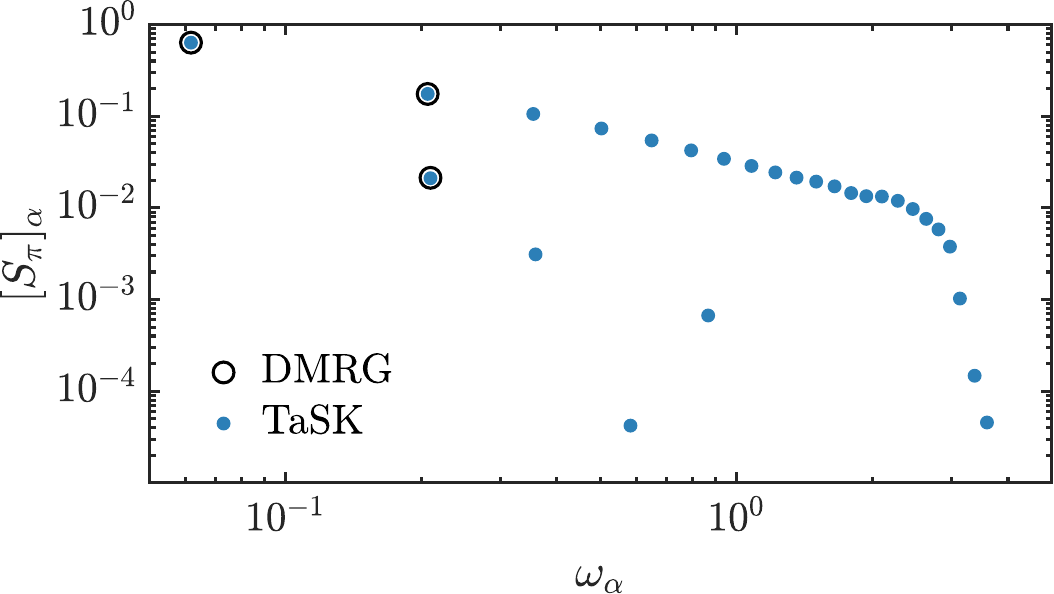} \vspace{-6mm}
 \caption{
 \label{fig:Heisenberg_task_vs_dmrg}
Discrete spectral data for the dynamical structure factor of the $\eLL = 64$-site Heisenberg chain at $k=\pi$. Blue dots are computed via \task, black circles by targeting low-energy states with DMRG. 
We obtained the \task\ data in Fig.~\ref{fig:HeisenbergPowerLaw} of the main text by binning the data shown here, as described near the end of Sec.~\ref{sec:Convergence_Spins}.}
\vspace{-4mm}
 \end{figure}

For sufficiently large $\Nkr$, some of the low-frequency spectral peaks develop a secondary satellite that is nearly degenerate with the main peak. Even though their weight is at least about an order of magnitude lower than that of the main peak, it is nonetheless substantial, unlike in the HSM case. Since we employ full reorthogonalization, ensuring that the obtained Krylov states are mutually orthogonal within numerical accuracy, we again conclude that these indeed correspond to excitations of $H^{1\perp}$, not Lanczos ghosts.

To check whether they are also 
eigenstates of the unprojected Hamiltonian $H$, we target the four lowest-energy triplets ($S=1$ sector, $4\cdot 3 = 12$ states in total) of the $\eLL = 64$ Heisenberg chain using DMRG. These are denoted $\ket{\Psi^{s}_{\alpha}}$, where $s \in \{1,0,-1\}$ denotes the $S^{z}$ eigenvalue, and are used to compute the corresponding spectral weights, $[S_{\pi}]_{\alpha} = |\langle \Psi^{0}_{\alpha}| 
\mathrm{S}^{z}_{\pi} \ket{\Psig}|^2$. Since one of these triplets has momentum different from $\pi$, its spectral weight is zero. Consequently, we only recover the first three spectral peaks of $S_{\pi}(\omega)$. 

The result of this computation is shown in Fig.~\ref{fig:Heisenberg_task_vs_dmrg}, where we compare it with the discrete \task\ data. Evidently, the three lowest-frequency spectral peaks coincide very well, including the satellite of the second main peak. This demonstrates that at least this satellite is a genuine excitation of $H$, and not an artifact of the tangent space projection. Therefore, in the postprocessing of spectral data for this model, we treat all excitations of this kind as physical. Specifically, to plot Fig.~\ref{fig:HeisenbergPowerLaw} of the main text, we proceeded in two steps: (i) binning and (ii) rescaling. We now describe these in turn. 

(i) \textit{Binning:} If a large-weight main peak has a small-weight satellite very close to it (within, say, $0.1 \widetilde \Delta$, where $\widetilde \Delta$ is the average peak spacing
in the vicinity of the main peak), we  replace the two by a single combined peak with weight 
$S_\binned = S_\main + S_\satellite$ and position 
\begin{align}
\omega_\binned  = 
\frac{S_\main \omega_\main +S_\satellite\omega_\satellite}{S_\main + S_\satellite} \, .
\end{align}
(ii) \textit{Rescaling:} For binned discrete spectral data of the 
form $\{\omega_\alpha, S_\alpha\}$, both the average peak spacing and the weight per peak scale as $1/\eLL$. To combine such 
spectra from different system sizes in a single plot (e.g.\ $\eLL = 64$, $128$ in Fig.~\ref{fig:HeisenbergPowerLaw}), we 
use rescaled, $\eLL$-independent weights 
per frequency interval [following Ref.~\onlinecite{vonDelftHolzner2011}, Sec.~IV (D)]:
we plot the points $\{\omega_\alpha, \tilde S_\alpha\}$, where $\tilde S_\alpha$ is the weight $S_\alpha$  
divided by the width of the frequency interval hosting it: 
\begin{align}
\tilde S_\alpha = S_\alpha/\Delta_\alpha, 
  \quad   \Delta_\alpha = \frac{\omega_{\alpha+1}-\omega_{\alpha-1}}{2}.
\end{align}
The reason for binning before rescaling is now apparent:  without binning, the distance between a satellite peak and the closest (or next-closest) main peaks would be anomalously small (or large), yielding 
anomalously large (or small)  weights $\widetilde S_\alpha$.  (An alternative to binning would be discarding  satellites and plotting only main peaks. That would  amount to discarding a small amount of spectral weight, thus weakly violating a spectral sum rule.)

\begin{figure}[t]
\includegraphics[width=1.04\linewidth]{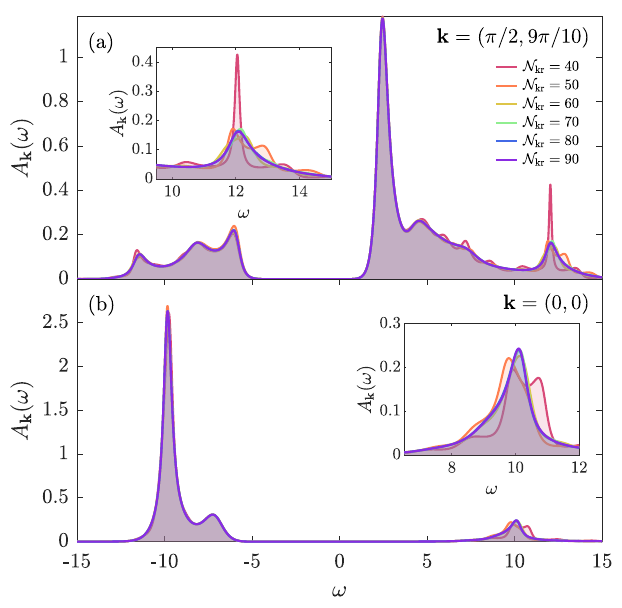} \vspace{-6mm}
 \caption{
 \label{fig:Convergence_Hubbard}
Convergence of $A_{\bk}(\omega)$ with the number of Krylov steps $\mathcal{N}_{\mathrm{kr}}$ for the $\tUn$ Hubbard model with $D^* = 2000$ for 
(a) $\bk=(\pi/2,9\pi/10)$ and (b) $\bk=\Gamma=(0,0)$.}
\vspace{-4mm}
 \end{figure}

\begin{figure*}[t]
\includegraphics[width=\textwidth]{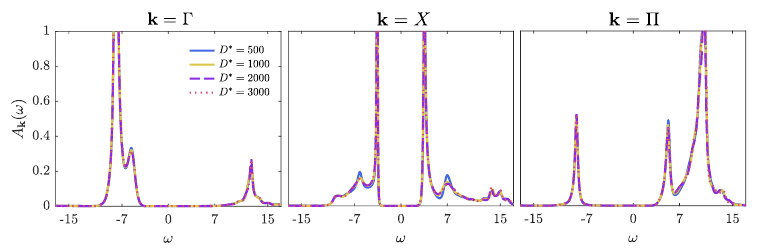} \vspace{-6mm}
 \caption{
 \label{fig:Convergence_BondDim}
Convergence of $A_\bk(\omega)$ with $D^*$ for the $t\textrm{-}t_n\textrm{-}U$ Hubbard model at the high symmetry points $\Gamma$, $X$ and $\Pi$.
For these plots, we used the broadening parameters  $\mathcal{N}_{\mr{CFE}} = 2$ and $\sigma = 0.5$.}

\vspace{-4mm}
 \end{figure*}

\subsection{Hubbard Model}
\label{sec:Convergence_Fermions}
 
Figure~\ref{fig:Convergence_Hubbard} demonstrates convergence of the broadened spectral function $A^\broadened(\omega)$ with $\Nkr$ for the 2D Hubbard model with density-assisted hopping. The parameters of the model ($U$, $t$, $t_n$) are identical to those specified in the main text. For the ground-state calculation, we retain $D^* = 2000$ $\mathrm{SU}(2)$ multiplets. Figures~\ref{fig:Convergence_Hubbard}(a) and 
\ref{fig:Convergence_Hubbard}(b) correspond to momentum points of $\bk=(\pi/2,9\pi/10)$ and $\bk=\Gamma=(0,0)$, respectively. 
For the Gaussian broadening of the residual function we set the broadening parameters (defined in Sec.~\ref{eq:residual-broadening}) to $\sigma = 0.5$ and $\mathcal{N}_{\mathrm{CFE}} = 2$, preserving the first 6 moments of $A_\kk(\omega)$. We use $\Nkr=50$ for the $\tU$ and $\tUSigma$ computations
in the main text [\Figs{fig:CylinderPlots}(a,b,e,f)], and 
$\Nkr=60$ for $\tUn$ [\Figs{fig:CylinderPlots}(c,d)],
which we find sufficient for a qualitative description of the dominant spectral features.

\begin{figure}[b]
\includegraphics[width=1.04\linewidth]{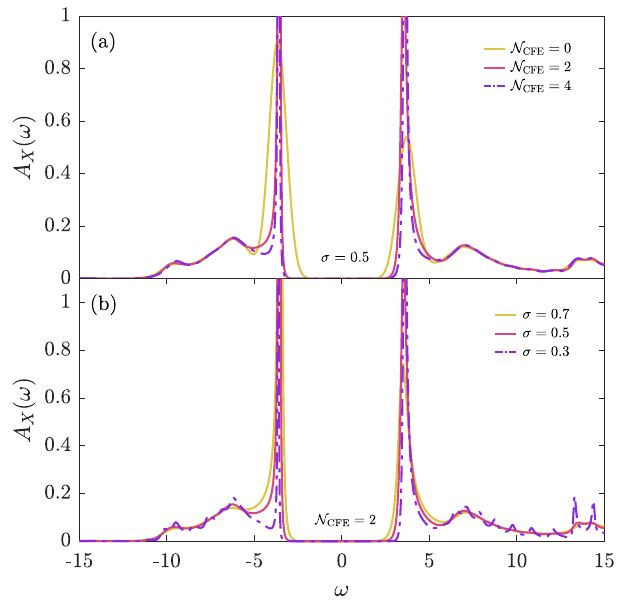} \vspace{-6mm}
 \caption{
 \label{fig:Broadening_Hubbard}
Comparison of $A_X(\omega)$ of the $\tUn$ Hubbard model for different choices of broadening parameters: (a) three values of $\mathcal{N}_{\mathrm{CFE}}$ at fixed $\sigma = 0.5$, and (b) three values of $\sigma$ at fixed $\mathcal{N}_{\mathrm{CFE}} = 2$.}
\vspace{-7mm}
 \end{figure}

Figure~\ref{fig:Convergence_BondDim} demonstrates convergence of the $\tUn$ Hubbard model with the number of SU(2) multiplets $D^*$ for $\kk = \Gamma$, $X$ and $\Pi$. We use $\Nkr=50$ and set the broadening parameters as above. For the computations in the main text, we used $D^* = 3000$ (discarded weight $\xi\sim \mc{O}(10^{-8})$) for the pure $\tU$ Hubbard model and (having realized that a smaller $D^\ast$ would have sufficed) proceeded with $D^* = 2000$ ($\xi\sim \mc{O}(10^{-8})$) for $\tUn$. As seen in Fig.~\ref{fig:Convergence_BondDim}, for the chosen values of the broadening parameters, spectral line cuts computed for $D^\ast = 1000$, $2000$ and $3000$ indeed coincide, and even for $D^* = 500$ we already recover the overall shape of the spectral function  remarkably well.

Finally, in Fig.~\ref{fig:Broadening_Hubbard} we show how different choices of broadening parameters, $\mathcal{N}_{\mathrm{CFE}}$ and $\sigma$, influence the resolution of the broadened spectral function $A^\broadened(\omega)$ for the $\tUn$ Hubbard model. It is to be expected that setting $\mathcal{N}_{\mathrm{CFE}}$ too high or similarly setting $\sigma$ too low results in an underbroadened $A^\broadened(\omega)$.
As an example, we compare $A^\broadened(\omega)$ at $\kk = X$ for different choices of $\sigma$ [Fig.~\ref{fig:Broadening_Hubbard}(a)] or $\mathcal{N}_{\mathrm{CFE}}$ [Fig.~\ref{fig:Broadening_Hubbard}(b)], while keeping the corresponding other broadening parameter fixed. From this, we conclude that choosing the Gaussian width $\sigma$ too small leads to a uniformly underbroadened spectrum, i.e.\ wiggles appear across the entire frequency range. 
Increasing $\mathcal{N}_{\mathrm{CFE}}$, on the other hand, has no substantial effect on the spectral continuum and mostly 
sharpens the main spectral peaks.
Nevertheless, too large $\mathcal{N}_{\mathrm{CFE}}$ eventually leads to overly sharp main peaks, and a false occurrence of sharp peaks in the continuum. In the extreme case, when $\mathcal{N}_{\mathrm{CFE}}$ matches the number of discrete spectral peaks, no broadening occurs at all. The general rule is that $\sigma$ controls the broadening of the incoherent background and should 
be chosen such that wiggles are broadened out while genuine structure is not blurred. At the same time, $\mathcal{N}_{\mathrm{CFE}}$ controls the sharpness of the more coherent features (the ``main peaks''), and should not be chosen too large to avoid the spurious occurrence of sharp peaks in the continuum. Therefore, within the scope of this work, we broaden the discrete spectral functions of the Hubbard model using 
$\sigma = 0.5$ and $\mathcal{N}_{\mathrm{CFE}} = 2$, which provides a good balance between spectral resolution and smoothness.

\begin{figure}[b]
\includegraphics[width=\linewidth]{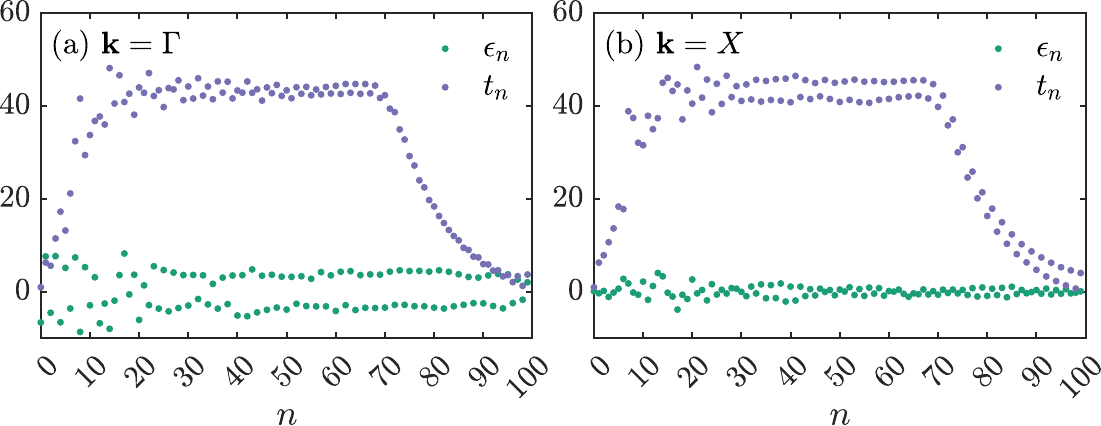} \vspace{-6mm}
 \caption{
 \label{fig:CFE_coefficients_Hubbard}
$n$-dependence of the CFE coefficients $\{\epsilon_n,t_n\}$ of $G_X(\omega)$ for the $\tU$ Hubbard model for $\bk = \Gamma$ and $X$.}
\vspace{-4mm}
 \end{figure}

We conclude this section by remarking that it would
be useful, in future work, to study the $n$-dependence of the CFE coefficients $\{\epsilon_n,t_n\}$ in some detail, for two reasons. First, if their $n$-dependence exhibits a regular pattern over a reasonably large range of $n$, one may consider extrapolating them to even larger $n$. That could yield improved resolution without the computational cost of performing more Lanczos iterations. Second, knowing 
their $n$-dependence could be exploited to fine-tune how we broaden the residual function, since broadening it in effect modifies the CFE coefficients for $n >  \mathcal{N}_\mathrm{CFE}$. 

For our Hubbard model data, the CFE coefficients do 
not show a very regular $n$-dependence, as exemplified 
in \Fig{fig:CFE_coefficients_Hubbard} for two fixed choices of $\bk$. We leave a systematic analysis of the $n$-dependence seen there to future work.

\changed{
\section{Additional benchmarks}
\label{sec:AdditionalBenchmarks}

In this section, we provide additional benchmark results illustrating the performance of TaSK for 
two types of spectral functions not analyzed in detail in the main text.
Section~\ref{sec:Heisenberg-k-not-pi/2}
considers spectral functions having gaps
(i.e.\ where dominant weights lie at frequencies away from $\omega=0$), showing that \task\ manages to resolve these reliably. Section~\ref{sup:current-susceptibility}, too, deals with a gapped spectrum, but with an additional challenge: it 
considers the correlator of the current operator, 
which is not purely local, and illustrates that the 1-site tangent space explored by \task\ is large enough to capture its dominant contributions.}

\changed{\subsection{Gapped spectra: Heisenberg model at  $k \neq \pi$}
\label{sec:Heisenberg-k-not-pi/2}}

\begin{figure}[t]
\includegraphics[width=0.9\linewidth]{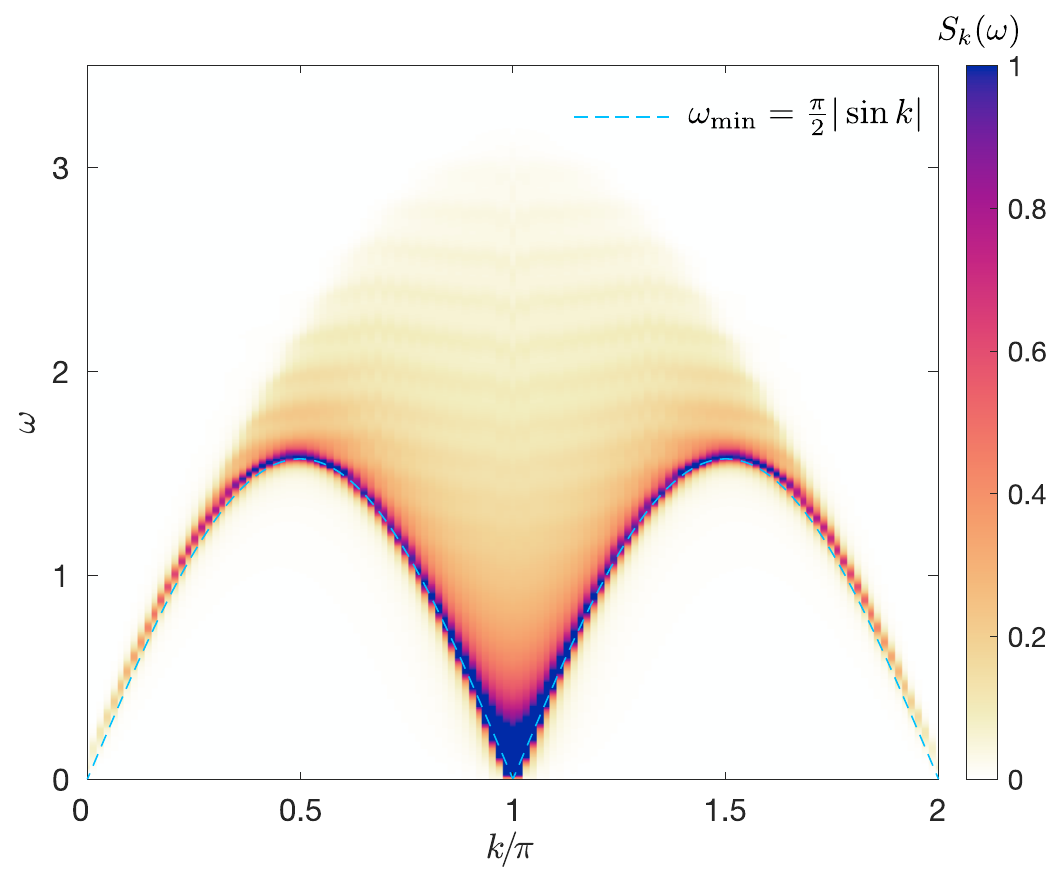} \vspace{-5mm}
\caption{
\label{fig:HeisenbergPineapple}
\changed{Dynamical structure factor $S_k(\omega)$ of the spin-$\frac{1}{2}$ Heisenberg chain, 
computed by \task\ as a function of $\omega$ and $k$, showing the full multi-spinon continuum. We used $\eLL = 64$, $D^\ast = 512$ and 
$\Nkr=100$. The discrete \task\ data was broadened using $\sigma = 0.08$ and $\mathcal{N}_{\mr{CFE}}=2$. Dashed blue line: the treshold frequency 
$\omega_\mathrm{min} = \frac{\pi}{2} |
\sin k|$.}}
\end{figure}

\changed{In the main text, we considered 
DSFs of two spin models (Haldane-Shastry and  
spin-$\tfrac{1}{2}$ Heisenberg chain)
at a choice of momentum, $k=\pi$, 
where the DSFs have no gap 
(see Figs.~\ref{fig:HaldaneShastry}, \ref{fig:HeisenbergPowerLaw}). Then, spectral weight  is dominantly concentrated near $\omega = 0$, where \task\ is expected to perform best. In this
subsection, we address the question:  how does \task\  perform for gapped spectra, where the dominant spectral features are located away from $\omega = 0$?

To this end, we consider the DSF 
$S_k (\omega)$ of the spin-$\tfrac{1}{2}$ Heisenberg chain at momentum values $k \neq \pi$. 
There, the spectrum is gapped, with the lower boundary given by the des Cloizeaux-Pearson dispersion relation~\cite{desCloizeaux1962}, $\omega_\mr{min} = \tfrac{\pi}{2} |\sin{k}|$.
This gapped behavior is well reproduced by \task, as seen in Figure~\ref{fig:HeisenbergPineapple},  
which shows the full multi-spinon continuum as a function of $\omega$ and $k$.} 

\changed{Next, we focus on the line-shape of $S_k(\omega)$
at $\pi/2$, shown in Figure~\ref{fig:HeisenbergPihalf} (the counterpart of 
\Fig{fig:HeisenbergPowerLaw} of the main text for $k=\pi$).
Solid dots show \task\ results 
for $\eLL = 64$ and $\eLL = 128$. 
Solid and dashed lines show exact Bethe-ansatz results \cite{Caux2006}, 
which exhibit a divergence at $\omega_\mathrm{min} =\pi/2$.
For above-gap frequencies, 
$\omega_\alpha > \omega_\mr{min}$, 
the discrete \task \ data is in excellent agreement with the Bethe-ansatz curves. Similarly to $k=\pi$, most frequencies $\omega_\alpha$ found by \task\ correspond to the full $2+4$-spinon contribution.
\changed{\task\ also yields some below-gap $(\omega_\alpha < \omega_\mathrm{min}$) excitations with very low weight.
These are not artifacts of the method but true excitations,  expected due to the finite system size; their weight will 
vanish for $\eLL \to \infty$.}

\begin{figure}[t]
 \includegraphics[width=1.025\linewidth]{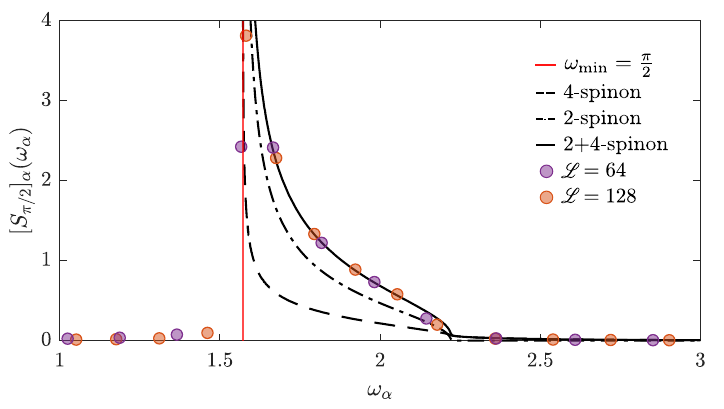} \vspace{-7mm}
 \caption{
 \label{fig:HeisenbergPihalf}
\changed{Dynamical structure factor of the spin-$\frac{1}{2}$ Heisenberg chain at $k = \pi/2$. We compare \task\ results for $D^\ast = 512$ and $\Nkr = 100$ and $ \Nkr = 140$ for $\eLL = 64$ and $\eLL = 128$, respectively, (symbols)
to exact Bethe ansatz results from Ref.~\onlinecite{Caux2006} (solid and dashed lines). Above $\omega_\mathrm{min} = \pi/2$ (solid red line), the exact thermodynamic-limit threshold frequency, the agreement is excellent. Sub-threshold excitations reflect finite-size effects.}}
 \end{figure} 

Finally, we note that when the \task\ result for $S_{\pi/2}(\omega)$  is Fourier-transformed to the time domain, 
it agrees very well with TDVP results, see Fig.~\ref{fig:TDVP_vs_TaSK_Heisenberg}(b) below.

Additional examples of \task-resolved dynamical correlation functions with spectral features away from $\omega = 0$ are given in the the next subsection, and in Ref.~\cite{Kovalska2026} on the $J_1$-$J_2$ Heisenberg model on the triangular lattice.

\bigskip 

\subsection{Non-local operators: current susceptibility} 
\label{sup:current-susceptibility}

In this section, we examine the performance of \task\ for dynamic correlators of \textit{non-local} operators, consisting of sums of terms having support on more than one site. 

As an example of a non-local operator we  consider the current operator defined on a chain or a cylinder (see below). We compute its spectral properties for an exactly solvable model,
which we call the ``non-interacting $cf$ model''. It features a dispersive $c$-band hybridizing locally with a flat $f$-band, 
\begin{align}
\label{eq:H_fc}
H = \sum_{\langle \br \br'\rangle} t_{\br \br'} c^{\dag}_{\br} c^{\pdag}_{\br'} + V \sum_{\br} (f^{\dag}_\br c^{\pdag}_\br + c^{\dag}_\br f^{\pdag}_\br) \, .
\end{align}
Here, $c^{\pdag}_\br$ and $f^{\pdag}_\br$ annihilate spinless fermions at site $\br$ in the dispersive and flat bands, respectively, and 
$t_{\br \br'}$ is the nearest-neighbor $c$-band hopping amplitude.  We
take  $t_{\br \br'} = -1$ for nearest neighbors and zero otherwise, and $V=0.5$. We consider the
case of half-filling, where the $cf$-model~\eqref{eq:H_fc} describes a band insulator. The $cf$-hybridization provides a source of scattering for the $c$-electrons and hence an intrinsic decay mechanism for the current carried by them. 

(Note in passing: The periodic Anderson model relevant for heavy-fermion physics is a spinful generalization  of this model that also involves strong $f$-band interactions. For present purposes, we consider the non-interacting case since then exact benchmarks are available, and we consider the spinless case to reduce computational costs.)

We study two different geometries: a one-dimensional chain of 
$\eLL=60$ sites (writing $\br = \ell)$, and a
cylinder of $\eLL_x \times \eLL_y = 10 \times 4$
sites (writing $\br = (x,y))$. We define the corresponding $c$-electron current operators as%
\begin{subequations}
\begin{align}
\label{eq:current-1D}
J^{x}_\mr{chain} & = -\mr{i} \sum_{\ell = 1}^{\eLL-1} (c^{\dag}_{\ell} c^{\pdag}_{\ell+1} - c^{\dag}_{\ell+1} c^{\pdag}_{\ell}) \, ,
\\
J^{x}_\mr{cyl} & = 
-\mr{i} \sum_{x < \seLL_x} \big(c^{\dag}_{(x,y)} c^{\pdag}_{(x+1,y)} - c^{\dag}_{(x+1,y)} c^{\pdag}_{(x,y)}\big) \, .
\end{align}
\end{subequations}
They are non-local operators, in that their terms create and destroy particles on two different sites: nearest neighbors in the 1D chain, or sites separated by $\eLL_y$ bonds along the MPS snaking around the cylinder.

Due to the non-local nature of $J^x$, 
the tangent-space projection of $J^{x} \ket{\Psig}$ is not exact,
\begin{align}
\ket{\Psi_0} = \mc{P}^{1\perp} J^{x} \ket{\Psig} \neq J^{x} \ket{\Psig} \, .
\end{align}
To benchmark how \task\ performs in such situations, we will compare \task\ results to exact ones for two quantities, current susceptibilities and current moments, that are 
both derived from the current correlator $S[J^x](\omega)$. 

We define the current susceptibility $\chi[J^x](\omega) = \chi'[J^{x}](\omega) - \mr{i}\pi \chi''[J^{x}](\omega)$ through
\begin{subequations}
\label{subeqs:chiJx}
\begin{align}
\chi''[J^{x}](\omega) &= S[J^{x}](\omega) - S[J^{x}](-\omega) \, ,
\\
\chi'[J^{x}](\omega) &= \mc{P}\int_{-\infty}^{\infty} \mr{d}\omega' \frac{\chi''(\omega')}{\omega - \omega'} \, .
\end{align}
\end{subequations}
To compute it, we use either \task\ or an exact treatment to obtain a discrete 
spectral representation of $S[J^x](\omega)$, then CFE-broaden the latter using $\sigma = 0.1$ and $\mathcal{N}_{\mr{CFE}}  = 2$, and subsequently compute $\chi[J^{x}](\omega)$ via \Eqs{subeqs:chiJx}.  
 
Current moments are defined as 
\label{subeq:current_sum_rules}
\begin{align}
\mu_n[J^x] &= \int_{-\infty}^{\infty} \mr{d}\omega \, \omega^n \, S[J^{x}](\omega) \, ,
\end{align}
and we denote the relative difference between \task\ and exact results by $\delta \mu_n = | \mu_n^\mr{TaSK}/\mu_n^\mr{exact} -1|$. We will focus on two moments: $\mu_0$, giving the total weight of the current spectrum, and $\mu_{-1}$, relevant for computing optical conductivities. 
\task\ approximates the total weight of the current spectrum, $\mu_0[J^x] = \bra{\Psig} J^{x} J^{x} \ket{\Psig}$, by 
\begin{align}
\mu_0[J^x] 
\overset{\mr{\task}}{\simeq}
 \bra{\Psig} J^{x} \mc{P}^{1\perp} J^{x} \ket{\Psig} \, .
\end{align}
Therefore, the 
difference $\delta \mu_0$ is a measure of how well the initial state $J^{x} \ket{\Psig}$ is approximated by the closest state in the tangent space. $\mu_0$ is not sensitive to the positions and relative weights of spectral peaks. By contrast, $\mu_{-1}$ \textit{is},  due to the factor $\omega^{-1}$. The latter implies that  
an accurate determination of $\mu_{-1}$
requires an accurate resolution of the current spectrum $S[J^x]$. 

\begin{figure}[!t]
\includegraphics[width=\columnwidth]{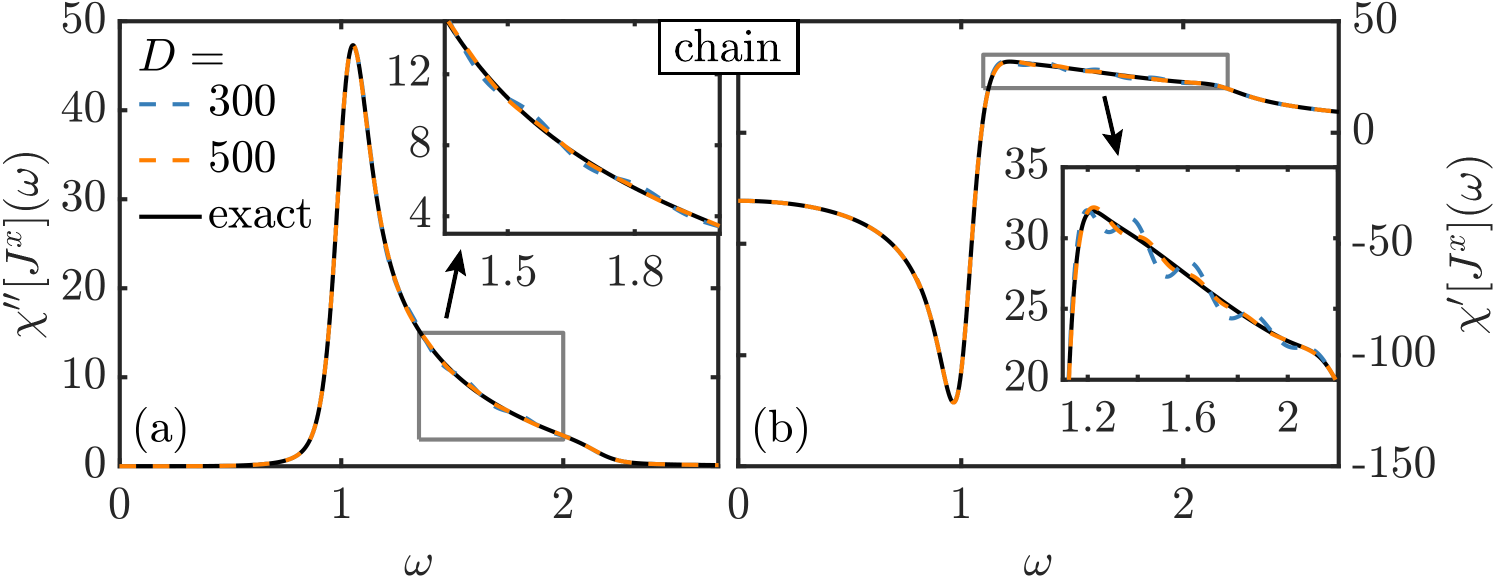} 
 \includegraphics[width=\columnwidth]{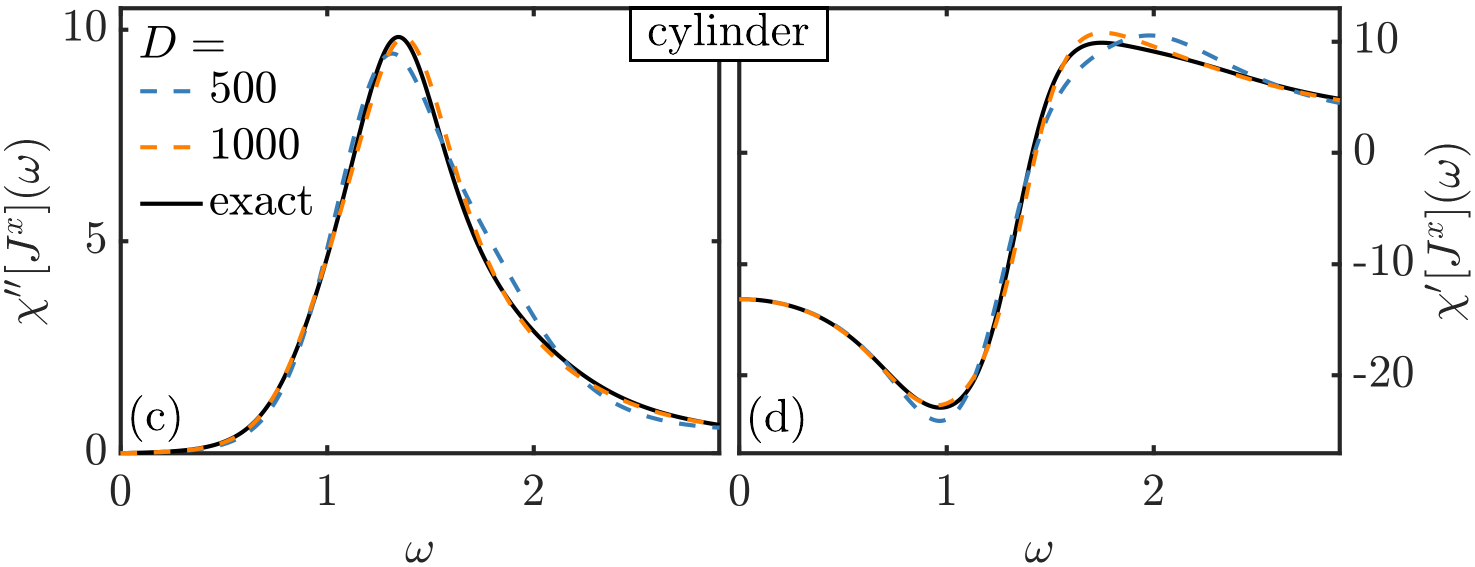} \vspace{-7mm}
 \caption{
 \label{fig:PAM_1d_J}
 \changed{Current susceptibility 
 $\chi[J^x](\omega)$ of 
 the non-interacting $cf$-model \eqref{eq:H_fc}, 
 for (a,b) a chain of $60$ sites, 
 and (c,d) a cylinder of $10 \times 4$ sites.
 (a,c) Spectral part  $\chi''$;  (b,d) real part $\chi'$. In each case, we compare \task\ results computed for two bond dimensions $D$ (blue, orange) to exact results (black). 
 Differences between \task\ and exact results decrease with increasing $D$. The insets to (a,b)
 illustrate this by zooming in on the intermediate-frequency regime $1.3<\omega< 2$. }}
 \end{figure} 
 
\begin{table}[!t]
\setlength{\tabcolsep}{7pt}
\begin{ruledtabular}
\begin{tabular}{rccc}
 chain 
& $D = 300$ 
& $D = 500$ 
& \mr{exact} \\[1mm] 
  $\mu_{0}[J^x]$ 
& 18.644677332 
& 18.644677227 
& 18.644677222 \\
  $\delta\mu_{0}[J^x]$ 
& $5.9 \times 10^{-9}$ 
& $2.4 \times 10^{-10}$ 
& --- \\[1mm]
  $\mu_{-1}[J^x]$ 
& 15.336113930 
& 15.336113832 
& 15.336113828 \\
  $\delta\mu_{-1}[J^x]$ 
& $6.6 \times 10^{-9}$ 
& $2.5 \times 10^{-10}$ 
& ---
\\[3mm]
cylinder 
& $D = 500$ 
& $D = 1000$ 
& \mr{exact} \\[1mm]
 $\mu_{0}[J^x]$ 
& 9.2752 
& 9.2116 
& 9.1885 \\
  $\delta\mu_{0}[J^x]$ 
& $9.4 \times 10^{-3}$ 
& $2.5 \times 10^{-3}$
& --- \\[1mm]
  $\mu_{-1}[J^x]$ 
& 6.5597 
& 6.5526 
& 6.5457 \\
  $\delta\mu_{-1}[J^x]$ 
& $2.1 \times 10^{-3}$
& $1.1 \times 10^{-3}$
& ---
\end{tabular}
\end{ruledtabular}
\caption{\label{tab:PAM1d_moments}
\changed{Moments $\mu_n[J^x]$ [Eq.~\eqref{subeq:current_sum_rules}] of the current spectrum of the $cf$-model \eqref{eq:H_fc}, 
for (top) a chain of $60$ sites, 
 and (bottom) a cylinder of $10 \times 4$ sites, comparing \task\ results for two
 bond dimensions to exact ones. $\delta \mu_n[J^x]$ denotes their relative difference.}} 
\end{table}

We have used \task\ with $\Nkr = 100$ to compute $S[J^{x}]$ for 
the 1D chain using bond dimensions  $D \in \{300,500\}$, and for the cylinder using $D \in \{500,1000\}$. 

Figure~\ref{fig:PAM_1d_J}
shows the corresponding results for $\chi$, 
comparing the spectral ($\chi''$) and  real ($\chi'$) parts from \task\ (dashed lines) to exact results (solid lines). For both  geometries, the agreement is good, both qualitatively (\task\  
does not yield spurious features not present in the exact result) and quantitatively: though some differences are visible at intermediate frequencies [magnified in insets of \Figs{fig:PAM_1d_J}(a,b)], these decrease with increasing $D$, as expected.

Table~\ref{tab:PAM1d_moments} compares
\task\ and exact results for the  moments $\mu_0[J^{x}]$ and $\mu_{-1}[J^{x}]$.
The agreement is remarkable for the chain 
($\delta \mu_n \sim \mathcal{O}(10^{-10})$ for $D=500$) 
and very good for the cylinder
($\delta \mu_n \sim \mathcal{O}(10^{-3})$ for $D=1000$). The accurate determination of $\mu_{-1}[J^{x}]$, in particular, demonstrates 
that \task\ can reliably compute conductivities on cylinders. Moreover, we note that $D = 1000$ is not very large for a 2-orbital system on a cylinder; errors can be expected to further systematically decrease when the bond dimension is increased further.
 
To summarize: we conclude that the 1-site tangent-space approximation works very well for computing the spectrum of the current operator
despite its non-local nature. Evidently, the 1-site tangent space is large enough to accurately capture its spectral properties --- even for a cylinder geometry and a gapped spectrum, two further potentially challenging features.}

\section{Time-domain comparison to TDVP}
\label{sec:Comparison_to_TDVP}

\changed{A standard approach for computing dynamical correlation functions is to compute them in the time domain and then Fourier-transform to the frequency domain. A leading MPS method for this purpose is the the time-dependent variational principle~(TDVP)
\cite{Haegeman2011,Haegeman2016,Paeckel2019,Li2024}. 
In this section, we benchmark \task\ against TDVP both in terms of accuracy and computational run-time. To this end, we Fourier-transform \task\ results for the 64-site spin-$\tfrac{1}{2}$ Heisenberg chain from the frequency domain to the time domain and compare them against TDVP results. We find excellent agreement while requiring significantly lower
run-times.}

We consider the time-dependent correlation function
\begin{align}
\label{eq:S_pi_t}
    S_{k}(t) = \langle \Psig | [\mathrm{S}^z_{k}]^{\dag} e^{-\mr{i} (H - \Eg ) t} \mathrm{S}^z_{k} | \Psig \rangle \, ,
\end{align}
where $ \mathrm{S}^z_{k} = \eLL^{-1/2} 
\sum_{\ell = 1}^{\, \scripteLL} e^{-\mr{i} k\ell} \mathrm{S}^z_{\ell}$, and $\mathrm{S}^z_{\ell}$ is the local spin-$z$ operator on site $\ell$. We compute $S_{k}(t)$ using two different approaches. 
First, we time-evolve
\begin{align}
\label{eq:sm_SzPsi_t}
    |\Psi(t) \rangle &= e^{-\mr{i} (H - \Eg ) t} \, 
    \mathrm{S}^z_{k} | \Psig \rangle \, ,
    \\
    S_{k}(t) &= \langle \Psig |[\mathrm{S}^z_{k}]^{\dag}|\Psi(t) \rangle
\end{align}
using single-site TDVP with controlled bond expansion~\cite{Li2024}.
We note that it is, in principle, somewhat more efficient to run both a forward and a backward time evolution and compute $S_{k}(t) = \langle \Psi (-t/2) |\Psi(t/2) \rangle$.
For the accuracy comparison done here, this is, however, not important. We use a third-order integrator (error of $\mc{O}(\delta t^4)$) and time step size $\delta t = 0.1$. 
We take 250 time steps, such that the largest time is 25. 
Truncation is done via a singular-value-decomposition threshold $\epsilon_{\mr{svd}} = 10^{-8}$. The largest bond dimension encountered with this threshold is $D^{\ast} = 502$ SU(2) multiplets, similar to $D^{\ast} = 512$ in \task. This TDVP result serves as a benchmark.

Second, we Fourier-transform our TaSK data to the time domain,
\begin{align}
\label{eq:sm_Spi_t_task}
    S_{k}(t) = \sum_\alpha |\langle \Psig 
    |[\mathrm{S}^z_{k}]^{\dag} |\Psi^{1\perp}_\alpha \rangle|^2 e^{-\mr{i} \omega_\alpha t} \, ,
\end{align}
where $\bigl (\omega_{\alpha},|\Psi^{1\perp}_\alpha\rangle\bigr )$ are eigenpairs of the Krylov Hamiltonian $H^{1\perp}$, computed via TaSK.
Equation~\eqref{eq:sm_Spi_t_task} is equivalent to approximating Eq.~\eqref{eq:sm_SzPsi_t} as 
\begin{align}
\label{eq:sm_SzPsi_t_task}
    |\Psi(t) \rangle &\simeq e^{-\mr{i} H^{1\perp} \!t} \, 
    \mathrm{S}^z_{k} | \Psig \rangle \, ,
\end{align}
and computing the exponential containing $H^{1\perp}$ using a global Lanczos scheme.
In contrast to TDVP, this global Lanczos scheme does not involve projector splitting or local dynamical adjustments of the tangent space used for the exponentiation. As a consequence, Eq.~\eqref{eq:sm_SzPsi_t_task} \textit{does not rely on small time steps}. This implies a substantial reduction of computational time relative to TDVP and is one of the main reasons why TaSK is computationally more efficient than TDVP for computing dynamical correlators.

\begin{figure}[bt!]
 \includegraphics[width=\linewidth]{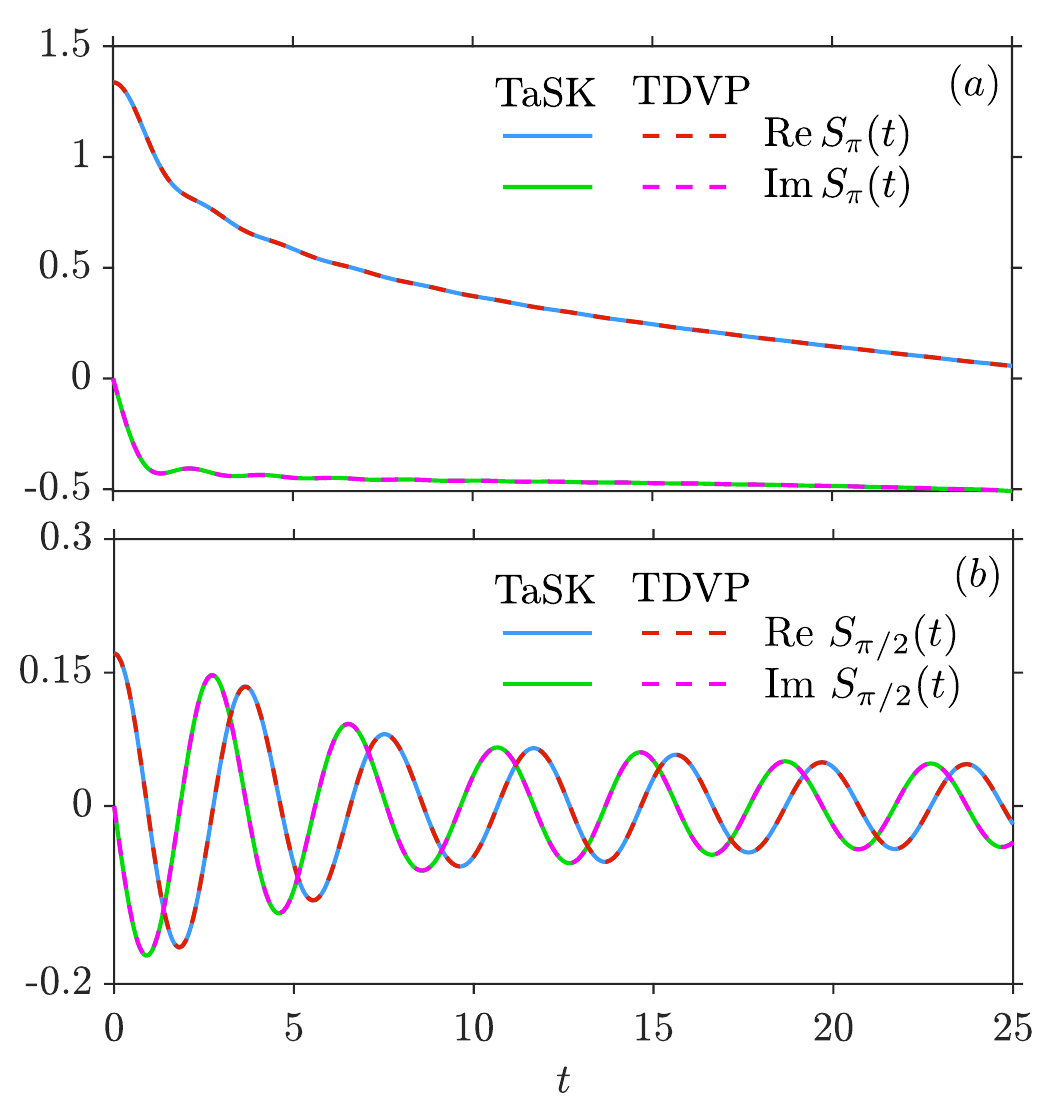}
 \vspace{-7mm}
 \caption{
 \label{fig:TDVP_vs_TaSK_Heisenberg}
Comparison of the time-dependent correlation function of a spin-$\tfrac{1}{2}$ Heisenberg chain ($\eLL = 64$), computed via TDVP (dashed) and \task\ (solid) \changed{for (a) $k=\pi$ and (b) $k=\pi/2$}.
We use a singular value threshold of $\epsilon_{\mr{svd}} = 10^{-8}$ for the TDVP, resulting in a bond dimension of at most $D^{\ast} = 502$. For \task\ computations, we set $\Nkr = 90$.}
\vspace{-4mm}
 \end{figure}

\changed{We have checked the consequences of the TaSK simplification of 
replacing \Eq{eq:sm_SzPsi_t} by \eqref{eq:sm_SzPsi_t_task} 
for two time-dependent correlators of the Heisenberg model. Figures~\ref{fig:TDVP_vs_TaSK_Heisenberg}(a,b) 
show  $S_\pi(t)$ and $S_{\pi/2}(t)$, respectively, computed} with either TaSK plus Fourier transformation (solid lines) or TDVP (dashed lines). Remarkably, said simplification has essentially no noticeable effect: the two results
align perfectly, and the deviation (not visible in the figure) is within the expected Trotter error of TDVP.
This handshake between \task\ and TDVP, whose comfort zones are low energies or short times, respectively, is very satisfying: it shows that \task\ not only captures the low-energy features of the spectrum accurately, but also describes well the high-energy features governing the short-time dynamics. 

\changed{However, the computational costs of the \task\ and TDVP approaches differ significantly. Since costs depend on detailed settings, we exemplify this for a specific case, namely the computations used to obtain the data in \ref{fig:TDVP_vs_TaSK_Heisenberg}(b) for $S_{\pi/2}(t)$. To compare the efficiency of the two algorithms, we measured the wall time (on a \changed{MacBook Pro with an Apple M1 Pro processor and 16 GB of unified memory}) of the most expensive computational step of each scheme,
averaged over all iterations. For TDVP, one third-order Trotter time step involving three half-sweeps  required $(36.8+41+36.8)\,s = 114$\,s on average, resulting in a total runtime for 250 time steps of $(250 \times 114\, \text{s}) \simeq 8$ hours. For \task, the most expensive step is the application of the Hamiltonian $\ket{\Psi_{i+1}} = H^{1\perp} \ket{\Psi_i}$ inside the Lanczos loop (see Sec.~\ref{sec:applyHamiltonian}), which required 57\,s on average. To achieve the same accuracy as TDVP we needed $\Nkr = 90$ \task \ iterations, resulting in a total runtime of 
$(90 \times 57\, \text{s}) \simeq 1.4$\,hours. Although the elementary 
steps of \task\ and TDVP, 
one Krylov step vs.\ one half-sweep, 
have runtimes similar within a factor of 2, \task\ requires much fewer elementary steps than TDVP, leading to an approximate factor-of-five overall speedup compared to TDVP.} 
 
\changed{For the model considered here, this speedup does not come at the expense of accuracy, as demonstrated in \Fig{fig:TDVP_vs_TaSK_Heisenberg}. More generally, we believe that 
for generic models (excluding situations discussed in \Sec{app:TaSK-limitations}), obtaining accurate  \textit{low-frequency} results will be easier/less costly with TaSK than with TDVP, where entanglement growth for large times
prevents access to very low frequencies.

To be fair, we note that this entanglement growth problem can be somewhat ameliorated using TDVP on a \underline{c}omplex time contour (cTDVP) \cite{Grundner2024,Paeckel2025}. A comparison of this approach to TaSK, in the spirit of \Fig{fig:TDVP_vs_TaSK_Heisenberg}, would be interesting but lies beyond the scope of this work. However, an indirect comparison of the low-frequency performance of cTDVP vs.\ \task\ is available 
for the specific case of the symmetric single-impurity Anderson model. Its dynamical properties 
can be computed using the numerical renormalization 
group (NRG) \cite{Bulla2008,Weichselbaum2012a}, which for low frequencies yields reliable, benchmark-quality results. Such results can also be 
obtained via \task, adapted to impurity models \cite{Picoli2026}. This model was recently used
in two publications, one benchmarking 
\mbox{cTDVP} vs.\ NRG \cite{Grundner2024}, the other TaSK vs.\ NRG \cite{Zacinskis2026}. Both benchmarks used the same model parameters (semicircular density of states, bandwidth $D=1$, interaction $U/D=2$) and had the same goal: 
verifying the zero-temperature Fermi-liquid behavior 
$\Sigma''(\omega) \sim \omega^2$ of the imaginary part of the self-energy down to ever lower frequencies. We can thus obtain a cTDVP vs.\ \task\ comparison by inspecting the published numerical benchmark results. These show
clean $\omega^2$-behavior down to 
$\omega/D$ as small as $5 \times 10^{-3}$ for 
cTDVP, $5 \times 10^{-4}$ for NRG, and 
$2 \times 10^{-5}$ for \task, see \cite{Grundner2024} Fig.~3(b), 
\cite{Zacinskis2026} Fig.~1(b), and 
\cite{Zacinskis2026} Fig.~1(a), respectively. These results were obtained with maximum 
multiplet dimensions of 
$D^{\ast} = 1500$ U(1)$\times$SU(2) multiplets for cTDVP, $D^{\ast}  = 4000$ SU(2)$\times$SU(2) multiplets for NRG, and $D^{\ast} = 198$ 
SU(2)$\times$U(2) multiplets for \task.
While the cTDVP results constitute a remarkable improvement compared to previous purely-real-time TDVP results, \task\ reaches more than two orders of magnitude lower in frequency while using very much smaller bond dimensions than both cTDVP and NRG.}

\medskip

\section{Second–order self-energy for the Hubbard model with density-assisted hopping}
\label{sec:si_selfenergy2_dah}

Sections~\ref{sec:si_selfenergy2_dah} to \ref{sec:SI_PH_asym_DAH} are devoted to the $\tUn$ Hubbard model---the nearest–neighbor Hubbard model augmented by a DAH term. They provide analytical details elucidating the origin of particle-hole asymmetry of our \task\ results for spectral quantities reported in the main text. 

In this section, we compute the self-energy of the 
$\tUn$ Hubbard model up to second order in the interaction, i.e.\ $\mc{O}(U^2,U t_n, t_n^2)$. 
The purpose of this analytical exercise is to gain an understanding of what kind of leading-order scattering processes lead to particle-hole asymmetry in the Hubbard model with DAH.
Readers interested only in the numerical procedure can skip this section.

We consider the  $\tUn$ Hamiltonian  
$H=H_t+H_U+H_n$,
\begin{subequations}
\label{eq:si_model}
\begin{align}
H_t&=-t\sum_{\langle \br \br' \rangle,\sigma}\!\left(c^\dagger_{\br \sigma}c_{\br' \sigma}+\text{h.c.}\right), \label{eq:si_model_t}\\
H_U&=U\sum_\br n_{\br \uparrow}n_{\br \downarrow}, \label{eq:si_model_U}\\
H_n&=-t_n\sum_{\langle \br \br' \rangle,\sigma}\!\left(c^\dagger_{\br \sigma}c_{\br' \sigma}+\text{h.c.}\right)\!\left(n_{\br \bar\sigma}+n_{\br' \bar\sigma}\right). \label{eq:si_model_n}
\end{align}
\end{subequations}
We choose the chemical potential $\mu$ such that the ground state is 
half-filled and describes a paramagnetic state. 
Specifically, half-filling implies $\langle n_{\bar{\sigma}} \rangle = 1/2$ and, if $t_n$ is treated within the Hartree-Fock approximation, $\mu = U/2$.

\subsection{Hartree-Fock treatment}
\label{subsec:HartreeFock}

We begin our analysis by showing that a simple Hartree-Fock (HF) treatment
is insufficient to yield particle-hole asymmetric features.
The HF self-energy has the form 
\begin{align}
\Sigma_{\kk}^{\mr{HF}} &= U \langle n_{\bar{\sigma}} \rangle - 4 \langle n_{\bar{\sigma}} \rangle t_n \gamma_{\kk}
\\
\gamma_{\kk} &= \cos k_x + \cos k_y \, .
\end{align}
At half-filling,  the HF propagator has the form
\begin{align}
G^{\mr{HF}}_{\kk}(\omega) &= \frac{1}{\omega^{+} + \mu - \epsilon_{\kk} - \Sigma_{\kk}^{\mr{HF}}} \, , 
\\ \nonumber
&= \frac{1}{\omega^{+} - \epsilon_{\kk} - 2 t_n \gamma_{\kk}} \, ,
\end{align}
with $\omega^+ \!=\! \omega + \mr{i} 0^+$ and free dispersion $\epsilon_{\kk} = -2 t \gamma_{\kk}$. $G^{\mr{HF}}_{\kk}(\omega)$
is particle-hole symmetric, with an effective nearest-neighbor single-particle hopping amplitude and dispersion
\begin{align}
t_{\mr{eff}} = t + t_n \, , \quad \epsilon^{\mr{eff}}_{\kk} = -2 t_{\mr{eff}} \gamma_{\kk} \, .
\end{align}
Therefore, the significant particle-hole 
\textit{asymmetries} cannot be explained by treating the DAH on the HF level. Even if the self-energy is computed to all orders in $U$, 
a HF treatment of DAH would yield a particle-hole symmetric spectrum. 
Understanding the particle-hole asymmetry requires at least second-order perturbation theory, which we outline below.

\subsection{Momentum space interaction vertices}
Fourier transforming the interaction terms 
$H_U$ and $H_{n}$ in \Eq{eq:si_model} yields
\begin{subequations}%
\begin{align}%
H_x & =\frac{1}{N}\sum_{\kk,\bq,\bp,\sigma}
\Vkq{x} \;
c^\dagger_{\kk+\bq \sigma}c^{\pdag}_{\kk\sigma}\;
c^\dagger_{\bp-\bq \bar\sigma}c^{\pdag}_{\bp\bar\sigma} \, , 
\\
\Vkq{U}& =U \,  , 
\quad 
\Vkq{n}  =-2 t_n\,[\gamma_{\kk}+\gamma_{\kk+\bq}]\, . 
\label{eq:si_vertex_n}
\end{align}%
\end{subequations}%
Hence, the total vertex entering second-order diagrams is
\begin{equation}
\Vkq{}=\Vkq{U}+\Vkq{n}
=U-2 t_n\,[\gamma_{\kk}+\gamma_{\kk+\bq}] .
\label{eq:si_vertex_total}
\end{equation}

\subsection{Second-order self-energy}
We now compute the second-order self-energy on top of the particle-hole symmetric HF solution. 
It is given by the usual ``sunset'' diagram, which describes an electron scattered by opposite-spin density fluctuations.
On the Matsubara axis, this gives 
\begin{align}
\Sigma^{(2)}_\kk(\mr{i}\omega_m) & =\frac{1}{\beta N}\sum_{n \bq}
\big[\Vkq{}\big]^2
\,G^{\mr{HF}}_{\kk+\bq}(\mr{i}\omega_m+\mr{i}\Omega_n)\,
\Pi^{\mr{HF}}_{\bq}(\mr{i}\Omega_n)  ,
\nonumber 
\\ 
\Pi^{\mr{HF}}_{\bq}(\mr{i}\Omega_n) & =-\frac{1}{\beta N}%
\sum_{n \bq} 
 G^{\mr{HF}}_{\bp}(\mr{i}\nu_n)\,G^{\mr{HF}}_{\bp+\bq}(\mr{i}\nu_n+\mr{i}\Omega_n)   .
\label{eq:si_bubble_matsu}
\end{align}
where $\Pi^\mr{HF}_\bq$ is a particle-hole bubble.
Analytic continuation, $\mr{i} \omega_m \to \omega^{+}$, yields the retarded real-frequency self-energy
\begin{align}
\label{eq:si_sigma_realaxis}
\Sigma^{(2)}_\kk(\omega) & =\frac{4 \pi^2}{N}\sum_{\bq}\big[\Vkq{}\big]^2
\int_{-\infty}^{\infty}\!\frac{d\Omega}{2\pi}\!\int_{-\infty}^{\infty}\!\frac{d\omega'}{2\pi} 
\\ & \qquad  
\frac{A^{\mr{HF}}_{\kk+\bq}(\omega')\,B^{\mr{HF}}_{\bq}(\Omega)\,[n_B(\Omega)+f(\omega')]}
{\omega^{+}-\omega'-\Omega}, 
\nonumber
\end{align}
with Fermi-Dirac and Bose-Einstein distributions $f(x)$ and $n_B(x)$, respectively, and HF spectral functions 
\begin{subequations}%
\begin{align}%
A^{\mr{HF}}_{\kk}(\omega) &= \delta(\omega-\epsilon^{\mr{eff}}_{\kk}) \, ,
\\
B^{\mr{HF}}_{\bq}(\Omega) & =-\frac{1}{\pi}\,\Im\Pi^{\mr{HF}}_{\bq}(\Omega) 
\\ \nonumber 
& = \frac{1}{N}\sum_{\bp}
\big[f(\epsilon^{\mr{eff}}_{\bp})-f(\epsilon^{\mr{eff}}_{\bp+\bq})\big]
\delta\big(\Omega+\epsilon^{\mr{eff}}_{\bp}-\epsilon^{\mr{eff}}_{\bp+\bq}\big) \, .
\end{align}%
\end{subequations}%
The spectral part of \Eq{eq:si_sigma_realaxis} is given by
\begin{flalign}
\label{eq:si_ImSigma}
& -\Im\Sigma^{(2)}_\kk(\omega)  =
\frac{\pi}{N^2}\sum_{\bq,\bp}\big[\Vkq{}\big]^2
\big[f(\epsilon^{\mr{eff}}_{\bp})-f(\epsilon^{\mr{eff}}_{\bp+\bq})\big]  \hspace{-1cm} & \\
& \nonumber 
\; \times \big[n_B(\epsilon^{\mr{eff}}_{\bp+\bq}\!-\epsilon^{\mr{eff}}_{\bp})
+f(\epsilon^{\mr{eff}}_{\kk+\bq})\big] 
\delta \big(\omega -\epsilon^{\mr{eff}}_{\kk+\bq}\!-\epsilon^{\mr{eff}}_{\bp+\bq}\!+\epsilon^{\mr{eff}}_{\bp}\big) .
\hspace{-1cm} & 
\end{flalign}

\subsection{Decomposition into $U^2$, $U t_n$, and $t_n^2$, and origin of leading-order particle-hole asymmetries}
To understand the origin of particle-hole asymmetries, we express the squared vertex \Eq{eq:si_vertex_total}
as%
\begin{subequations}%
\label{eq:si_vertex_expand}%
\begin{align}
\bigl[\Vkq{}\bigr]^2 &= \bigl[\Vkq{U}\bigr]^2 
+ 2 \Vkq{U} \Vkq{n}+  \bigl[\Vkq{n}\bigr]^2, \\
\bigl[\Vkq{U}\bigr]^2 &= U^2 ,
\\
2 \Vkq{U} \Vkq{n} &= -4U t_n\,(\gamma_{\kk}+\gamma_{\kk+\bq}) , 
\\
\bigl[\Vkq{n}\bigr]^2 &= 4 t_n^2\,(\gamma_{\kk}+\gamma_{\kk+\bq})^2.
\end{align}%
\end{subequations}%
This induces a decomposition of the 2nd-order self-energy,
\begin{align}
\Sigma^{(2)}_\kk(\omega)=
\Sigma^{U^2}_\kk(\omega) + \Sigma^{U t_n}_\kk(\omega) + \Sigma^{t_n^2}_\kk(\omega) \, ,
\label{eq:si_sigma_split}
\end{align}
obtained by  replacing the square of the full vertex $\bigl[\Vkq{}\bigr]^2$ in Eq.~\eqref{eq:si_sigma_realaxis} by
$\bigl[\Vkq{U}\bigr]^2$, $2 \Vkq{U} \Vkq{n}$ or $\bigl[\Vkq{n}\bigr]^2$, respectively. Using  $\gamma_{\kk + \boldsymbol{\Pi}} = -\gamma_{\kk}$, where  $\boldsymbol{\Pi} = (\pi,\pi)\jvdx{^T}$,
one may verify that both $\Sigma^{U^2}_\kk(\omega)$ and $\Sigma^{t_n^2}_\kk(\omega)$ are particle-hole symmetric, 
\begin{align}
\Sigma^{U^2}_\kk(\omega) = \Sigma^{U^2}_{\kk + \boldsymbol{\Pi}}(-\omega) \, , \quad \Sigma^{t_n^2}_\kk(\omega) = \Sigma^{t_n^2}_{\kk + \boldsymbol{\Pi}}(-\omega) \, ,
\end{align}
whereas  $\Sigma^{U t_n}_\kk(\omega)$ is \textit{not}.
 
A simpler and more intuitive way to see this is to check how the Hamiltonian Eq.~\eqref{eq:si_model} behaves under particle-hole transformations. 
Both the single-particle hopping $H_t$ and the Hubbard interaction (plus chemical potential) $H_U - U/2 \sum_{i\sigma} n_{i\sigma}$ are invariant under particle-hole transformations.
However, the DAH term transforms as
\begin{align}
H_n \to &-2 t_n\sum_{\langle ij\rangle,\sigma}\!\left(c^\dagger_{i\sigma}c_{j\sigma}+\text{h.c.}\right)
\\ \nonumber
&+t_n\sum_{\langle ij\rangle,\sigma}\!\left(c^\dagger_{i\sigma}c_{j\sigma}+\text{h.c.}\right)\!\left(n_{i\bar\sigma}+n_{j\bar\sigma}\right),
\end{align}
i.e.\ it becomes a single-particle hopping term with amplitude $-2t_n$ plus a DAH term with \textit{opposite} sign. 
This sign flip means that the corresponding vertex flips sign under a particle-hole transformation, $\Vkq{n} \to -\Vkq{n}$. The single-particle hopping term ensures that the effective hopping amplitude in the HF propagator does not change under particle-hole transformations at half-filling:
\begin{align}
t & \to t^{\mr{ph}} = t + 2 t_n , \quad 
t_n  \to t_n^{\mr{ph}} = - t_n , \\
t_{\mr{eff}} & \to  t^{\mr{ph}} + t_n ^{\mr{ph}} = t + t_n  
= t_{\mr{eff}} \, . 
\end{align}
Thus, $G_\bk^\mr{HF}(\omega)$ is particle-hole symmetric, as stated before.

From that discussion, we immediately see that all diagrams (assuming expansion in terms of HF propagators, and not counting the vertices in HF diagrams) containing an even number of $V_n$ vertices
are particle-hole symmetric, while those with an odd number of $V_n$ vertices are particle-hole \textit{anti-}symmetric, for instance 
\begin{align}
\Sigma^{U t_n}_\kk(\omega) = -\Sigma^{U t_n}_{\kk + \boldsymbol{\Pi}}(-\omega) \, .
\end{align}
At second order, the particle-hole asymmetry at half-filling thus arises via an 
``cooperative'' 
$t_n$-$U$ scattering event,
where a particle-hole pair is emitted via $H_n$, followed by re-absorption via $H_U$, or vice versa.

\section{Approximate self-energy correction from density-assisted hopping}
\label{sec:si_DAH_correction}

Guided by Sec.~\ref{sec:si_selfenergy2_dah}, we construct a simple DAH‑induced correction to the self‑energy starting from a known Hubbard‑model solution. 
The idea is to replace one of the two Hubbard vertices in the Hubbard self‑energy by the external‑leg part of the DAH vertex to expose the origin of the particle–hole asymmetry introduced by $t_n$.

In Sec.~\ref{sec:si_selfenergy2_dah}, we have seen that at the lowest order beyond Hartree-Fock, particle-hole asymmetry arises from mixed $U$-$t_n$ scattering from the Hubbard interaction and DAH. 
However, second-order perturbation theory describes a Fermi liquid and is not able to capture the Mott physics of the half-filled Hubbard model. 
To make progress on DAH in the presence of Mott physics, we use the fact that 
(i) a full solution to the 
$\teffU$ Hubbard model, which has single-particle hopping $t_{\mr{eff}} = t+t_n$ but no  DAH, is equivalent to (ii) a solution of the $\tUn$ Hubbard model with DAH treated in the HF approximation, while $U$ is treated non-perturbatively to all orders.

However, approximation (ii) (HF-for-$t_n$ plus all-orders-in-$U$) does not capture particle-hole asymmetry. To go beyond (ii), we return to (i),  but now add
to the self-energy a beyond-HF estimate of the effect of DAH. To this end, we first
note that the 
full $\tUn$ 
self-energy can be expressed in terms of bare incoming and outgoing scattering vertices, contracted with a six-point propagator $R$ [cf.\ Eq.~(19) of Ref.~\onlinecite{Dickhoff2004}],
\begin{flalign}
\Sigma_{\kk}(\omega) &= \Sigma^{\mr{HF}}_{\kk} + V \! \circ R \circ V  &
\\
\nonumber
V \! \circ R \circ V &= \sum_{\bp_1 \bq_1 \bp_2\bq_2} \raisebox{-8.0mm}{\includegraphics[width=0.55\linewidth]{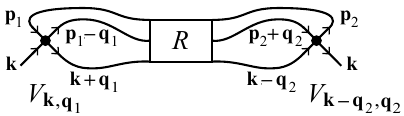}} &
\\ \label{eq:si_Sigma_sixpt}
& =
\sum_{\bp_1 \bq_1 \bp_2\bq_2} \hspace{-3mm} V_{\kk, \bq_1}  
R_{\kk \bp_1  \bp_2, \bq_1 \bq_2}(\omega) V_{\kk-\bq_2,\bq_2}  . \hspace{-1cm} &
\end{flalign}
Here, $ \Vkq{}=\Vkq{U}+\Vkq{n}$ is the bare vertex of \Eq{eq:si_vertex_total}.
We now exploit the fact that we can evaluate Eq.~\eqref{eq:si_Sigma_sixpt} for the 
$\teffU$ Hubbard model using \task.
Its vertex $\Vkq{U}$ does not depend on the internal or external momenta, i.e.
\begin{align}
V^U \! \circ R^\teffU  
\! \circ V^U = U^2  \sum_{\bp_1 \bq_1} \sum_{\bp_2\bq_2} 
R^\teffU_{\kk\bp_1 \bp_2,\bq_1\bq_2}(\omega) \, . 
\end{align}
Further, the DAH vertex $\Vkq{n}
= -2 t_n\,[\gamma_{\kk}+\gamma_{\kk+\bq}]$
has a part that does not depend on the internal momenta,
$\widetilde{V}^n_{\bk,\bq} 
 = -2 t_n\, \gamma_{\kk}$ (and the same is true for $V_{\bk-\bq,\bq}$). We can therefore evaluate the cross terms
involving $\widetilde{V}^n$ and $V^U$,
\begin{align}
\widetilde{V}^n \! \circ R^{\teffU} \circ V^U &=  
V^U \! \circ R^{\teffU} \circ \widetilde{V}^n 
\\ \nonumber
&= -2 \frac{t_n \gamma_{\kk}}{U}  V^U \! \circ R^{\teffU} \circ V^U \, ,
\end{align}
which captures some of the first-order-in-$t_n/U$ corrections. The corresponding correction term
to the self-energy is 
\begin{align}
\label{eq:si_SEcorrection}
\delta \Sigma_{\kk}(\omega) = -4 \frac{t_n \gamma_{\kk}}{U }  (\Sigma^{\teffU}_{\kk}(\omega) - \Sigma^{\mr{HF}}_{\kk}) \, .
\end{align}
This is \Eq{eq:SE_correction_term} of the main text.

It should be acknowledged, though, that this expression is at best
a rough estimate of the beyond-HF effects of DAH. We now discuss two aspects
that it neglects. 

First: There is no strict reason why we can ignore the $\propto \gamma_{\kk + \bq}$ momentum-transfer part of $V_n$. That said, we expect its contribution
to be smaller than that of the $\propto \gamma_\bk$ part, since 
 $R^\teffU$ can be expected to become approximately local deep in a Mott insulator. (This statement is exact for a
Hubbard atom, or for a Mott insulator described by dynamical mean-field theory.) If $R^\teffU$ is approximately local, it depends only weakly on the transfer momenta $\mathbf{q}_1$ and $\mathbf{q}_2$,
such that
\begin{align}
R^\teffU_{\kk\bp_1\bp_2,\bq_1\bq_2}(\omega) &\simeq 
\overline{R}^\teffU_{\kk\bp_1\bp_2}(\omega) = \frac{1}{N^2}\sum_{\mathbf{q}_1 \mathbf{q}_2} R^\teffU_{\kk\bp_1\bp_2,\bq_1\bq_2}(\omega) \, .
\end{align}
As a result,
\begin{align}
\sum_{\bq_1\bq_2} \gamma_{\kk+\bq_1} R^\teffU_{\kk\bp_1\bp_2,\bq_1\bq_2}(\omega) 
 \simeq 
 \sum_{\bq_1} \gamma_{\kk+\bq_1} N  \overline{R}^\teffU_{\kk\bp_1\bp_2}(\omega) = 0 \, ,
\end{align}
since $\sum_{\bq_1} \gamma_{\kk+\bq_1} = 0$. However, since $\Sigma_{\kk}^\teffU(\omega)$ is momentum-dependent, $R^\teffU$ is clearly not completely local, i.e.\ the above argument 
is not quantitative.

Second: There is also no strict reason why we should ignore first-order in $t_n/U$ changes to $R$.
Therefore, Eq.~\eqref{eq:si_SEcorrection} clearly neglects order $t_n/U$ terms.

To summarize: we regard \Eq{eq:si_SEcorrection} as a simplified approximation that is not quantitatively reliable, but which does give some insight into the 
origin of the particle-hole asymmetries we observe in a full, non-perturbative treatment of DAH using \task\  in the main text.

\section{Modification of the Mott pole dispersion due to density-assisted hopping}
\label{sec:SI_PH_asym_DAH}

In this section, we show that the self-energy pole dispersion of a Mott insulator in the presence of DAH contains terms that exceed the nearest-neighbor range of the DAH term itself. We accomplish this using a high-frequency expansion of the Green's function.
We first review this expansion and the result of Ref.~\onlinecite{Wagner2023}, 
which identified the connection between the Mott pole dispersion and the single-particle dispersion for the case of a $\tU$ Hubbard model. Our approach, 
while close in spirit to
that of Ref.~\onlinecite{Wagner2023}, reveals that the pole dispersion can be thought of as the \textit{effective dispersion of a dressed fermion}~\cite{Pangburn2025}.
This dressed fermion is exactly local for a pure Hubbard interaction, and \textit{becomes non-local in the presence of DAH}. 
This non-locality causes the range of the pole dispersion to become larger in the presence of DAH.

Mott insulators have a pole in their self-energies that (in finite dimensions) is generically dispersive. Generally, the Green's function and self-energy can be written as
\begin{align}
G_{\kk}(z) & =  \frac{1}{ z + \mu -\epsilon_{\kk} - \Sigma_{\kk}(z)} \,,
\\
\Sigma_{\kk}(z) & = \Sigma^{\rm HF}_{\kk} + \frac{\Delta_{\kk}}{ z-\xi_{\kk} - R_{\kk}(z)} \,,
\label{eq:SI_Sig_pole_ansatz}
\end{align}
Here, $\epsilon_{\kk} $ is the free dispersion and $\mu$ the chemical potential. The  parameters $\Sigma^{\rm HF}_{\kk}$, $\Delta_{\kk}\ge 0$ and $\xi_{\kk}$ are fixed by equal-time correlators, and the residual function $R_{\kk}(z)$ scales as
 $\mc{O}(z^{-1})$ for large $|z|$. 
$\xi_\bk$ is the ``dispersion of the self-energy pole''. 

Following the standard continued-fraction construction~\cite{Viswanath1994,Pelz2025}, we define the interaction-dressed creation operator
\begin{subequations}
\label{eq:SI_q_def}
\begin{align}
\widetilde q^\dagger_{i\sigma} &= [H,c^\dagger_{i\sigma}] -  \sum_j c^\dagger_{j\sigma} \langle\{c^{\pdag}_{j\sigma},[H,c^\dagger_{i\sigma}]\}\rangle \, ,
\\
\widetilde q^\dagger_{\kk\sigma} &= \frac{1}{\sqrt N}\sum_j \mr e^{-\mr{i}\kk \cdot \mathbf r_j}\,\widetilde q^\dagger_{j\sigma}
\\ \nonumber
&= [H,c^\dagger_{\kk\sigma}]  -  c^\dagger_{\kk\sigma} \langle\{c^{\pdag}_{\kk\sigma},[H,c^\dagger_{\kk\sigma}]\}\rangle \, .
\end{align}
\end{subequations}
By construction $\langle\{c_{\kk\sigma}^{\pdag},\widetilde q^\dagger_{\kk\sigma}\}\rangle=0$, i.e.\ $\widetilde q^\dagger_{\kk\sigma}$ is orthogonal to $c_{\kk\sigma}^{\dag}$ w.r.t.\ the operator scalar product $(A|B) = \langle \{ A^{\dag},B\}\rangle$.
We call operators orthogonal to single-particle operators ``single-particle irreducible'', since they determine the dynamical part of the self-energy~\cite{Kugler2022,Pelz2025}.

The parameters $\Sigma^{\rm HF}_{\kk}$, $\Delta_{\kk}\ge 0$ and $\xi_{\kk}$ are given by
\begin{subequations}
\label{eq:SI_Delta_xi_defs}
\begin{align}
\label{eq:SI_SigmaHF}
\Sigma^{\rm HF}_{\kk} + \epsilon_{\kk} - \mu &= \big\langle\{c^{\pdag}_{\kk\sigma},[H,c^\dagger_{\kk\sigma}]\}\big\rangle
\\
\label{eq:SI_Deltak}
\Delta_{\kk} &= \big\langle\{\widetilde q^{\pdag}_{\kk\sigma},\widetilde q^\dagger_{\kk\sigma}\}\big\rangle \, ,
\\
\label{eq:SI_xik}
\xi_{\kk} \Delta_{\kk} &= \big\langle\{\widetilde q^{\pdag}_{\kk\sigma},[H,\widetilde q^\dagger_{\kk\sigma}]\}\big\rangle \, .
\end{align}
\end{subequations}
 Equations~\eqref{eq:SI_SigmaHF} and~\eqref{eq:SI_xik} can be thought of 
as matrix-elements of the Liouvillian superoperator, defined as $\mc{L} O = [H,O]$, w.r.t.\ $c^\dagger_{\kk\sigma}$ or $\widetilde q^\dagger_{\kk\sigma}$, respectively,
while Eq.~\eqref{eq:SI_Deltak} is the norm of $\widetilde q^\dagger_{\kk\sigma}$. 

The corresponding real-space versions are obtained via Fourier-transform,
\begin{subequations}
\begin{align}
\Delta_{ij} &= \frac{1}{N} \sum_{\kk} \Delta_{\kk} \, \mr{e}^{-\mr{i} \kk (\mathbf{r}_i - \mathbf{r}_j)} = \big\langle\{\widetilde q^{\pdag}_{i\sigma},\widetilde q^\dagger_{j\sigma}\}\big\rangle
\\
[\xi \Delta]_{ij} &= \sum_{\ell} \xi_{i\ell} \Delta_{\ell j} 
= \frac{1}{N} \sum_{\kk} \xi_{\kk}\Delta_{\kk} \, \mr{e}^{-\mr{i} \kk (\mathbf{r}_i - \mathbf{r}_j)}  \nonumber \\
& = \big\langle\{\widetilde q^{\pdag}_{i\sigma},[H,\widetilde q^\dagger_{j\sigma}]\}\big\rangle \, .
\end{align}
\end{subequations}

\subsection{High-frequency expansion and identification of $\Delta_{\kk}$ and $\xi_{\kk}$}

The formulas above can be obtained through a high-frequency expansion of the Green's function,
\begin{align}
G_{\kk}(z) = \frac{1}{z}\sum_{n=0}^{\infty} \frac{M^{(n)}_{\kk}}{z^n} \, .
\end{align}
For our interests, we need this expansion up to and including $n=3$.
\begin{widetext}
Using $E_{\kk}\equiv \varepsilon_{\kk}+\Sigma^{\rm HF}_{\kk}-\mu$, we expand the Green's function, in powers of $1/z$:
\begin{align}
\label{eq:SI_Gz_expand}
G_{\kk}(z)
&=\frac{1}{z - E_{\kk} -\frac{\Delta_{\kk}}{z-\xi_{\kk}-R_{\kk}(z)}}
=\frac{1}{z}\frac{1}{1-\frac{E_{\kk}}{z}-\frac{\Delta_{\kk}}{z^{2}}-\frac{\Delta_{\kk}\xi_{\kk}}{z^{3}}
+\mc O(z^{-4})} \, ,
\\ \nonumber
&=\frac{1}{z}
+\frac{E_{\kk}}{z^{2}}
+\frac{E_{\kk}^{2}+\Delta_{\kk}}{z^{3}}
+\frac{E_{\kk}^{3}+2E_{\kk}\Delta_{\kk}+\Delta_{\kk}\xi_{\kk}}{z^{4}}
+\mc O(z^{-5}) \, .
\end{align}
\end{widetext}
The Green's function is fully determined by $E_{\kk}$, $\Delta_{\kk}$ and $\xi_{\kk}$ up to order $z^{-4}$. 

Alternatively, the high-frequency expansion can also be obtained from equations of motion for the retarded Green's function,
\begin{align}
G_{\kk}(t) &= -\mr{i}\,\theta(t)\,
 \langle\{c^{\pdag}_{\kk\sigma}(t),c^{\dagger}_{\kk\sigma}\}\rangle 
 \\ \nonumber
&=  -\mr{i}\,\theta(t)\,
 \langle\{c^{\pdag}_{\kk\sigma},c^{\dagger}_{\kk\sigma}(-t)\}\rangle \, .
\end{align}
For $\Im \, z>0$, we perform the high-frequency expansion of the integral
\begin{align}
\label{eq:SI_Gt_highz}
G_{\kk}(z)=\int_{0}^{\infty} \! \mr{d}t\,\mr{e}^{\mr{i}zt}G_{\kk}(t)
=\sum_{n=0}^{\infty}\frac{ \partial_t^{\,n} G_{\kk}(t)\!\mid_{t = 0^+} }{(\mr{i}z)^{n+1}} \, ,
\end{align}
which follows from repeated usage of
\begin{align}
\int_{0}^{\infty} \! \mr{d}t\,\mr{e}^{\mr{i}zt} f(t) = \left[\frac{\mr{e}^{\mr{i}zt}}{\mr{i} z} f(t) \right]_{0^+}^{\infty} - \frac{1}{\mr{i}z} \int_{0}^{\infty} \! \mr{d}t\,\mr{e}^{\mr{i}zt} \partial_t f(t) \, .
\end{align}

The expansion coefficients can be written as equal-time correlation functions,
\begin{align}
M^{(n)}_{\kk} &= (-\mr{i})^{n+1} \partial_t^{n} G_{\kk}(t) 
\\ \nonumber
&= (-1)^m \langle\{ \mc{L}^m c^{\pdag}_{\kk}, \mc{L}^{n-m} c^{\dag}_{\kk}\} \rangle \, , \;\; m \leq n \, ,
\end{align}
with the Liouvillian $\mc{L} O = [H,O]$.

Comparing the expansion coefficients from Eqs~\eqref{eq:SI_Gz_expand} and \eqref{eq:SI_Gt_highz}, we find
\begin{subequations}
\label{eq:SI_moments_vs_E_D_xi}
\begin{align}
M^{(0)}_{\kk} & = \langle\{c^{\pdag}_{\kk\sigma},c^{\dag}_{\kk\sigma}\}\rangle = 1 \, ,
\\
M^{(1)}_{\kk} &= \langle\{c^{\pdag}_{\kk\sigma},[H,c^{\dag}_{\kk\sigma}]\}\rangle = \langle\{[c^{\pdag}_{\kk\sigma},H],c^{\dag}_{\kk\sigma}\}\rangle = E_{\kk} \, ,
\\
M^{(2)}_{\kk} &= \langle\{[c^{\pdag}_{\kk\sigma},H],[H,c^{\dag}_{\kk\sigma}]\}\rangle = 
\langle\{\widetilde q^{\pdag}_{\kk\sigma},\widetilde q^\dagger_{\kk\sigma}\}\rangle + E_{\kk}^2
\nonumber
\\ 
&= E_{\kk}^2 + \Delta_{\kk}  \, ,
\\ \nonumber
M^{(3)}_{\kk} &= \langle\{[c^{\pdag}_{\kk\sigma},H],[H,[H,c^{\dag}_{\kk\sigma}]]\}\rangle
\\ \nonumber
&= \langle\{\widetilde{q}^{\pdag}_{\kk\sigma} + E_{\kk} c^{\pdag}_{\kk},[H,\widetilde{q}^{\dag}_{\kk\sigma} + E_{\kk} c^{\dag}_{\kk}]\}\rangle
\\ \nonumber
&= \langle\{\widetilde{q}^{\pdag}_{\kk\sigma},[H,\widetilde{q}^{\dag}_{\kk\sigma}]\}\rangle
+ E_{\kk} \underbrace{\langle\{\widetilde{q}^{\pdag}_{\kk\sigma},[H,c^{\dag}_{\kk}]\}\rangle}_{= \Delta_{\kk}}
\\ \nonumber
&+ E_{\kk} \underbrace{\langle\{c^{\pdag}_{\kk},[H,\widetilde{q}^{\dag}_{\kk\sigma}]\}\rangle}_{= \Delta_{\kk}}
+ E^2_{\kk} \langle\{c^{\pdag}_{\kk},[H,c^{\dag}_{\kk}]\}\rangle
\\ 
&= E_{\kk}^3+2E_{\kk}\Delta_{\kk}+\Delta_{\kk}\xi_{\kk},
\end{align}
\end{subequations}
where we used $\langle\{c_{\kk\sigma}^{\pdag},\widetilde q^\dagger_{\kk\sigma}\}\rangle=0$ to eliminate mixed terms and the relation
$\langle\{A,[H,B]\}\rangle = \langle\{[A,H],B\}\rangle$, which follows from trace cyclicity. Equations~\eqref{eq:SI_Delta_xi_defs} directly follow from Eqs.~\eqref{eq:SI_moments_vs_E_D_xi}.

\subsection{Pure Hubbard model (no DAH): local $\widetilde q$ and nearest-neighbor $\xi$}

Consider now the Hamiltonian
\begin{subequations}
\begin{align}
H &= \sum_\sigma T_\sigma + H_U
\\
H_U &= U\sum_i n_{i\uparrow}n_{i\downarrow} - \mu \sum_{i\sigma} n_{i\sigma}
\\
T_\sigma &=\sum_{ij} t_{ij}\,c^\dagger_{i\sigma}c^{\pdag}_{j\sigma} \, ,
\end{align}
\end{subequations}
with chemical potential $\mu = U/2$, nearest-neighbor hopping $t_{ij}$, and therefore half-filling, $\langle n_{i\sigma} \rangle = \tfrac{1}{2}$ (assuming no ordering). 
Then, we find
\begin{subequations}
\begin{align}
[H,c^\dagger_{i\sigma}] &= c^\dagger_{i\sigma}(U n_{i\bar\sigma}-\mu) + \sum_j t_{ji} c^\dagger_{j\sigma}
\\
\widetilde q^\dagger_{i\sigma} &= U c^\dagger_{i\sigma} \underbrace{\big(n_{i\bar\sigma}-\langle n_{i\bar\sigma}\rangle\big)}_{\equiv \delta n_{i\bar\sigma}}.
\end{align}
\end{subequations}
At half-filling, $\langle n_{i\bar\sigma}\rangle=\tfrac12$ and $(n_{i\bar\sigma}-\tfrac12)^2=\tfrac14$, hence
\begin{equation}
\Delta_{\kk}=\langle\{\widetilde q^{\pdag}_{\kk\sigma},\widetilde q^\dagger_{\kk\sigma}\}\rangle
=U^2\big\langle(n_{i\bar\sigma}-\tfrac12)^2\big\rangle=\frac{U^2}{4}.
\label{eq:SI_Delta_U}
\end{equation}
To obtain $\xi_{\kk}$, we compute
\begin{subequations}
\begin{align}
[H_U,\widetilde q^\dagger_{i\sigma}] &= U\,c^\dagger_{i\sigma}(Un_{i\bar\sigma}-\mu)\delta n_{i\bar{\sigma}}
\\ \nonumber
&= U\,c^\dagger_{i\sigma}[(U - \mu - U\langle n_{i\bar\sigma} \rangle)n_{i\bar\sigma} + \mu \langle n_{i\bar\sigma} \rangle] \, ,
\\
[T_\sigma,\widetilde q^\dagger_{i\sigma}] &= U\sum_j t_{ji}\,c^\dagger_{j\sigma} \delta n_{i\bar{\sigma}} \, ,
\\
[T_{\bar\sigma},\widetilde q^\dagger_{i\sigma}] &= U\,c^\dagger_{i\sigma}\,[T_{\bar\sigma},n_{i\bar\sigma}] = U\,c^\dagger_{i\sigma}\,X_{i\bar{\sigma}} \, ,
\\
X_{i\bar{\sigma}} &= [T_{\bar\sigma},n_{i\bar\sigma}] 
\\ \nonumber
&=\sum_\ell \big(t_{i\ell}\,c^\dagger_{i\bar\sigma}c^{\pdag}_{\ell\bar\sigma}-t_{\ell i}\,c^\dagger_{\ell\bar\sigma}c^{\pdag}_{i\bar\sigma}\big).
\end{align}
\end{subequations}
Assuming particle-hole symmetry, the corresponding contributions are
\begin{subequations}
\label{eq:SI_xi_pieces_U}
\begin{align}
\big\langle\{\widetilde q^{\pdag}_{j\sigma},[H_U,\widetilde q^\dagger_{i\sigma}]\}\big\rangle
&=\frac{U^2}{4} (U - \mu - U\langle n_{i\bar\sigma} \rangle) = 0 \, ,
\\
\label{eq:SI_Hubbard_txi}
\big\langle\{\widetilde q^{\pdag}_{j\sigma},[T_\sigma,\widetilde q^\dagger_{i\sigma}]\}\big\rangle
&=U^2\,t_{ji}\,\big\langle\delta n_{j\bar\sigma}\,\delta n_{i\bar\sigma}\big\rangle,\\
\label{eq:SI_Hubbard_txibar}
\big\langle\{\widetilde q^{\pdag}_{j\sigma},[T_{\bar\sigma},\widetilde q^\dagger_{i\sigma}]\}\big\rangle
&= U^2 \delta_{ij} \langle\,\delta n_{j\bar\sigma}\,X_{i\bar\sigma} \rangle
\\ \nonumber
&+ U^2 \langle c^\dagger_{i\sigma}\,[X_{i\bar\sigma},\delta n_{j\bar\sigma}]\,c^{\pdag}_{j\sigma} \rangle \, .
\end{align}
The commutator in Eq.~\eqref{eq:SI_Hubbard_txibar} evaluates to
\begin{align}
[X_{i\bar\sigma},\delta n_{j\bar\sigma}]
&=\sum_{\ell}\big(t_{\ell i}(\delta_{\ell j}-\delta_{ij})\,c^\dagger_{\ell\bar\sigma}c_{i\bar\sigma}
\\ \nonumber
&\quad - t_{i\ell}(\delta_{ij}-\delta_{\ell j})\,c^\dagger_{i\bar\sigma}c_{\ell\bar\sigma}\big)
\\ \nonumber
&=
\begin{cases}
\displaystyle
t_{ji}\,c^\dagger_{j\bar\sigma}c_{i\bar\sigma}
+ t_{ij}\,c^\dagger_{i\bar\sigma}c_{j\bar\sigma}, & j\neq i,\\[6pt]
\displaystyle
 -\sum\limits_{\ell}\Big(t_{\ell i}\,c^\dagger_{\ell\bar\sigma}c_{i\bar\sigma}
+ t_{i\ell}\,c^\dagger_{i\bar\sigma}c_{\ell\bar\sigma}\Big), & j=i .
\end{cases}
\end{align}
Inserting this in Eq.~\eqref{eq:SI_Hubbard_txibar}, we get
\begin{align}
\big\langle\big\{\widetilde q_{j\sigma},[T_{\bar\sigma},\widetilde q^\dagger_{i\sigma}]\big\}\big\rangle
&=
\\ \nonumber
U^2\,\delta_{ij}\big[&\sum_{\ell}\big(
t_{\ell i}\,\langle \delta n_{i\bar\sigma}\,c^\dagger_{\ell\bar\sigma}c_{i\bar\sigma}\rangle
- t_{i\ell}\,\langle \delta n_{i\bar\sigma}\,c^\dagger_{i\bar\sigma}c_{\ell\bar\sigma}\rangle
\big)
\\ \nonumber
-&\sum\limits_{\ell}\!\big(
t_{\ell i}\,\langle n_{i\sigma}\,c^\dagger_{\ell\bar\sigma}c_{i\bar\sigma}\rangle
+ t_{i\ell}\,\langle n_{i\sigma}\,c^\dagger_{i\bar\sigma}c_{\ell\bar\sigma}\rangle
\big) \big]
\\ \nonumber
+ U^2 (\delta_{ij}-1) \big[& t_{ji}\,\langle c^\dagger_{i\sigma}c_{j\sigma}\,c^\dagger_{j\bar\sigma}c_{i\bar\sigma}\rangle
+ t_{ij}\,\langle c^\dagger_{i\sigma}c_{j\sigma}\,c^\dagger_{i\bar\sigma}c_{j\bar\sigma}\rangle \big] \, .
\end{align}
\end{subequations}

\begin{figure}[bt!]
 \includegraphics[width=\linewidth]{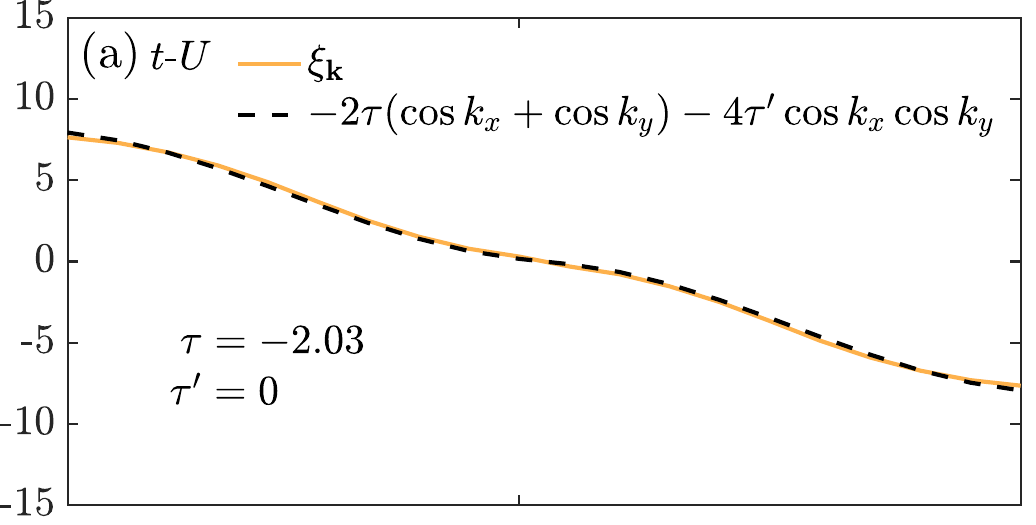}
 \includegraphics[width=\linewidth]{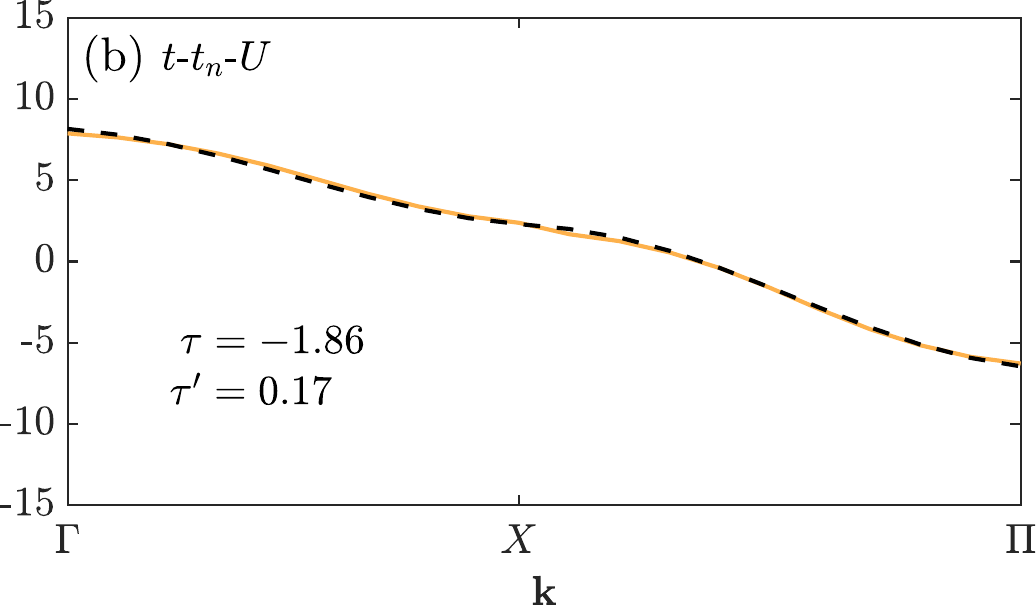}
 \vspace{-6mm}
 \caption{
 \label{fig:xi_fit_t-U}
Self-energy pole dispersion $\xi_{\kk}$ of (a) the $\tU$ and (b) the $\tUn$ model, with parameters as in the main text. The solid orange line is obtained from \task, the black dashed line is a fit. 
}
\vspace{-4mm}
 \end{figure}

Only Eqs.~\eqref{eq:SI_Hubbard_txi} and~\eqref{eq:SI_Hubbard_txibar} 
generate
non-local contributions. Since $\Delta$ in Eq.~\eqref{eq:SI_Delta_U} is $\kk$-independent (local), $[\xi\Delta]_{ij} = \xi_{ij} \Delta$. 
Hence, for $i \neq j$,
\begin{align}
\label{eq:xi_ij_poledispersion}
\xi_{ij} = 4 t_{ij} \Bigl(\langle \delta n_{i\bar{\sigma}} \delta n_{j\bar{\sigma}} \rangle  + \langle c^\dagger_{i\sigma}\,[X_{i\bar\sigma},\delta n_{j\bar\sigma}]\,c^{\pdag}_{j\sigma} \rangle \Bigr) \, ,
\end{align}
so the pole dispersion inherits the range of the bare hopping: $\xi_{ij} \propto t_{ij}$ and vanishes when $t_{ij} = 0$, i.e.\ $\xi_{ij}$ does not exceed the range of the latter.

If the hopping $t_{ij}$ is nearest-neighbor, we therefore expect
\begin{align}
\xi_{\kk} = -2 \tau (\cos k_x + \cos k_y) \, ,
\end{align}
where $\tau$ is given by Eq.~\eqref{eq:xi_ij_poledispersion} for $i,j$ nearest-neighbors. 
We further expect $\tau < 0$ because of antiferromagnetic correlations~\cite{Wagner2023}. 
Figure~\ref{fig:xi_fit_t-U}(a) shows $\xi_{\kk}$ (solid line) extracted from our discrete $\tU$ data via continued fraction expansion, along the same momentum path as that taken in Fig.~\ref{fig:CylinderPlots}(a), 
together with a fit (dashed line) using the ansatz $-2 \tau (\cos k_x + \cos k_y) - 4 \tau' \cos k_x \cos k_y$, with $\tau'$ a possible next-nearest neighbor amplitude. 
We find $\tau = -2.03 < 0$ and, as expected, $\tau' = 0$.

\subsection{Adding density-assisted hopping: nonlocal $\widetilde q$ and emergent next-nearest neighbor term in $\xi_{\kk}$}
\label{subsec:addingDAH}

We now augment the Hubbard model with a DAH term,
\begin{equation}
T_{n\sigma}=\sum_{ij} t^n_{ij}\,c^\dagger_{i\sigma}c^{\pdag}_{j\sigma}\,(n_{i\bar\sigma}+n_{j\bar\sigma}).
\end{equation}
The $\widetilde{q}$ operator acquires
additional contributions from DAH,
\begin{align}
[T_{n\sigma},c^{\dagger}_{i\sigma}] &= \sum_{j} t^n_{ji} c^{\dagger}_{j\sigma} (n_{i\bar{\sigma}} + n_{j\bar{\sigma}}) 
\\
[T_{n\bar{\sigma}},c^{\dagger}_{i\sigma}] &= \sum_{j} t^n_{ji} (c^{\dagger}_{j\bar{\sigma}} c^{\pdag}_{i\bar{\sigma}} + \mr{h.c.}) c^{\dagger}_{i\sigma}
\\
\label{eq:qOp_DAH}
\widetilde{q}^{\dagger}_{i\sigma} &=  U c^{\dagger}_{i\sigma} \delta n_{i\bar{\sigma}}
+ \sum_{j} t^n_{ji} c^{\dagger}_{j\sigma} (\delta n_{i\bar{\sigma}} + \delta n_{j\bar{\sigma}}) 
\\ \nonumber
&+ \sum_{j} t^n_{ji} (c^{\dagger}_{j\bar{\sigma}} c^{\pdag}_{i\bar{\sigma}} - \langle c^{\dagger}_{j\bar{\sigma}} c^{\pdag}_{i\bar{\sigma}} \rangle + \mr{h.c.}) c^{\dagger}_{i\sigma} \, .
\end{align}
Equation~\eqref{eq:qOp_DAH} shows that $\tilde q^\dagger_{i\sigma}$ becomes
\emph{nonlocal} 
in the presence of DAH:
(i) a creation operator $c^\dagger_{j\sigma}$ on site $j$ multiplied by a density operator $\delta n_{i\bar\sigma}$ at site $i$,
and (ii) an intersite spin-exchange term $(c^\dagger_{j\bar\sigma}c_{i\bar\sigma}+\text{h.c.})\,c^\dagger_{i\sigma}$.

The non-local nature of $\widetilde{q}^{\dagger}_{i\sigma}$ results in non-local contributions to $\Delta$ that are of order $t^{n}/U$, 
\begin{align}
\Delta_{ij} = \frac{U^2}{4} (\delta_{ij} + \mc{O}(t^{n}/U)) \, .
\end{align}
To isolate the key effects on the range of $\xi$, we focus on the large local part of $\Delta$ in the discussion below. Taking into account its non-locality would reinforce the conclusions.

Crucially, the non-locality of $\widetilde{q}_{i\sigma}^\dagger$ produces
contributions to $[\xi\Delta]_{ij}$ (and therefore also $\xi_{ij}$) that extend beyond the range of either
$t_{ij}$ or $t^{n}_{ij}$. The spatial range of $[\xi\Delta]_{ij}$ is determined by
(i) the range of $\tilde{q}_{i\sigma}^\dagger$ and (ii) how the Liouvillian
$\mc{L} = [H, \bullet]$ acts to ``move'' fermionic operators--through either
the single-particle hopping or the DAH process, whichever is
longer-ranged (both are nearest-neighbor in the present case).
Consequently, $\xi_{ij}$ acquires terms proportional to
$t^{n}_{i\ell} t_{\ell j}$; unlike in the $\tU$ model,
\textit{the range of $\xi$ exceeds the range of $t$}.

For the $\tUn$ model studied in the main text, where both $t$ and $t_n$ are nearest-neighbor, we expect that $\xi_{ij}$ contains \textit{next}-nearest neighbor terms.
As for the $\tU$ model, we extract the pole dispersion $\xi_{\kk}$ via a continued fraction expansion of the discrete spectral function, and fit it with the ansatz
\begin{align}
\xi_{\kk} = -2 \tau (\cos k_x + \cos k_y) - 4 \tau' \cos k_x \cos k_y \, .
\end{align}
The result of this is shown in Fig.~\ref{fig:xi_fit_t-U}(b), and we find $\tau = -1.86$ and $\tau' = 0.17$, i.e.\ the next-nearest neighbor term $\tau'$ is indeed sizeable.
We have also tried including a next-to-next nearest neighbor term $- 2 \tau'' (\cos 2 k_x + \cos 2 k_y)$ (not shown). 
We find that $\tau'' = 0.05$, while $\tau$ and $\tau'$ are unaffected.

\end{document}